\documentclass[runningheads]{llncs}

\usepackage{multirow}
\usepackage{eccv}

\usepackage{eccvabbrv}
\usepackage{pifont}
\newcommand{\cmark}{\ding{51}}%
\newcommand{\xmark}{\ding{55}}%
\newcommand{\myparagraph}[1]{\smallskip\noindent\textbf{#1}}

\usepackage{graphicx}
\usepackage{booktabs}

\usepackage[accsupp]{axessibility}  

\usepackage{hyperref}

\usepackage{orcidlink}
\usepackage{soul}

\begin{document}

\title{Semi-supervised Segmentation of Histopathology Images with Noise-Aware Topological Consistency} 

\titlerunning{Topological Consistency for Semi-supervised Segmentation}

\author{Meilong Xu\thanks{Email: meixu@cs.stonybrook.edu.}\inst{1} \and Xiaoling Hu\inst{2} \and Saumya Gupta\inst{1} \and Shahira Abousamra\inst{1} \and Chao Chen\inst{1}}

\authorrunning{M.~Xu et al.}

\institute{Stony Brook University, Stony Brook, NY, USA \and 
Athinoula A. Martinos Center for Biomedical Imaging, \\
Massachusetts General Hospital and Harvard Medical School, Boston, MA, USA\\
}

\maketitle

\begin{abstract}
In digital pathology, segmenting densely distributed objects like glands and nuclei is crucial for downstream analysis. Since detailed pixel-wise annotations are very time-consuming, we need semi-supervised segmentation methods that can learn from unlabeled images. Existing semi-supervised methods are often prone to topological errors, \textit{e.g.}, missing or incorrectly merged/separated glands or nuclei. To address this issue, we propose \textit{TopoSemiSeg}, the first semi-supervised method that learns the topological representation from unlabeled histopathology images. 
The major challenge is for unlabeled images; we only have predictions carrying noisy topology.
To this end, we introduce a noise-aware topological consistency loss to align the representations of a teacher and a student model. By decomposing the topology of the prediction into signal topology and noisy topology, we ensure that the models learn the true topological signals and become robust to noise.
Extensive experiments on public histopathology image datasets show the superiority of our method, especially on topology-aware evaluation metrics. Code is available at \url{https://github.com/Melon-Xu/TopoSemiSeg}.
\end{abstract}

\setcounter{footnote}{0}

\section{Introduction}
\label{sec:intro}
In digital pathology, histopathology images can provide crucial insights for clinical diagnoses and treatment planning. Pathologists can make diagnosis and prognosis decisions by studying the morphology of glands/nuclei and their spatial arrangements. For example, assessing gland morphology can help pathologists determine different stages of colon cancer~\cite{fleming2012colorectal} and prostate cancer~\cite{montironi2005gleason}.
Traditionally, this would rely on manual annotations by pathologists, which is costly, time-consuming, and error-prone.
To alleviate this burden, deep learning methods have been proposed to automatically segment the objects of interest~\cite{wang2022uctransnet, isensee2021nnu, cao2022swin, zhou2018unet++, graham2019mild}. However, despite their satisfactory performances, these methods still rely on a large amount of high-quality annotations, which is expensive and requires a lot of domain expertise. 

\begin{figure}[t]
    \centering
    \begin{subfigure}{0.215\textwidth}
        \includegraphics[width=\linewidth]{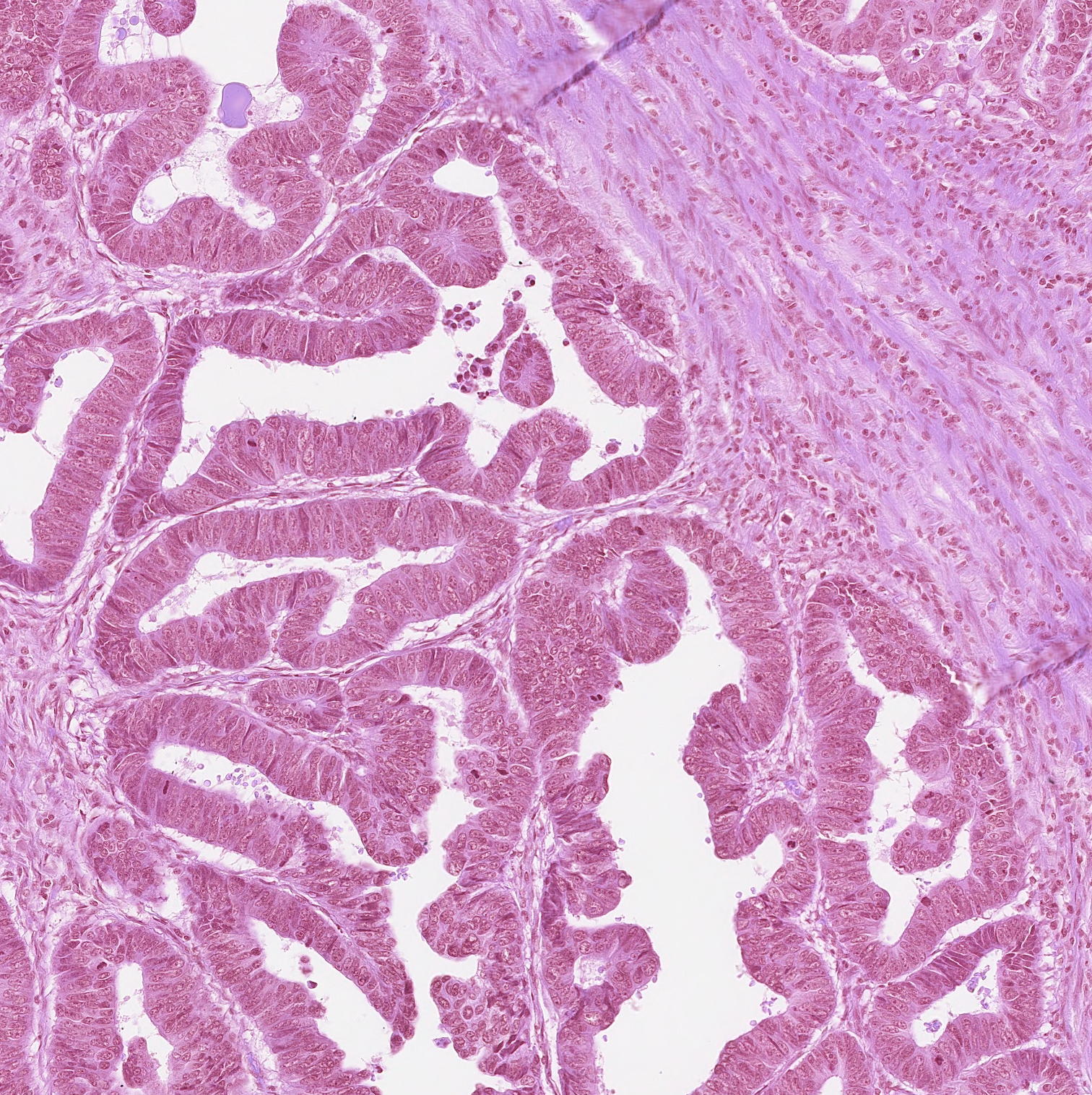}
        \caption{Input}
        \label{fig:teaser_sub1}
    \end{subfigure} 
    \begin{subfigure}{0.215\textwidth}
        \includegraphics[width=\linewidth]{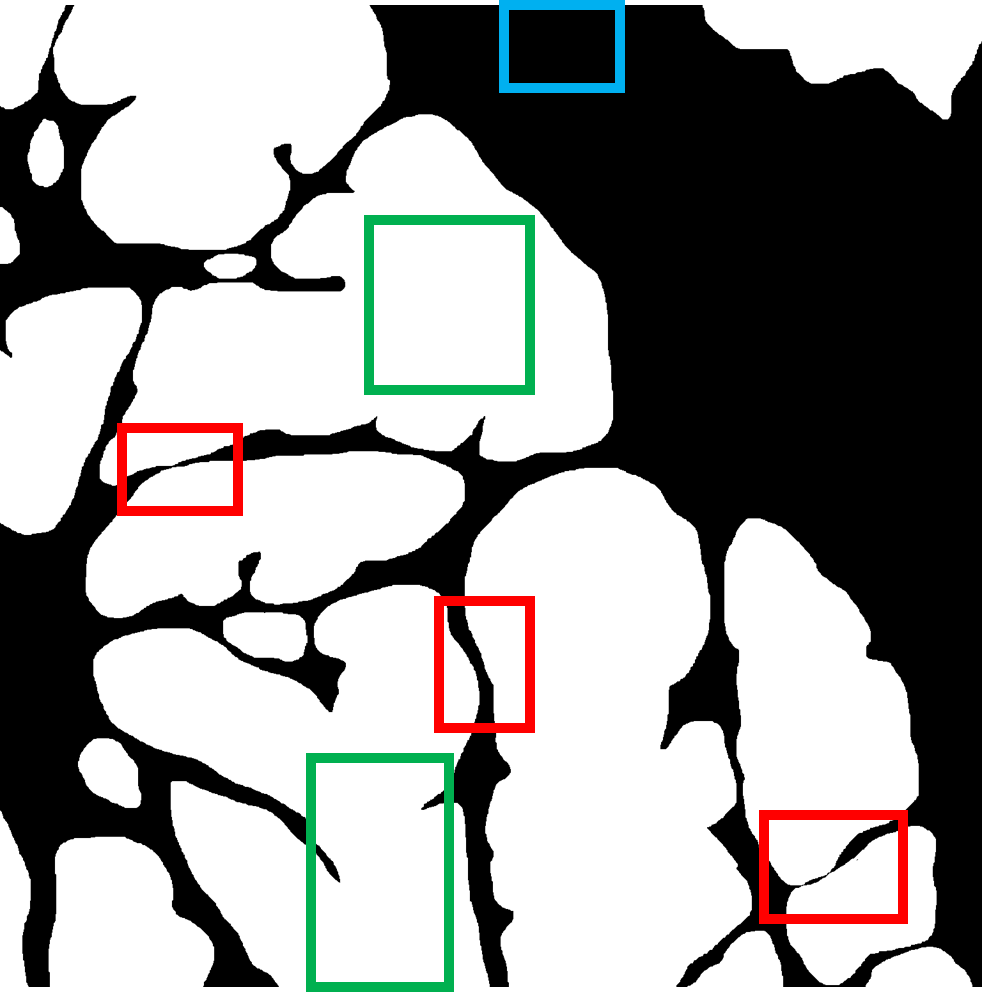}
        \caption{GT}
        \label{fig:teaser_sub2}
    \end{subfigure}
    \begin{subfigure}{0.215\textwidth}
        \includegraphics[width=\linewidth]{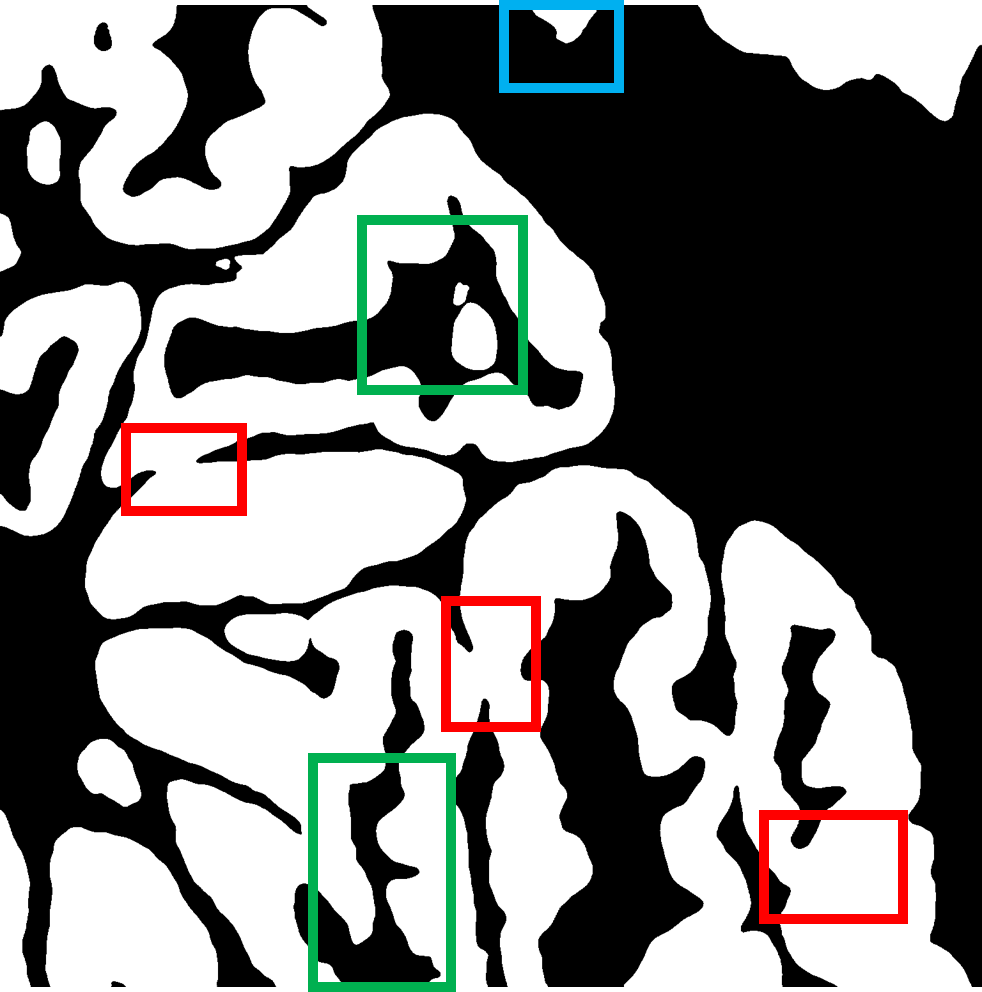}
        \caption{XNet~\cite{zhou2023xnet}}
        \label{fig:teaser_sub3}
    \end{subfigure}
    \begin{subfigure}{0.215\textwidth}
        \includegraphics[width=\linewidth]{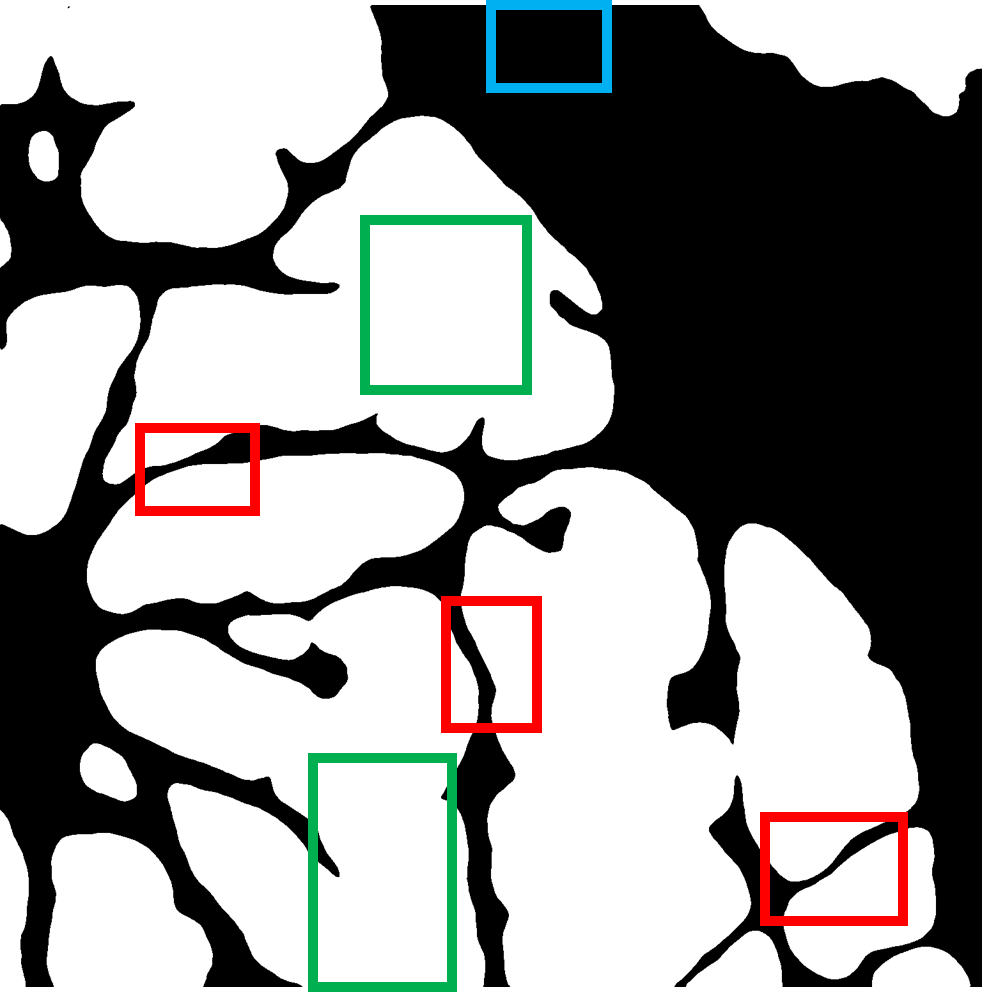}
        \caption{Ours}
        \label{fig:teaser_sub4}
    \end{subfigure}
    \caption{\small
    Illustration of the significance of topological correctness in gland segmentation. \textbf{(a)} an input image. \textbf{(b)} ground truth GT. \textbf{(c)} the result of SoTA semi-supervised segmentation method~\cite{zhou2023xnet} 
    devoid of any topological regularization. 
    \textbf{(d)} our segmentation result.
    For the regions within boxes, the SoTA's result has errors that, while minor at the pixel level, significantly alter the semantic interpretation.
    The \textcolor{red}{\textbf{red}} boxes indicate prediction errors such as incorrectly merging adjacent glands, the \textcolor{CornflowerBlue}{\textbf{blue}} box indicates false positive gland predictions, and the \textcolor{ForestGreen}{\textbf{green}} boxes indicate the false negative holes in glands.
    These errors affect the pathologist's decision and analysis.
    }
    \label{fig:Motivation_sample}
\end{figure}

One of the dominant schemes to reduce the cost of annotation is semi-supervised learning (SemiSL)~\cite{zhang2022boostmis, zhang2022discriminative, li2023calibrating, sohn2020fixmatch,jeong2019consistency,jin2022semi, vu2019advent,fang2020dmnet,wu2022mutual}. 
SemiSL leverages a small group of labeled data along with a large amount of unlabeled data to train a model. By harvesting the rich information in the unlabeled data, they can perform as well as fully-supervised methods.
While existing works mostly focus on pixel-wise accuracy, limited progress has been made to address topological correctness. At regions where glands or nuclei are densely distributed, the model tends to make topological errors such as mistakenly merged/separated glands, or missing components. 
See~\cref{fig:Motivation_sample} for an illustration. Even a strong semi-supervised method like~\cite{zhou2023xnet} still fails to properly maintain glands' topological correctness, as highlighted by the boxed regions. Such topological errors significantly change their morphological measures. Similar issues also occur in the nuclei segmentation task due to dense distribution characteristics. If not addressed properly, these topological errors will significantly impact downstream analysis. 

\begin{figure}[t]
\centering
    \includegraphics[width=0.8\linewidth]{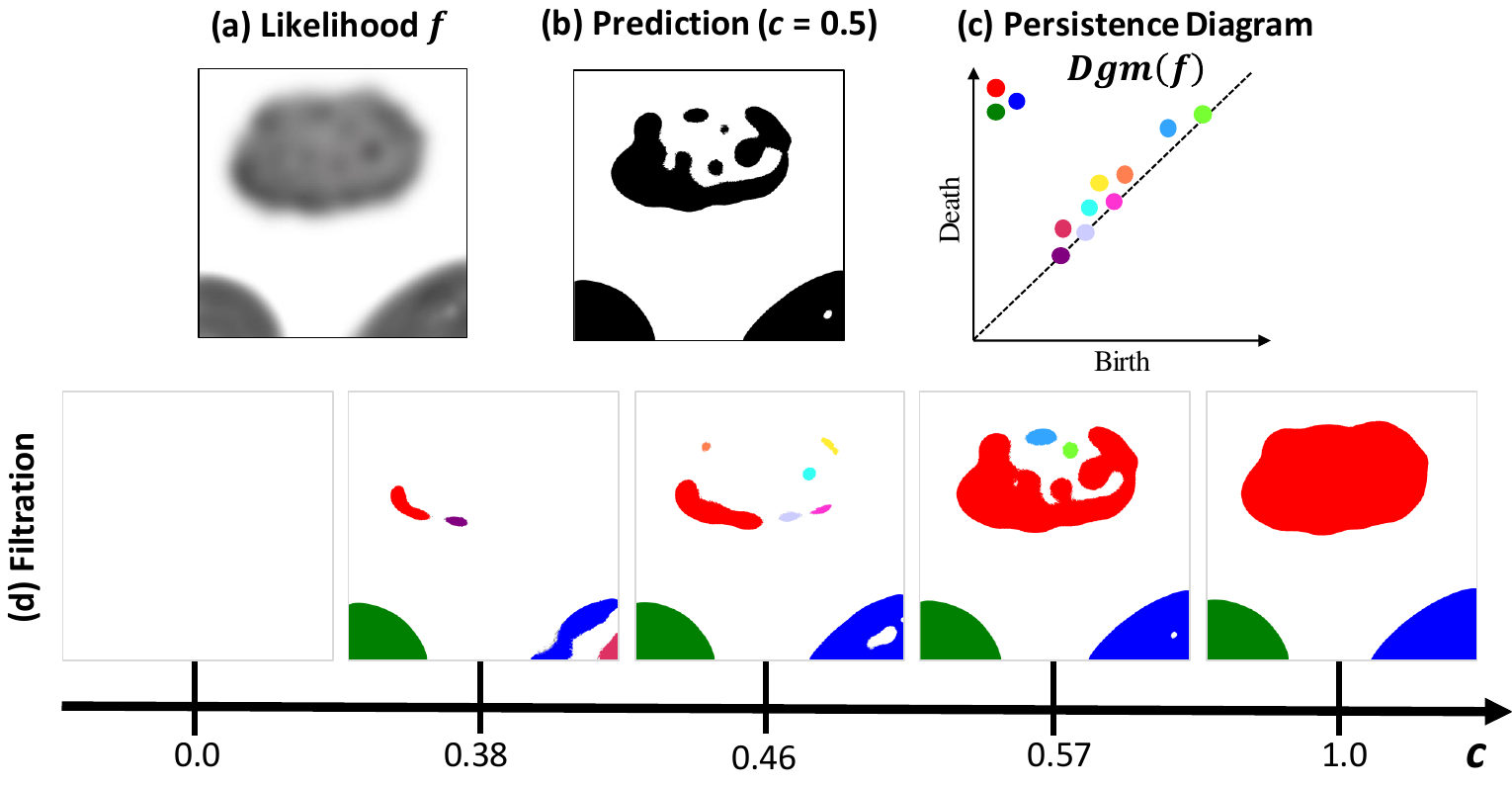}
    \caption{\small
    \textbf{(a)} A predicted likelihood map $f$, \textbf{(b)} the binary prediction, and \textbf{(c)} the corresponding persistence diagram $Dgm(f)$, which tends to be noisy.
    In \textbf{(d)}, consider the filtration for different values of threshold $c$. Notice that there are three true, or \textit{signal}, structures, denoted by colors \textcolor{red}{\textbf{red}}, \textcolor{OliveGreen}{\textbf{green}}, and \textcolor{blue}{\textbf{blue}}, which persist across the range of $c$. Hence the dots corresponding to these structures are located at the upper-left corner of $Dgm(f)$. The remaining colors denote several \textit{noisy} structures which persist for a short range of $c$, and thus their dots appear closer to the diagonal. Note that we only show 0-dim persistent dots referring to connected components in $Dgm(f)$.
    }
    \label{fig:illustration}
\end{figure}

In this paper, we investigate how to help models learn the correct topological characteristics from unlabeled data. Traditional methods~\cite{hu2019topology, hu2020topology, hu2022structure, stucki2023topologically, clough2020topological, wang2022ta} solve the segmentation problem by encoding topological properties as constraints during training. 
However, these methods require ground-truth annotations and a clean topology to train. Therefore, they are not able to exploit unlabeled data.

To address this issue, we propose \textit{the first topology-aware solution for semi-supervised segmentation of histopathology images}. Unlike the clean topology of the ground truth (GT), the topology of the predictions is noisy and contains spurious structures (see \cref{fig:illustration}). 
These \textbf{noisy structures} may be the holes inside large glands or small islands that are false positives. They will oscillate through training and significantly distract the learning from concentrating on the \textbf{true topological signals} (the structures we want to preserve).

To this end, based on the teacher-student framework~\cite{tarvainen2017mean}, 
we propose to enforce noise-aware topological consistency between the predictions of different augmented inputs. More specifically, we propose to decompose the topological structures of a potentially noisy prediction into \emph{signal topology} and \emph{noisy topology}. This can be achieved by decomposing the topological features, formalized as the \emph{persistence diagram} (PD)~\cite{edelsbrunner2022computational}, into signal and noise. We only enforce the signal topology of the teacher and the student's prediction to be consistent.  This is achieved by a \emph{signal topology consistency loss} that matches the signal topological features using the Wasserstein distance~\cite{cohen2005stability, cohen2010lipschitz}. Meanwhile, for the noisy topological features, we introduce a \emph{noisy topology removal loss}, based on a theoretical measure called \emph{total persistence}~\cite{cohen2010lipschitz}. It aggregates the saliency of all noisy topological structures. Minimizing it essentially removes all these noisy structures. Combining the proposed \emph{signal topology consistency loss} and \emph{noisy topology removal loss} with the classic pixel-wise consistency loss, our method achieves the desired goal and ensures the student model learns the robust topological representations from the unlabeled data.

We note that the method in~\cite{hu2019topology} (which we refer to as TopoLoss) also designed a topology-aware loss based on the persistence diagram. However, this method is designed for a fully-supervised setting and relies heavily on the clean persistence diagrams of the GTs. 
The clean GTs facilitate the straightforward matching of the persistent dots. In contrast, SemiSL encounters a unique issue where we only have noisy predictions of the unlabeled data, containing a lot of spurious structures. A non-trivial solution is required to address the challenge of matching two noisy PDs.
Hence, the previous TopoLoss cannot adapt to the teacher-student network, where we are forced to compare the topology of noisy predictions.

We evaluate the proposed method by conducting experiments on 
three public histopathology image datasets. 
The results show that our method outperforms other SoTA semi-supervised methods on both pixel- and topology-wise evaluation metrics, across $10\%$ and $20\%$ labeled data settings.
In summary, our contributions are as follows:

\begin{itemize}
    \item We propose the first topology-aware semi-supervised framework that enforces topological consistency in segmenting densely distributed objects of interest in histopathology images.
    \item We propose a learning strategy that produces robust topological representations from the noisy topological feature space of the unlabeled images.
    \item Through extensive experiments on multiple histopathology imaging datasets, we show that our method effectively improves the segmentation quality both pixel- and topology-wise.
\end{itemize}

\section{Related work}
\myparagraph{Segmentation with limited annotations.} 
To address the scarcity of annotated data, semi-supervised learning (SemiSL) has emerged as a pivotal methodology in medical image segmentation~\cite{jiao2022learning}. 
The primary schemes in this domain encompass pseudo-labeling~\cite{yao2022enhancing, zhang2022discriminative, seibold2022reference}, consistency learning~\cite{huang2022semi, li2020transformation, ouali2020semi} and entropy minimization~\cite{wu2022cross, grandvalet2004semi, berthelot2019mixmatch}. 
Pseudo-labeling-based methods aim to generate pseudo-labels for unlabeled data, which are then used to train the model further. 
To improve the quality of pseudo-labels, Wang \textit{et al.}~\cite{wang2022ssa} propose a confidence-aware module to 
select pseudo labels with high confidence. Some works try to refine the pseudo-labels by morphological methods~\cite{thompson2022pseudo} or by adding additional refinement networks~\cite{zhang2022discriminative, shi2021inconsistency}. 
By learning better representations that pull similar samples together and push dissimilar ones apart, contrastive learning is also applied in SemiSL~\cite{you2022simcvd,you2023rethinking,basak2023pseudo}.

Another main scheme in SemiSL is consistency learning, which emphasizes consistent predictions under various perturbations. 
Different perturbations at input or feature level are proposed to make the model robust~\cite{li2020transformation, li2021dual}. Also, most of these methods are the variants of Mean-Teacher framework~\cite{tarvainen2017mean}, which encourages invariant predictions for perturbed inputs, 
like combining with uncertainty~\cite{yu2019uncertainty} and calculating different levels of consistency~\cite{luo2022semi, chen2021mtans}.

\myparagraph{Topology-aware image segmentation. }
There are existing methods enforcing segmentation to have correct topology~\cite{hu2019topology, hu2020topology, shit2021cldice, clough2020topological, yang2021topological, hu2022structure, gupta2022learning, wang2022ta, stucki2023topologically, wang2020topogan}. 
These methods compare the predictions and ground truth (GT) in terms of their topology, using differentiable loss functions based on tools such as persistent homology~\cite{hu2019topology,clough2020topological,stucki2023topologically}, discrete Morse theory~\cite{hu2020topology,hu2023learning,gupta2023topology}, homotopy warping~\cite{hu2022structure}, topological interactions~\cite{gupta2022learning}, and centerline comparison~\cite{shit2021cldice,wang2022ta}.
Despite the success of these topology-aware segmentation methods, they rely heavily on well-annotated, topologically correct labels, as well as the explicit topological information extracted from these labels. 
These methods are not suitable for a semi-supervised setting with limited annotations.
Clough \textit{et al}.~\cite{clough2020topological} assume a fixed topology for input data and use a topology-preserving loss in a semi-supervised setting.
However, their assumption is too strong and cannot adapt to histopathology images, where at different locations we have different numbers of glands/nuclei.
Our work aims to break such limitations by unearthing essential topological information from the vast amount of unlabeled images.

\section{Proposed Method}
In this section, we first provide an overview of our proposed method in \cref{subsec:stf}. Then, we give a brief introduction to the background of persistent homology in~\cref{subsec:persistent_homology}. Finally, we introduce our TopoSemiSeg in~\cref{subsec:topo_unsup_reg}.

In SemiSL, we have a small set 
of labeled training samples and a much larger set 
of unlabeled samples. Let $\mathcal{D}_L=\{(x^L_1, y_1), (x^L_2, y_2),...,(x^L_{N_L}, y_{N_L})\}$ be the dataset of $N_L$ labeled samples, and 
$\mathcal{D}_U=\{x^U_{1}, x^U_{2}, ... , x^U_{N_U}\}$ be the unlabeled dataset of $N_U$ images, where
$N_L<<N_U$. $x_i^U$ denotes 
the $i$-th unlabeled image and $x_i^L$ denotes the $i$-th labeled image with its corresponding pixel-wise label $y_i$. 

The objective of SemiSL is to unearth the rich information within the unlabeled data, accompanied by limited guidance from labeled data. Most existing works only consider pixel-wise accuracy, ignoring the importance of topological correctness. Here, we take both of them into consideration.

\begin{figure*}[t]
    \centering
    \includegraphics[width=.95\textwidth]{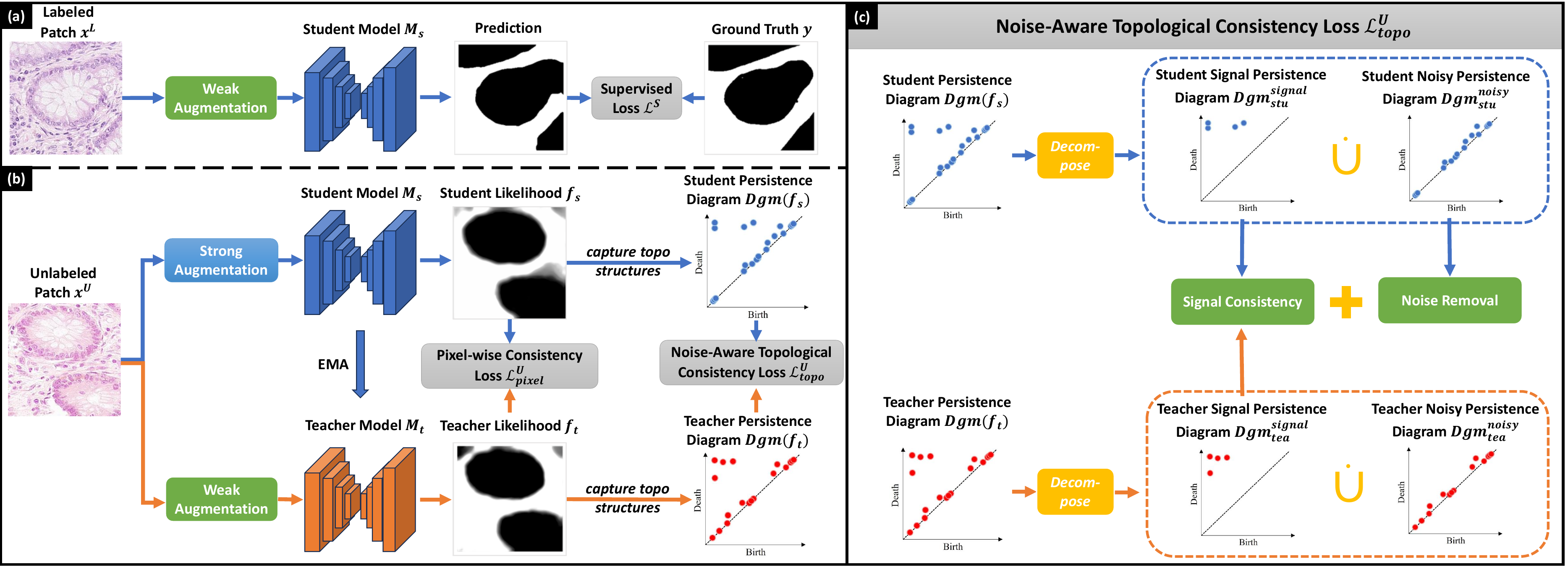}
    \caption{An overview of our method. (\textbf{a}) denotes the labeled workflow. The student model learns from labeled images via the supervised loss $\mathcal{L}^{S}$. (\textbf{b}) denotes the unlabeled workflow. The student model learns from unlabeled images using $\mathcal{L}^{U}$, which consists of pixel-wise consistency loss $\mathcal{L}^{U}_{\text{pixel}}$ and noise-aware topological consistency loss $\mathcal{L}^{U}_{\text{topo}}$. (\textbf{c}) shows the details of our proposed noise-aware topological consistency loss $\mathcal{L}^{U}_{\text{topo}}$, which encompasses our decomposition and optimal matching strategy, resulting in signal topology consistency loss $\mathcal{L}^{U}_{\text{topo-cons}}$ and noisy topology removal loss $\mathcal{L}^{U}_{\text{topo-rem}}$.
    }
    \label{fig:overall_pipeline}
\end{figure*}

\begin{figure}[t]
\centering
    \includegraphics[width=0.8\linewidth]{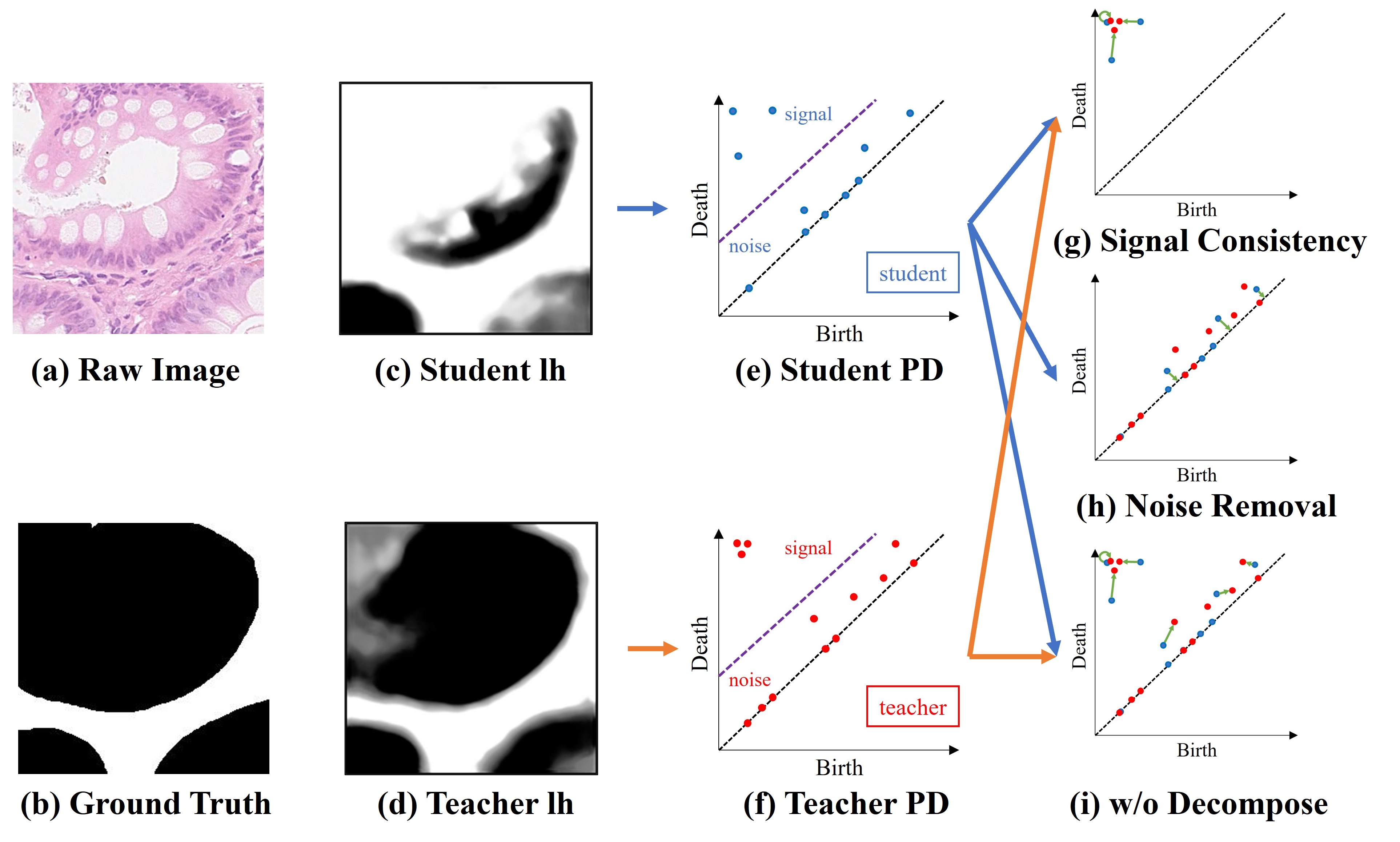}
    \caption{\small
    Inituition of our decomposition and matching strategy. (\textbf{a}) the raw image. (\textbf{b}) the ground truth, included for reference. (\textbf{c}) the student likelihood (lh). 
    (\textbf{d}) the teacher likelihood. 
    (\textbf{e}) decomposition of the persistence diagram of the student likelihood. The \textcolor{violet}{\textbf{purple}} line demonstrates the decomposition. (\textbf{f}) decomposition of the persistence diagram (tPD) of teacher likelihood. (\textbf{g}) the consistency between the signal topology. \textcolor{LimeGreen}{\textbf{Green}} arrows show the matching process. (\textbf{h}) the noisy topology removal process. (\textbf{i}) the matching process without decomposition. }
    \label{fig:inituition}
\end{figure}

\subsection{Overview of the Method}
\label{subsec:stf}
\cref{fig:overall_pipeline} provides an overview of our method. 
We adopt the popular teacher-student framework~\cite{tarvainen2017mean} in SemiSL. This framework contains two networks -- a student and a teacher -- with identical architecture. We denote the student network as $M_s$, parameterized by $\theta_s$, and the teacher network as $M_t$, parameterized by $\theta_t$. The student network learns from the teacher network. It is trained by minimizing the supervised loss $\mathcal{L}^{S}$ on the labeled data and the unsupervised loss $\mathcal{L}^{U}$ on the unlabeled data. More details can be found in \cref{fig:overall_pipeline}(a) and (b). The overall training objective is formulated as 
\begin{equation}
\mathcal{L} = \mathcal{L}^{S} + \mathcal{L}^{U}
\label{final_loss}
\end{equation}

To make full use of limited annotations, $\mathcal{L}^{S}$ is defined as the combination of cross-entropy loss ($\ell_{CE}$) and Dice loss ($\ell_{Dice}$)~\cite{sudre2017generalised} between the predictions and the labels: 
\begin{equation}
\mathcal{L}^{S}(D_L,M_s)=
\sum_{i=1}^{N_L}\left[\lambda^L_1 \ell_{CE}(M_s(x_i^L), y_i)+\lambda^L_2\ell_{Dice}(M_s(x_i^L), y_i) \right]
\nonumber
\label{supervised_loss}
\end{equation}
where $\lambda^L_{\ast}$
are adjustable weights. 

For unlabeled data, we apply strong ($A_{strong}$) and weak ($A_{weak}$) augmentations before passing them as input to the student and teacher networks, respectively. The unsupervised loss enforces the consistency between predictions of the student and teacher models. It consists of two loss terms: pixel-wise consistency loss ($\mathcal{L}^{U}_{\text{pixel}}$) and the noise-aware topological consistency loss $\mathcal{L}^{U}_{\text{topo}}$. 
\begin{equation}
\mathcal{L}^{U} = \lambda^U_{1}\mathcal{L}^{U}_{\text{pixel}} + 
\lambda^U_{2}\mathcal{L}^{U}_{\text{topo}} 
\label{unsupervised_loss}
\end{equation}
where $\lambda^U_{\ast}$ are adjustable weights.

We formulate the pixel-wise consistency loss as the cross-entropy (CE) loss between the outputs of the student and teacher models:
\begin{small}
\begin{equation}
\mathcal{L}^{U}_{\text{pixel}}(D_U,M_s,M_t) = \sum_{i=1}^{N_U}\ell_{CE}(M_s(A_{strong}(x_i^U)),M_t(A_{weak}(x_i^U))) 
\label{pixel-wise_consistency_loss}
\end{equation}
\end{small}
The second part of $\mathcal{L}^{U}$ is the noise-aware topological consistency loss $\mathcal{L}^{U}_{\text{topo}}$, which is crucial for learning a robust topological representation from the unlabeled data. It will be explained in detail the next subsection.

During the training phase, the student network's parameters $\theta_s$ are updated by minimizing the overall loss (\cref{final_loss}). We update the teacher model's parameters $\theta_t$ based on the student model's parameters using exponential moving average (EMA)~\cite{tarvainen2017mean}. In particular, at the $(\tau + 1)^{\text{th}}$ epoch, $\theta_t$ is updated as $\theta_t(\tau+1)=\alpha\theta_t(\tau)+(1-\alpha)\theta_s(\tau+1)$
where $\alpha$ is the EMA decay controlling the update rate.

\subsection{Background: Persistent Homology}
\label{subsec:persistent_homology}
In algebraic topology~\cite{munkres1984elements}, 
persistent homology~\cite{edelsbrunner2002topological} has emerged as a powerful tool for analyzing the topology of various kinds of real-world data, including images. It tracks the evolution of all topological structures, such as connected components and loops. All the topological structures and their birth/death times are captured in a so-called \textit{persistence diagram}, providing a multi-scale topological representation (See~\cref{fig:illustration}).

A persistence diagram (PD) consists of multiple dots in a 2-dimensional plane, referred to as \textit{persistent dots}. 
Each persistent dot $p\in Dgm(f)$ represents a topological structure. Its two coordinates denote the birth and death filtration values for the corresponding topological structure, i.e., $p=(b, d)$, where $b=birth(p)$ and $d=death(p)$. More details are in the Supplementary.

\subsection{Noise-aware Topological Consistency Loss}
\label{subsec:topo_unsup_reg}
We propose a noise-aware topological consistency loss to ensure that the teacher and the student models make consistent predictions in terms of topology. Given the likelihood maps of both the teacher and the student models, $f_t$ and $f_s$, we first compute the persistence diagrams, $Dgm(f_t)$ and $Dgm(f_s)$. However, directly comparing the two diagrams is not desirable. As shown in \cref{fig:inituition}, 
without supervision, both the student persistence diagram and the teacher persistence diagram are quite noisy. Direct comparison of the two diagrams can create a lot of unnecessary matching between the noisy structures. This will cause inefficiency in learning, and can potentially even derail the whole training.

To address this challenge, we propose to decompose $Dgm(f_s)$ and $Dgm(f_t)$ into signal and noise parts. The signal part is used to enforce teacher-student consistency via a signal topology consistency loss. The noise part will be removed through a novel noisy topology removal loss.

\myparagraph{Signal-Noise Decomposition of a Persistence Diagram.}
We would like to decompose a diagram into signal and noise parts.
However, in reality, without the ground truth, the decomposition cannot be guaranteed to be accurate. Hence, we use the classic measure of \emph{persistence} to decide whether a dot in the persistence diagram is a signal or noise. 

For a persistent dot $p\in Dgm(f)$, its persistence is simply its life span, i.e., the difference between its death and birth time: $per(p) = death(p) - birth(p)$.
Persistence is a good heuristic approximating the significance of a topological structure; the greater the persistence, the longer the structure exists through filtration, and the more likely the structure is a true signal. This is theoretically justified. The celebrated stability theorem \cite{cohen2005stability,cohen2010lipschitz} implies that low-persistence dots are much easier to be ``shed off'' through perturbation of the input function $f$.

Formally, using a predetermined threshold $\phi$, we decompose $Dgm(f)$ into disjoint signal and noise persistence diagrams based on the persistence:
\begin{eqnarray*}
    Dgm(f) &= Dgm(f)^{signal} \dot{\bigcup} Dgm(f)^{noise}\\
    Dgm(f)^{signal} &= \{p \in Dgm(f) \: | \: per(p) > \phi\} \label{signal_pd}\\
    Dgm(f)^{noise} &= \{p \in Dgm(f) \: | \: per(p) \leq \phi\}  \label{noise_pd}
\end{eqnarray*}
where $\dot{\bigcup}$ denotes the disjoint union. We apply the same decomposition to both teacher and student model outputs, acquiring their signal and noise parts respectively. 
The threshold $\phi$ is tuned empirically. These signal/noisy diagrams for the outputs will be used for $\mathcal{L}^{U}_{\text{topo}}$ in~\cref{unsupervised_loss}.

\myparagraph{Signal Topology Consistency Loss.}
After the decomposition of both persistence diagrams, we obtain $Dgm_{stu}^{signal}$ and $Dgm_{tea}^{signal}$ representing the meaningful topological signals. Our first topology-aware loss is to ensure the two signal diagrams are the same. Similar to previous topological losses~\cite{hu2019topology}, we will use the classic Wasserstein distance between the two diagrams. 
Note: for any diagram $Dgm(g)$, we regard it as the generalized persistence diagram\footnote{A generalized persistence diagram is a countable multiset of points in $\mathbb{R}^2$ along with the diagonal $\Delta = \{(b,d) \: | \: b=d\}\}$, where each dot on the diagonal has infinite multiplicity.}.

\begin{definition}[Wasserstein distance between PDs~\cite{cohen2010lipschitz}]
Given two diagrams $Dgm(g)$ and $Dgm(h)$, the $p$-th Wasserstein distance between them is defined as:\footnote{For ease of exposition, we change the original formulation and use the 2-norm instead of infinity norm for $||x-\gamma(x)||$. The difference is bounded by a $\sqrt{2}/2$ constant factor.}
\begin{equation}
    W_p(Dgm(g), Dgm(h))=\left( \underset{\gamma\in \Gamma}{inf}\sum\limits_{x \in Dgm(g)}{||x-\gamma(x)||}^p \right)^{\frac{1}{p}}
    \nonumber
\end{equation}
where $\Gamma$ represents all bijections from $Dgm(g)$ to $Dgm(h)$. 
\end{definition}

See \cref{fig:inituition}(g) and (h) for an illustration. The Wasserstein distance essentially finds an optimal matching between dots of the two diagrams. Unmatched dots are matched to their projection on the diagonal line. The distance is computed by aggregating over distance between all the matched pairs of dots. The optimal matching, as well as the distance, can be computed using either the classic Hungarian method, or more advanced algorithms~\cite{lacombe2018large,kerber2016geometry}. 

Next, we write the signal topology consistency loss in terms of the student's likelihood map, $f_s$.
Denote by $\gamma^*$ the optimal matching between $Dgm_{stu}^{signal}$ and $Dgm_{tea}^{signal}$. Each student persistent dot $p_{stu}^{signal} \in Dgm_{stu}^{signal}$ is matched to either a teacher persistent dot, or its projection on the diagonal. 
We can now formulate our signal topology consistency loss $\mathcal{L}_{\text{topo-cons}}^{U}$ as the squared distance between every student signal dot and its match:
\begin{equation}
     \mathcal{L}_{\text{topo-cons}}^{U} = \sum\limits_{p \in Dgm_{stu}^{signal}}||p-\gamma^\ast(p)||^2
     \label{eq:topo-cons-v1}
\end{equation}
We still have to write the loss in terms of the student likelihood map. Note that in persistent homology, the birth and death times of every persistent dot are the function values of certain critical points. See Supplementary for more details and illustrations. For each 0-dimensional persistent dot $p$ in a student diagram, the birth is at a local maxima $x^b_p$ and the death is at a saddle point $x^d_p$, formally, $birth(p) = f_s(x^b_p)$ and $death(p) = f_s(x^d_p)$. Substituting into \cref{eq:topo-cons-v1}, we have 
\begin{multline}    
     \mathcal{L}_{\text{topo-cons}}^{U}(f_s) = \sum\limits_{p \in Dgm_{stu}^{signal}}
     \{[f_s(x^b_p)-birth(\gamma^\ast(p))]^2 
      + [f_s(x^d_p)-death(\gamma^\ast(p))]^2\}
     \label{eq:topo-cons-v2}
\end{multline}
which can be optimized with respect to the student network's parameters $\theta_s$.

\myparagraph{Noisy Topology Removal Loss.} 
So far, we have introduced how to decompose the diagram and how the signal part of the diagrams can be used to enforce topological consistency. We also introduce a loss to remove the noisy topology from the student likelihood map. This turns out to be very powerful in practice: by removing the topological noise, we can stabilize the output of the student network, and eventually also stabilize the teacher network via EMA.

Our noisy topology removal loss is based on the concept of \textit{Total Persistence}, which essentially measures the total amount of information a diagram carries. By minimizing the total persistence of the noise diagram, we are effectively removing all noise dots.
\begin{definition}[Total Persistence~\cite{cohen2010lipschitz}]
Given a persistence diagram, $Dgm(g)$, the $p$-th total persistence is :
\begin{equation}
    P_{total}(Dgm(g)) = \sum\limits_{x\in Dgm(g)} [death(x)-birth(x)]^p
\end{equation}
\end{definition}
Similar to the consistency loss, we can define the loss in terms of the student likelihood map as follows:
\begin{equation}
    \mathcal{L}_{\text{topo-rem}}^{U}(f_s) = 
    P_{total}(Dgm_{stu}^{noise}) = 
    \sum\limits_{p \in Dgm_{stu}^{noise}}
    \left[f_s(x^b_p) 
      - f_s(x^d_p)\right]^2
      \label{eq:topo-rem-v1}
\end{equation}
The final noise-aware topological consistency loss $\mathcal{L}^{U}_{\text{topo}}$ becomes the sum of the two topology-aware loss terms: $\mathcal{L}^{U}_{\text{topo-cons}}$ and $\mathcal{L}^{U}_{\text{topo-rem}}$,
\begin{equation}
    \mathcal{L}^{U}_{\text{topo}} = \mathcal{L}^{U}_{\text{topo-cons}} + \mathcal{L}^{U}_{\text{topo-rem}} \label{twoloss}
\end{equation}
See Supplementary for the illustration of the differentiability of these two losses.

\section{Experiments}
\label{sec:exp}
We conduct extensive experiments on three public and widely used histopathology image datasets. We compare our method against SoTA semi-supervised segmentation methods on both pixel- and topology-wise evaluation metrics.

\noindent\textbf{Implementation details} are in the Supplementary.

\begin{table}[ht]
\centering
\scriptsize
\caption{
Quantitative results on three histopathology image datasets. We compare our method with several state-of-the-art semi-supervised medical image segmentation methods on two settings of $10\%$ and $20\%$ labeled data. The best results are highlighted in \textbf{bold}, and * indicates that the method is re-implemented by ourselves.} 
\resizebox{\linewidth}{!}{
\begin{tabular}{lccccccccl}
\hline
\multirow{2}{*}{Dataset} & \multirow{2}{*}{Labeled Ratio (\%)} & \multirow{2}{*}{Method} & \multicolumn{3}{c}{Pixel-Wise} &  & \multicolumn{3}{c}{Topology-Wise} \\ \cline{4-6} \cline{8-10} 
 &  &  & Accuracy $\uparrow$ & Dice\_Obj $\uparrow$ & IoU $\uparrow$ &  & Betti Error $\downarrow$ & Betti Matching Error $\downarrow$ & \multicolumn{1}{c}{VOI} $\downarrow$ \\ \hline
\multicolumn{1}{l|}{\multirow{14}{*}{CRAG}} & \multicolumn{1}{c|}{\multirow{7}{*}{10\%}} & MT~\cite{tarvainen2017mean} & 0.862 & 0.821 & 0.713 & & 2.238 & 62.250 & 0.977 \\
\multicolumn{1}{l|}{} & \multicolumn{1}{c|}{} & EM~\cite{vu2019advent} & 0.834 & 0.789 & 0.688 &  & 2.178 & 80.100 & 1.027\\
\multicolumn{1}{l|}{} & \multicolumn{1}{c|}{} & UA-MT~\cite{yu2019uncertainty} & 0.874 & 0.837 & 0.728 &  & 1.703 & 66.450 & 0.947\\
\multicolumn{1}{l|}{} & \multicolumn{1}{c|}{} & HCE*~\cite{jin2022semi_miccai} & 0.891 & 0.862 & 0.773 &  & 1.286 & 35.530 & 0.861 \\
\multicolumn{1}{l|}{} & \multicolumn{1}{c|}{} & URPC~\cite{luo2022semi} & 0.872 & 0.829 & 0.728 &  & 1.732 & 74.600 & 0.883 \\
\multicolumn{1}{l|}{} & \multicolumn{1}{c|}{} & XNet~\cite{zhou2023xnet} & 0.895 & 0.872 & 0.781 &  & 0.578 & 15.050 & 0.773\\
\multicolumn{1}{l|}{} & \multicolumn{1}{c|}{} & \textbf{TopoSemiSeg} & \textbf{0.905} & \textbf{0.884} & \textbf{0.798} &  & \textbf{0.227} & \textbf{10.475} & \textbf{0.758} \\ \cline{2-10} 
\multicolumn{1}{l|}{} & \multicolumn{1}{c|}{\multirow{7}{*}{20\%}} & MT~\cite{tarvainen2017mean} & 0.887 & 0.858 & 0.759 &  & 2.603 & 99.025 & 0.867 \\
\multicolumn{1}{l|}{} & \multicolumn{1}{c|}{} & EM~\cite{vu2019advent} & 0.903 & 0.869 & 0.776 &   & 1.933 & 75.225 & 0.798 \\
\multicolumn{1}{l|}{} & \multicolumn{1}{c|}{} & UA-MT~\cite{yu2019uncertainty} & 0.895 & 0.859 & 0.765 &  & 1.822 & 70.850 & 0.829 \\
\multicolumn{1}{l|}{} & \multicolumn{1}{c|}{} & HCE*~\cite{jin2022semi_miccai} & 0.910 & 0.881 & 0.809 &   & 0.875 & 17.400 & 0.769 \\
\multicolumn{1}{l|}{} & \multicolumn{1}{c|}{} & URPC~\cite{luo2022semi} & 0.881 & 0.849 & 0.744 &   & 2.489 &  99.500 & 0.912 \\
\multicolumn{1}{l|}{} & \multicolumn{1}{c|}{} & XNet~\cite{zhou2023xnet} & 0.907 & 0.883 & 0.792 &   & 0.422 & 10.900 & 0.735\\
\multicolumn{1}{l|}{} & \multicolumn{1}{c|}{} & \textbf{TopoSemiSeg} & \textbf{0.912} & \textbf{0.898} & \textbf{0.820} &   & \textbf{0.226} & \textbf{8.575} & \textbf{0.709} \\  \cline{2-10} 
\multicolumn{1}{l|}{} & \multicolumn{1}{c|}{\multirow{1}{*}{100\%}} & Fully-supervised & 0.945 & 0.928 & 0.869 &  & 0.149 & 5.650 & 0.547 \\ \hline

\multicolumn{1}{l|}{\multirow{14}{*}{GlaS}} & \multicolumn{1}{c|}{\multirow{7}{*}{10\%}} & MT~\cite{tarvainen2017mean} & 0.815 & 0.790 & 0.671 &   & 2.392 & 31.125 & 1.079 \\
\multicolumn{1}{l|}{} & \multicolumn{1}{c|}{} & EM~\cite{vu2019advent} & 0.833 & 0.819 & 0.708 &   & 1.431 & 19.188 & 1.051 \\
\multicolumn{1}{l|}{} & \multicolumn{1}{c|}{} & UA-MT~\cite{yu2019uncertainty} & 0.728 & 0.845 & 0.829 &    & 2.086 & 26.650 & 1.018 \\
\multicolumn{1}{l|}{} & \multicolumn{1}{c|}{} & HCE*~\cite{jin2022semi_miccai} & 0.859 & 0.852 & 0.762 &  & 0.631 & 11.950 & 0.953 \\
\multicolumn{1}{l|}{} & \multicolumn{1}{c|}{} & URPC~\cite{luo2022semi} & 0.829 & 0.849 & 0.751 &   & 1.155 & 19.588 & 0.968 \\
\multicolumn{1}{l|}{} & \multicolumn{1}{c|}{} & XNet~\cite{zhou2023xnet} & 0.871 & 0.874 & 0.786 &   & 0.843 & 14.238 & 0.917 \\
\multicolumn{1}{l|}{} & \multicolumn{1}{c|}{} & \textbf{TopoSemiSeg} & \textbf{0.890} & \textbf{0.878} & \textbf{0.797} &   & \textbf{0.551} & \textbf{8.300} & \textbf{0.811} \\ \cline{2-10} 
\multicolumn{1}{l|}{} & \multicolumn{1}{c|}{\multirow{7}{*}{20\%}} & MT~\cite{tarvainen2017mean} & 0.870 & 0.863 & 0.771 &   & 2.126 & 29.963 & 0.925 \\
\multicolumn{1}{l|}{} & \multicolumn{1}{c|}{} & EM~\cite{vu2019advent} & 0.861 & 0.865 & 0.776 &   & 1.255 & 17.275 & 0.841 \\
\multicolumn{1}{l|}{} & \multicolumn{1}{c|}{} & UA-MT~\cite{yu2019uncertainty} & 0.874 & 0.866 & 0.781 &   & 1.123 & 18.038 & 0.869 \\
\multicolumn{1}{l|}{} & \multicolumn{1}{c|}{} & HCE*~\cite{jin2022semi_miccai} & 0.864 & 0.871 & 0.779 &   & 0.871 &  16.213 & 0.824 \\
\multicolumn{1}{l|}{} & \multicolumn{1}{c|}{} & URPC~\cite{luo2022semi} & 0.876 & 0.878 & 0.794 &   & 0.759 & 14.350 & 0.837 \\
\multicolumn{1}{l|}{} & \multicolumn{1}{c|}{} & XNet~\cite{zhou2023xnet} & 0.886 & 0.884 & 0.804 &   & 0.735 & 10.188 & 0.816 \\
\multicolumn{1}{l|}{} & \multicolumn{1}{c|}{} & \textbf{TopoSemiSeg} & \textbf{0.896} & \textbf{0.895} & \textbf{0.818} &   & \textbf{0.510} & \textbf{9.825} & \textbf{0.808} \\ 
\cline{2-10} 
\multicolumn{1}{l|}{} & \multicolumn{1}{c|}{\multirow{1}{*}{100\%}} & Fully-supervised & 0.920 & 0.917 & 0.853 &   & 0.473 & 7.125 & 0.686 \\ \hline

\multicolumn{1}{l|}{\multirow{14}{*}{MoNuSeg}} & \multicolumn{1}{c|}{\multirow{7}{*}{10\%}} & MT~\cite{tarvainen2017mean} & 0.889 & 0.748 & 0.607 &   & 10.210 & 292.857 & 0.874\\
\multicolumn{1}{l|}{} & \multicolumn{1}{c|}{} & EM~\cite{vu2019advent} & 0.901 & 0.757 & 0.612 &  & 10.339 & 257.071 & 0.844\\
\multicolumn{1}{l|}{} & \multicolumn{1}{c|}{} & UA-MT~\cite{yu2019uncertainty} & 0.898 & 0.741 & 0.594 &  & 10.227 & 255.428 & 0.862\\
\multicolumn{1}{l|}{} & \multicolumn{1}{c|}{} & HCE*~\cite{jin2022semi_miccai} & 0.882 & 0.761 & 0.617 &   & 14.210 & 377.928 & 0.890\\
\multicolumn{1}{l|}{} & \multicolumn{1}{c|}{} & CCT~\cite{ouali2020semi} & 0.892 & 0.766 & 0.624 &  & 8.063 & 225.500 & 0.839\\
\multicolumn{1}{l|}{} & \multicolumn{1}{c|}{} & URPC~\cite{luo2022semi} & 0.896 & 0.774 & 0.633 &   & 6.829 & 214.428 & 0.863 \\
\multicolumn{1}{l|}{} & \multicolumn{1}{c|}{} & \textbf{TopoSemiSeg} & \textbf{0.909} & \textbf{0.783} & \textbf{0.646} & & \textbf{6.661} & \textbf{196.357} & \textbf{0.789}\\ \cline{2-10}
\multicolumn{1}{l|}{} & \multicolumn{1}{c|}{\multirow{7}{*}{20\%}} & MT~\cite{tarvainen2017mean} & 0.896 & 0.767 & 0.624 &   & 12.522 & 246.786 & 0.873\\
\multicolumn{1}{l|}{} & \multicolumn{1}{c|}{} & EM~\cite{vu2019advent} & 0.905 & 0.777 & 0.637 &    & 7.160 & 198.571 & 0.805\\
\multicolumn{1}{l|}{} & \multicolumn{1}{c|}{} & UA-MT~\cite{yu2019uncertainty} &  0.904 & 0.772 & 0.632 &  & 9.406 & 246.857 & 0.826\\
\multicolumn{1}{l|}{} & \multicolumn{1}{c|}{} & HCE*~\cite{jin2022semi_miccai} & 0.899 & 0.771 & 0.642 &   & 13.330 & 311.143 & 0.829\\
\multicolumn{1}{l|}{} & \multicolumn{1}{c|}{} & CCT~\cite{ouali2020semi} & 0.903 & 0.785 & 0.648 &   & 7.977 & 207.857 & 0.832\\
\multicolumn{1}{l|}{} & \multicolumn{1}{c|}{} & URPC~\cite{luo2022semi} & \textbf{0.909}& 0.779 & 0.639 &  & 5.325 & 193.429 & 0.788 \\
\multicolumn{1}{l|}{} & \multicolumn{1}{c|}{} & \textbf{TopoSemiSeg} & 0.908 & \textbf{0.793} & \textbf{0.653} &  & \textbf{4.250} &  \textbf{188.642} & \textbf{0.787} \\ \cline{2-10}
\multicolumn{1}{l|}{} & \multicolumn{1}{c|}{\multirow{1}{*}{100\%}} & Fully-supervised &  0.929 & 0.817 & 0.702 &  & 2.491 & 142.429 & 0.657\\ \hline
\end{tabular}}
\label{tab:quant}
\end{table}

\subsection{Datasets}
\label{sec:data}
We evaluate our proposed method on \textbf{Colorectal Adenocarcinoma Gland (CRAG)}~\cite{graham2019mild}, \textbf{Gland Segmentation in Colon Histology Images Challenge (GlaS)}~\cite{sirinukunwattana2017gland}, and \textbf{Multi-Organ Nuclei Segmentation (MoNuSeg)}~\cite{kumar2019multi}. More details are provided in the Supplementary.

\subsection{Evaluation Metrics}
\label{sec:evalmet}
We select three widely used pixel-wise evaluation metrics, \textbf{Object-level Dice coefficient (Dice\_{Obj})}~\cite{xie2019deep}, \textbf{Intersection over Union (IoU)} and \textbf{Pixel-wise accuracy}. Meanwhile, topology-relevant metrics measure the structural accuracy. Hence, we also select three topological evaluation metrics, \textbf{Betti Error}~\cite{hu2019topology}, \textbf{Betti Matching Error}~\cite{stucki2023topologically}, and \textbf{Variation of Information (VOI)}~\cite{meilua2003comparing}. More details are provided in the Supplementary.

\begin{figure*}[t]
    \centering
    \begin{subfigure}{0.115\textwidth}
        \includegraphics[width=\linewidth]{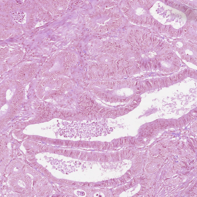}
    \end{subfigure} 
    \begin{subfigure}{0.115\textwidth}
        \includegraphics[width=\linewidth]{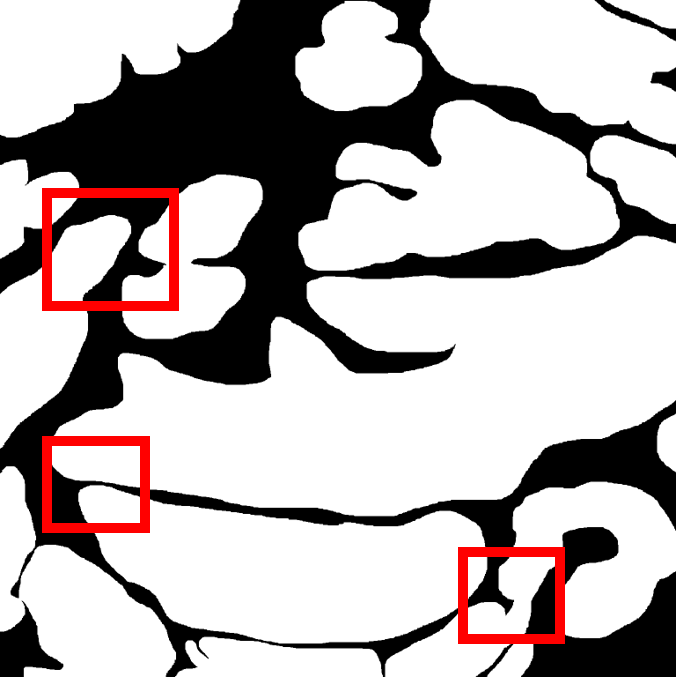}
    \end{subfigure}
    \begin{subfigure}{0.115\textwidth}
        \includegraphics[width=\linewidth]{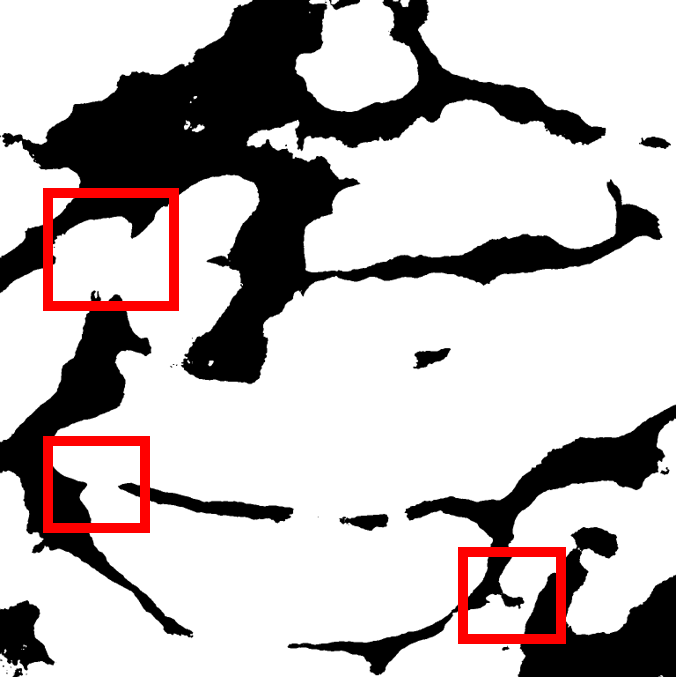}
    \end{subfigure}
    \begin{subfigure}{0.115\textwidth}
        \includegraphics[width=\linewidth]{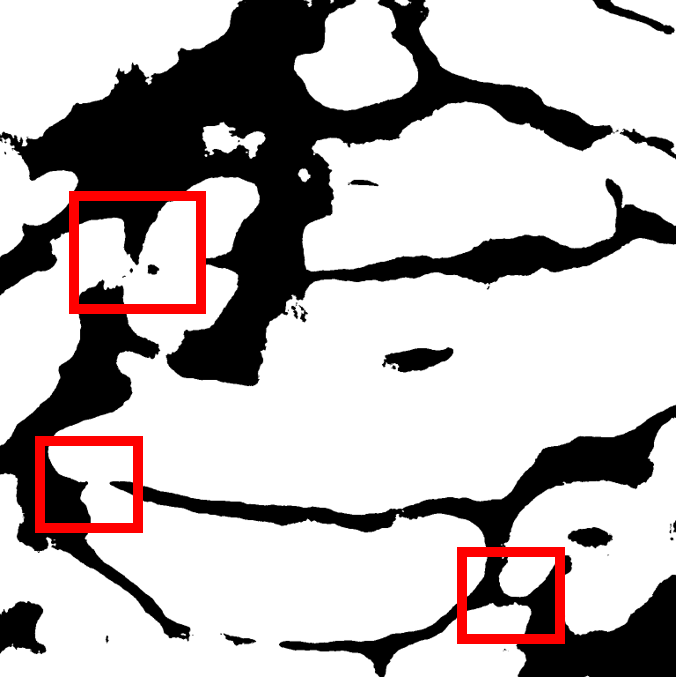}
    \end{subfigure}
    \begin{subfigure}{0.115\textwidth}
        \includegraphics[width=\linewidth]{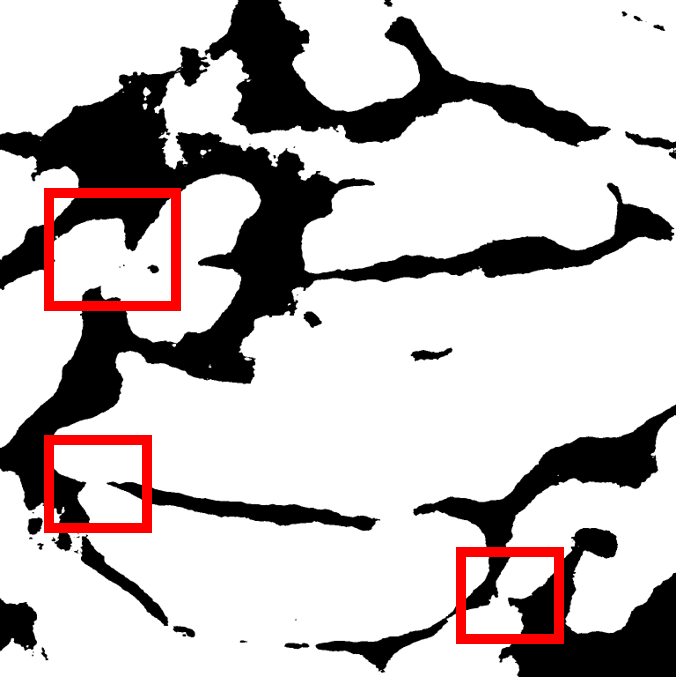}
    \end{subfigure}
    \begin{subfigure}{0.115\textwidth}
        \includegraphics[width=\linewidth]{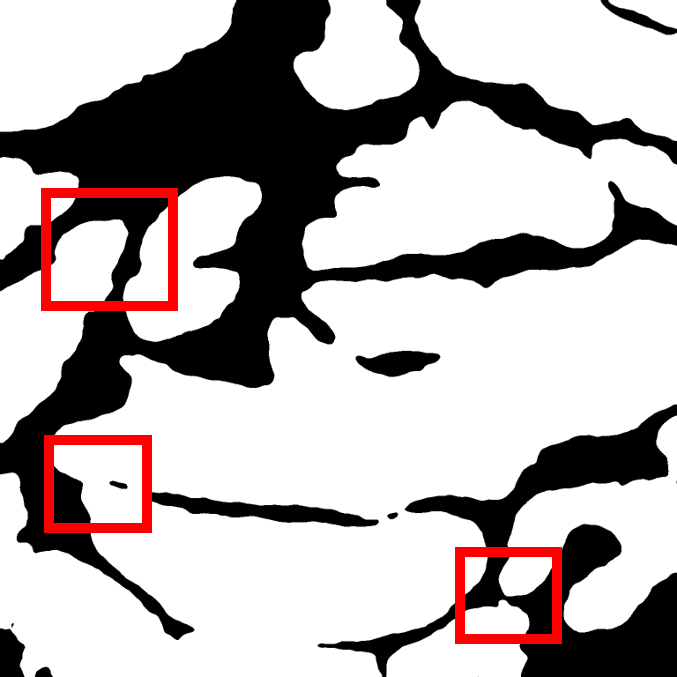}    
    \end{subfigure} 
    \begin{subfigure}{0.115\textwidth}
        \includegraphics[width=\linewidth]{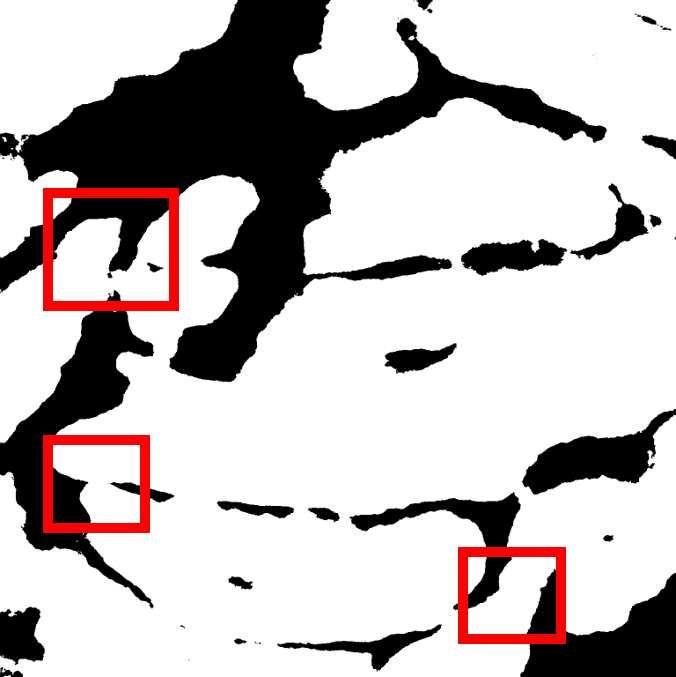}
    \end{subfigure}
    \begin{subfigure}{0.115\textwidth}
        \includegraphics[width=\linewidth]{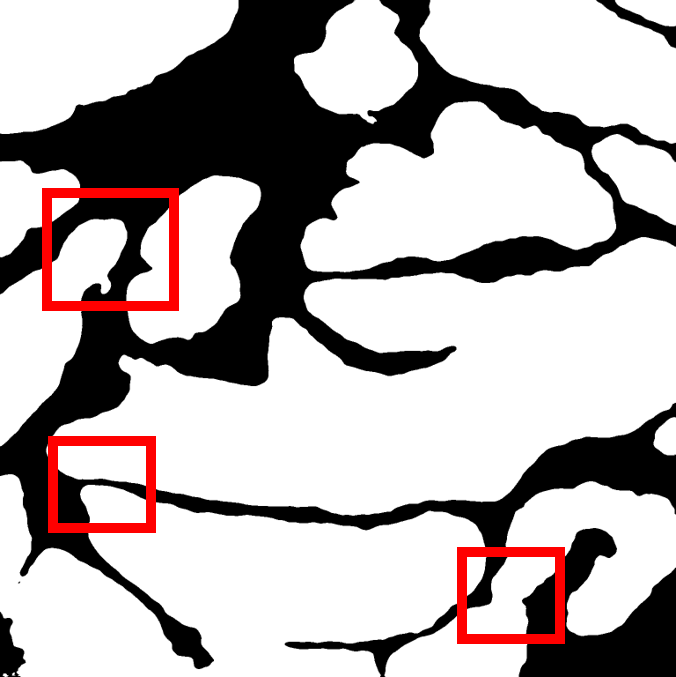}
    \end{subfigure}

    \begin{subfigure}{0.115\textwidth}
        \includegraphics[width=\linewidth]{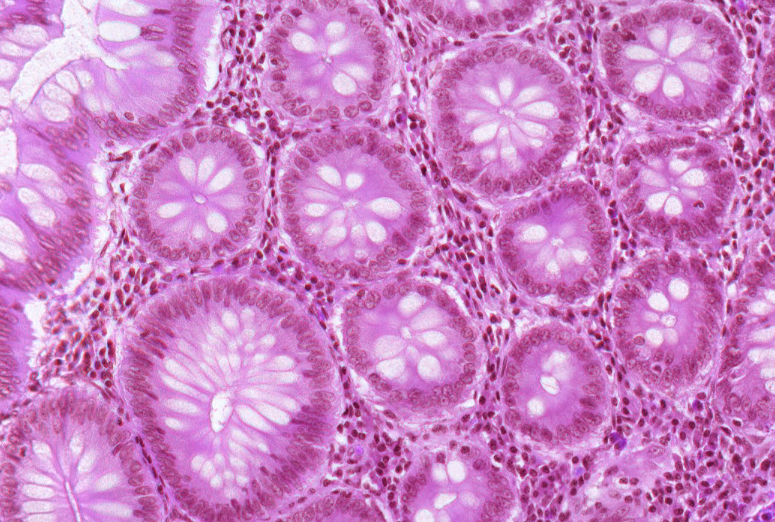}
    \end{subfigure}
    \begin{subfigure}{0.115\textwidth}
        \includegraphics[width=\linewidth]{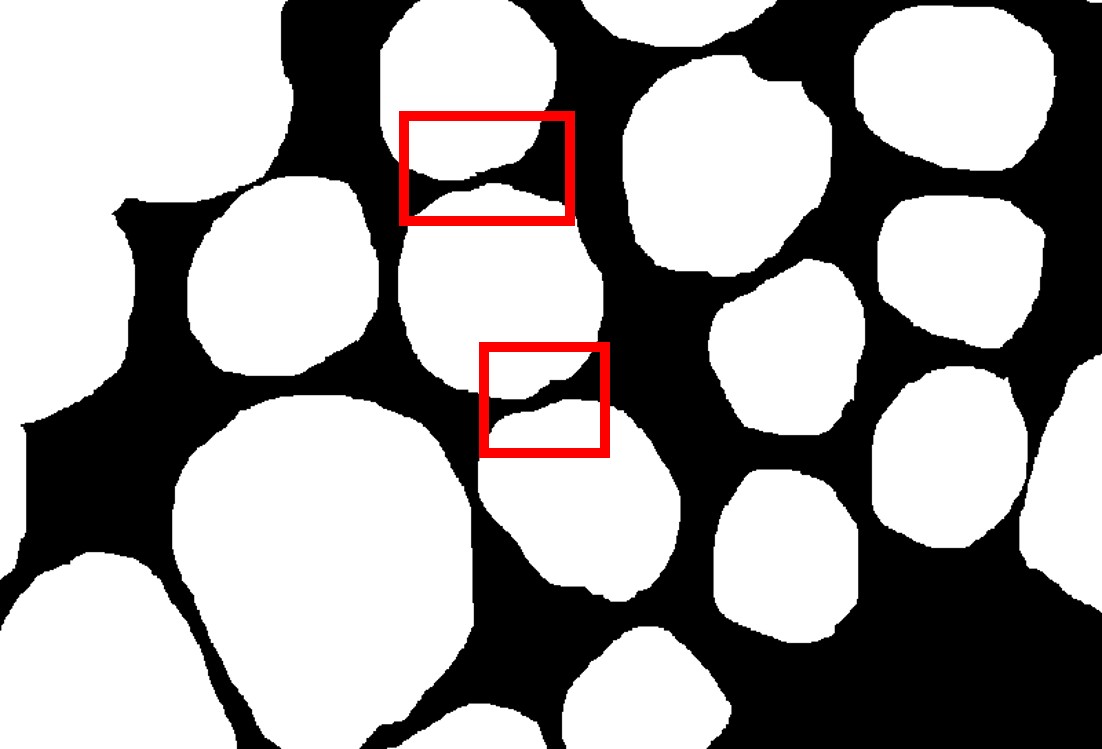}
    \end{subfigure} 
    \begin{subfigure}{0.115\textwidth}
        \includegraphics[width=\linewidth]{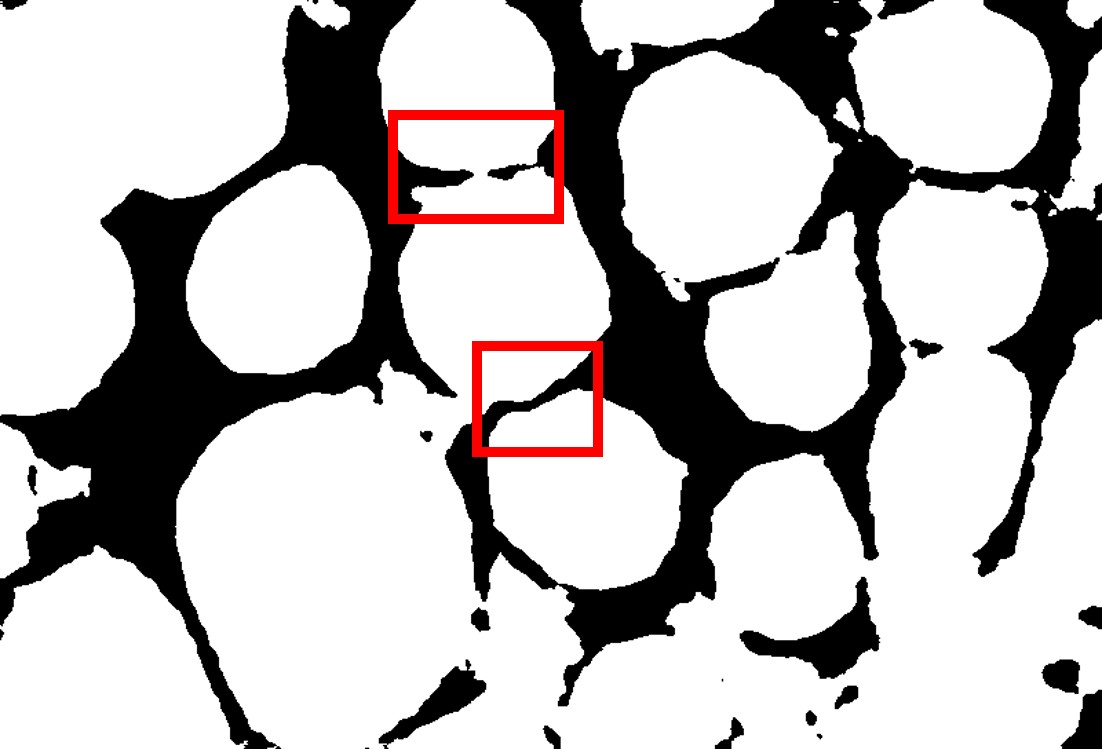}
    \end{subfigure} 
    \begin{subfigure}{0.115\textwidth}
        \includegraphics[width=\linewidth]{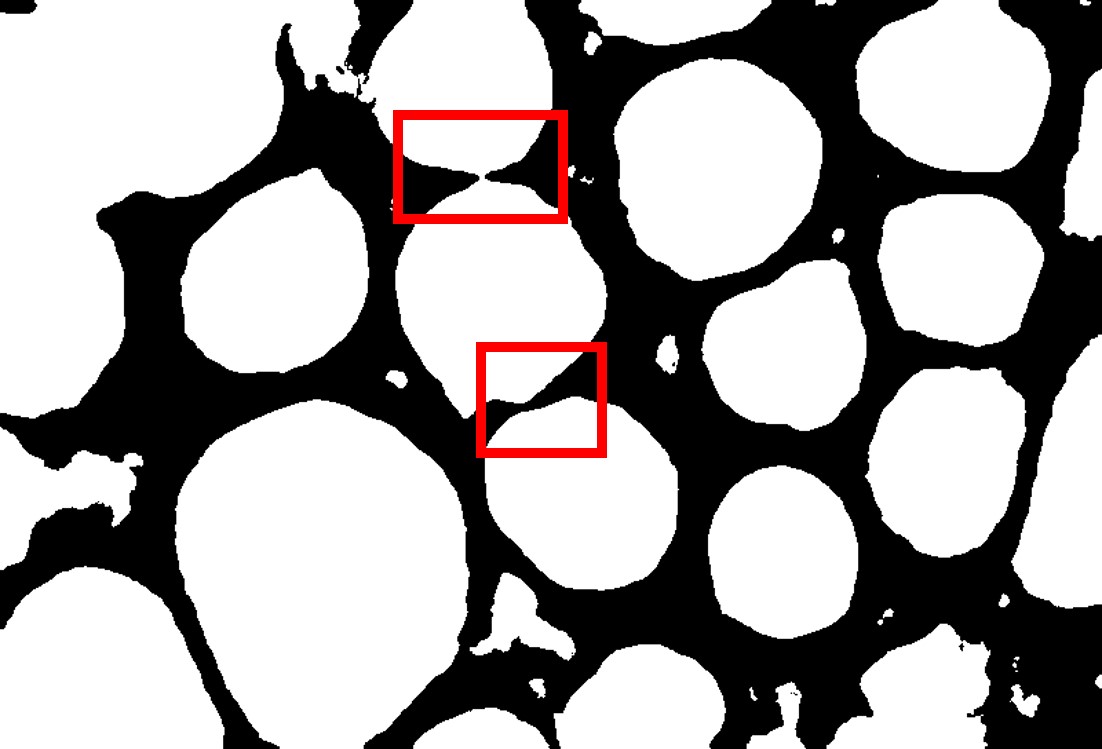}
    \end{subfigure} 
    \begin{subfigure}{0.115\textwidth}
        \includegraphics[width=\linewidth]{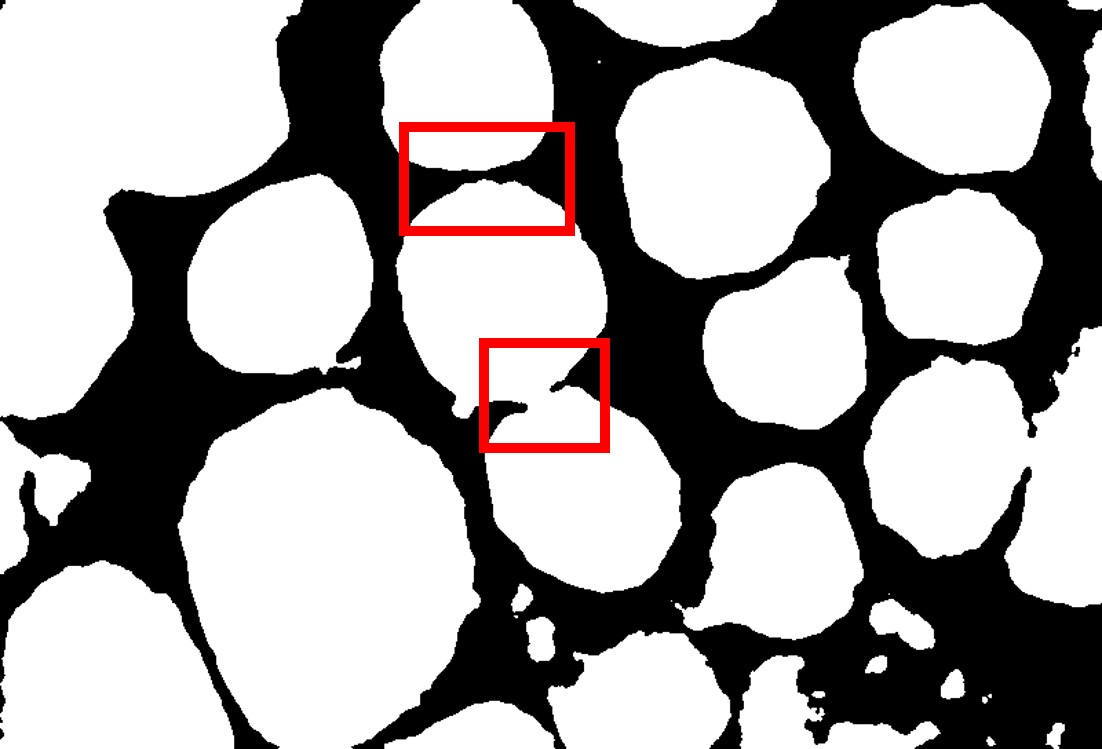}
    \end{subfigure} 
    \begin{subfigure}{0.115\textwidth}
        \includegraphics[width=\linewidth]{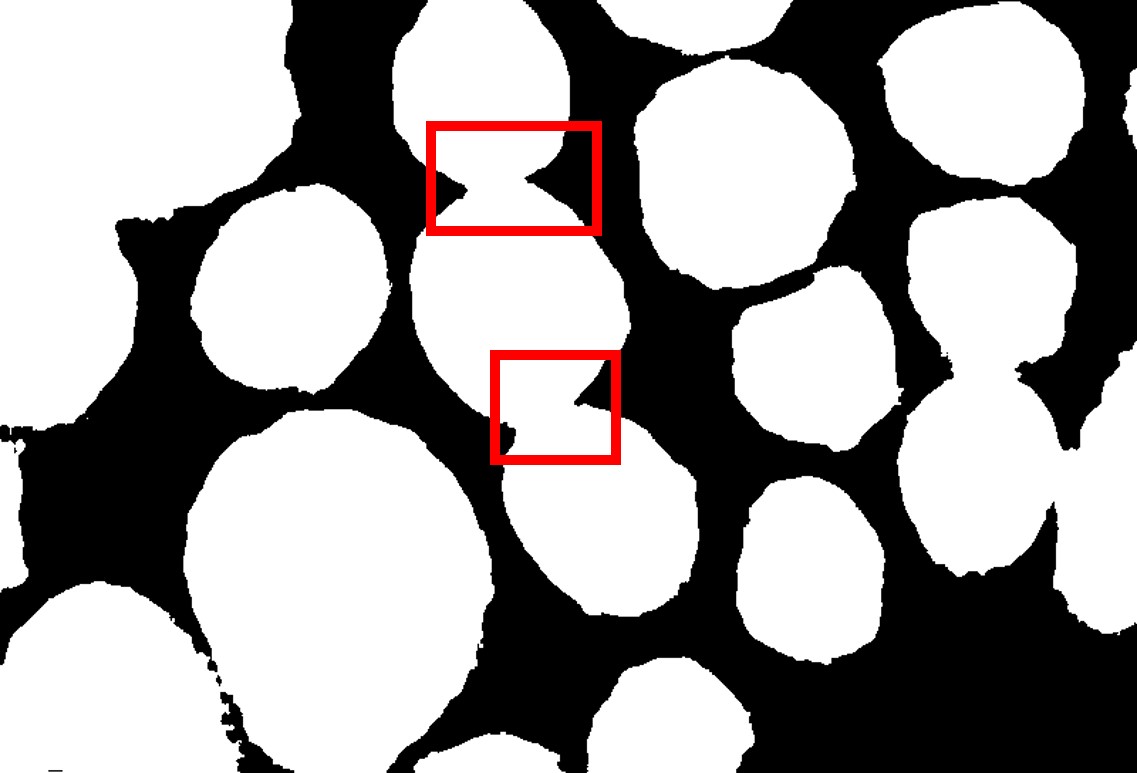}
    \end{subfigure} 
    \begin{subfigure}{0.115\textwidth}
        \includegraphics[width=\linewidth]{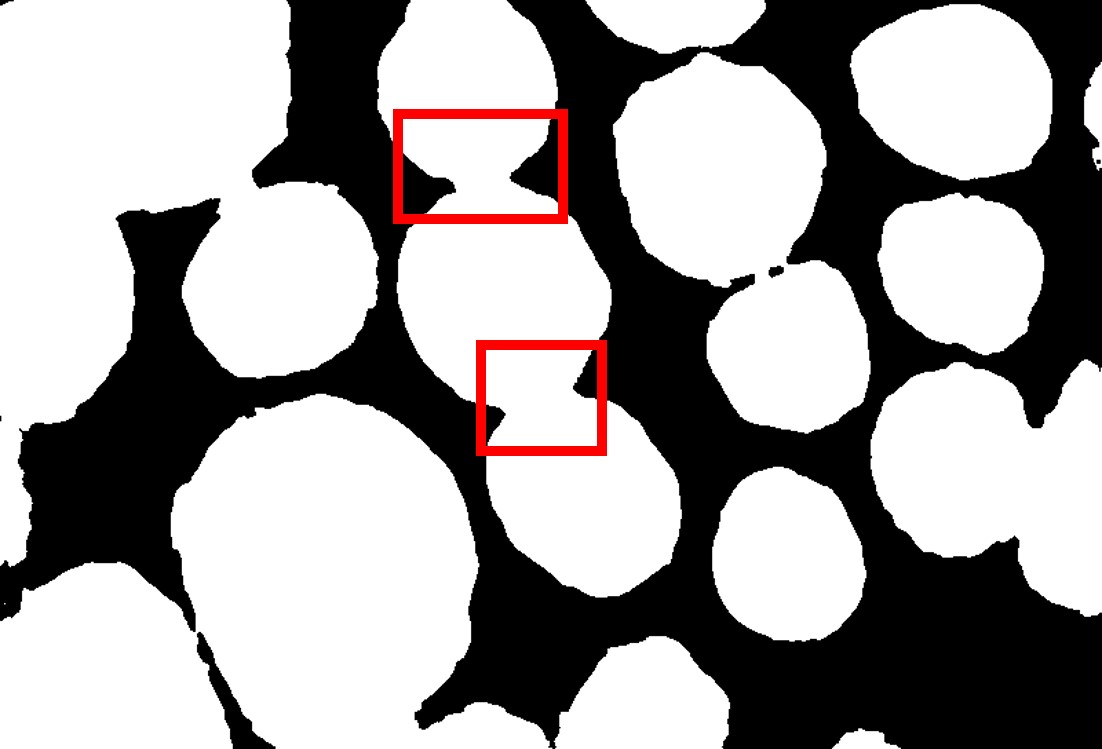}
    \end{subfigure} 
    \begin{subfigure}{0.115\textwidth}
        \includegraphics[width=\linewidth]{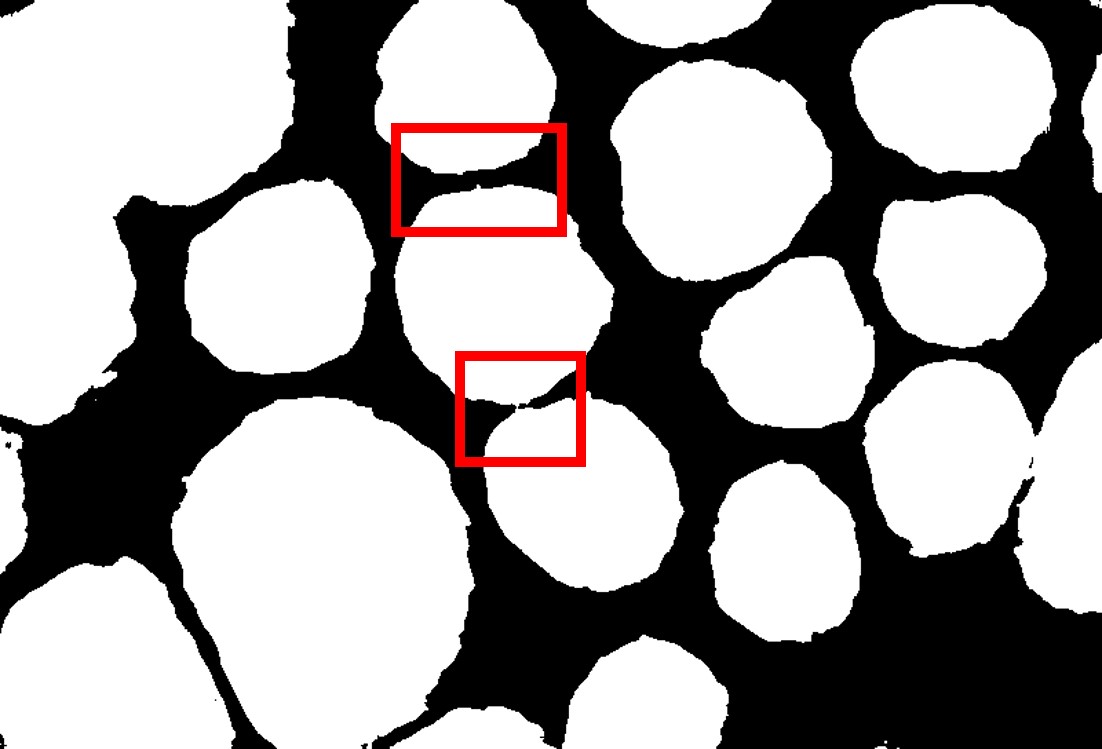}
    \end{subfigure} 

    \begin{subfigure}{0.115\textwidth}
        \includegraphics[width=\linewidth]{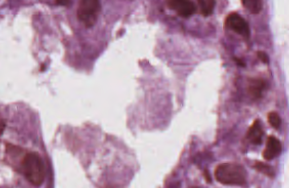}
    \end{subfigure}
    \begin{subfigure}{0.115\textwidth}
        \includegraphics[width=\linewidth]{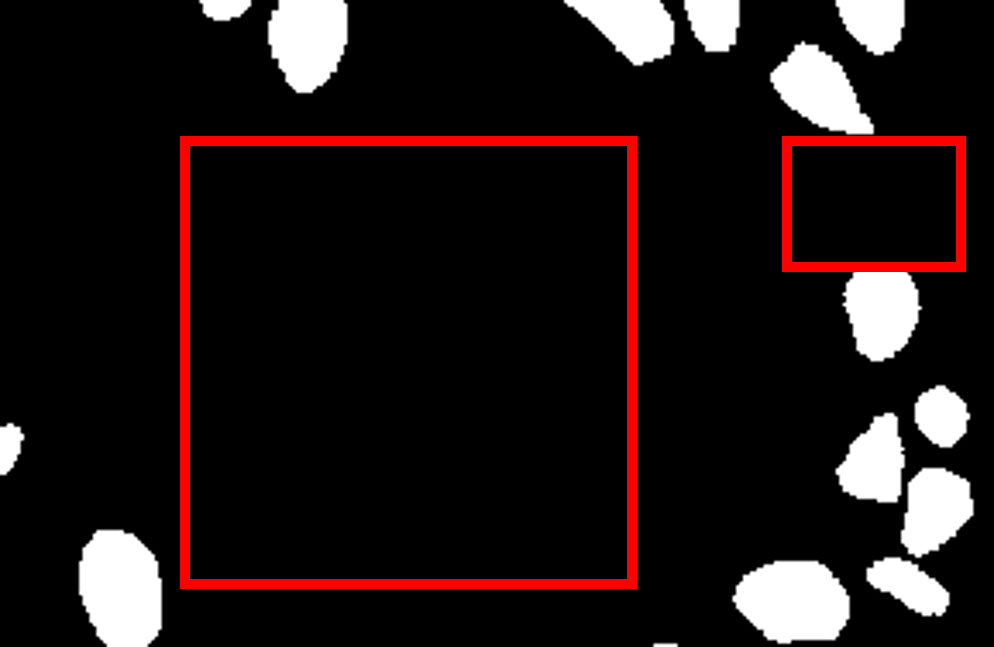}
    \end{subfigure} 
    \begin{subfigure}{0.115\textwidth}
        \includegraphics[width=\linewidth]{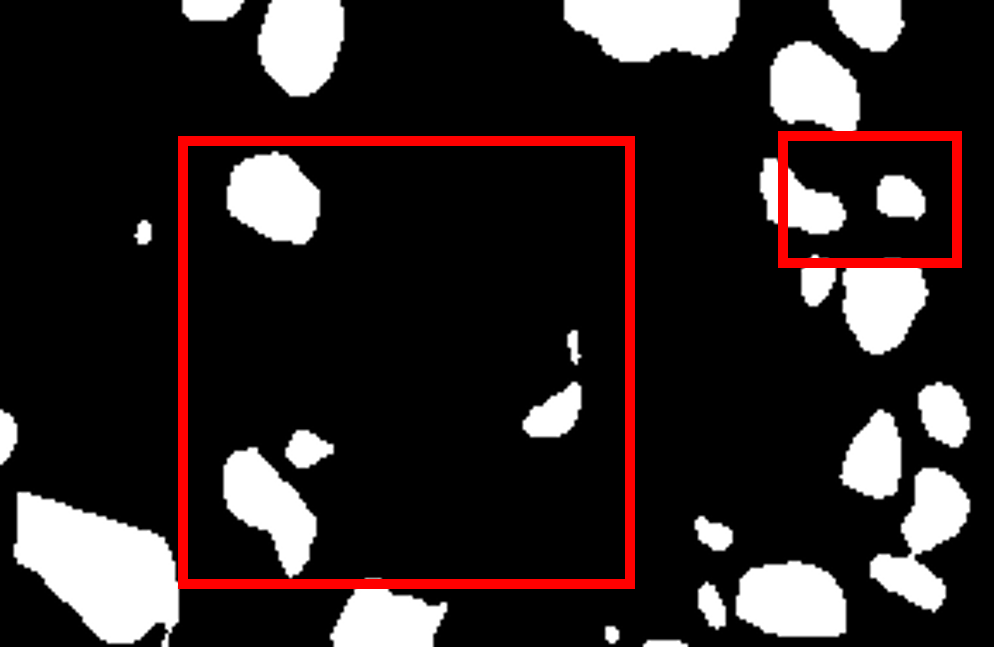}
    \end{subfigure} 
    \begin{subfigure}{0.115\textwidth}
        \includegraphics[width=\linewidth]{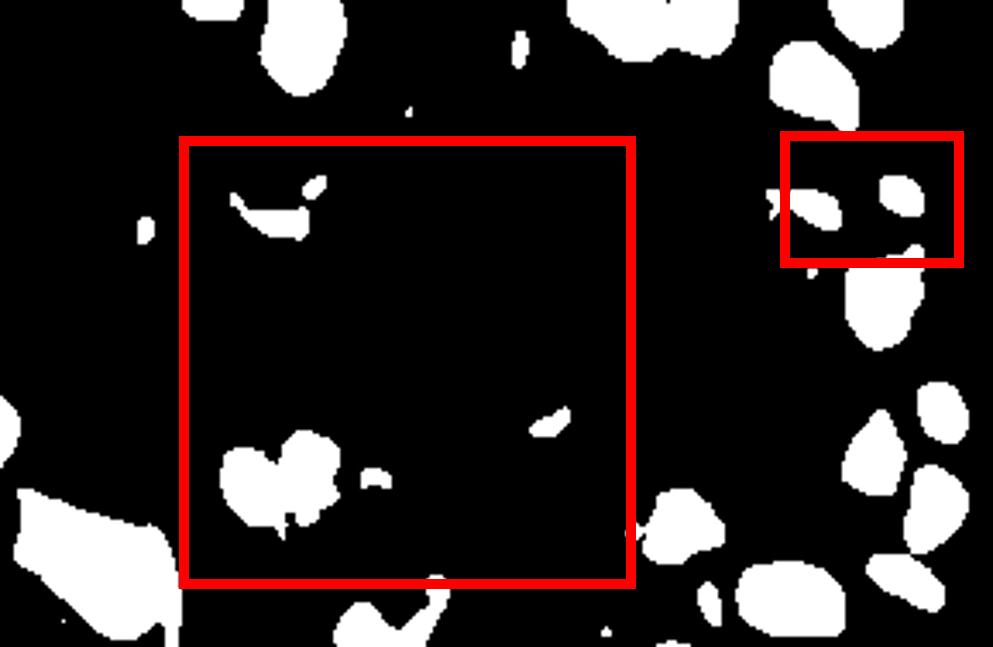}
    \end{subfigure} 
    \begin{subfigure}{0.115\textwidth}
        \includegraphics[width=\linewidth]{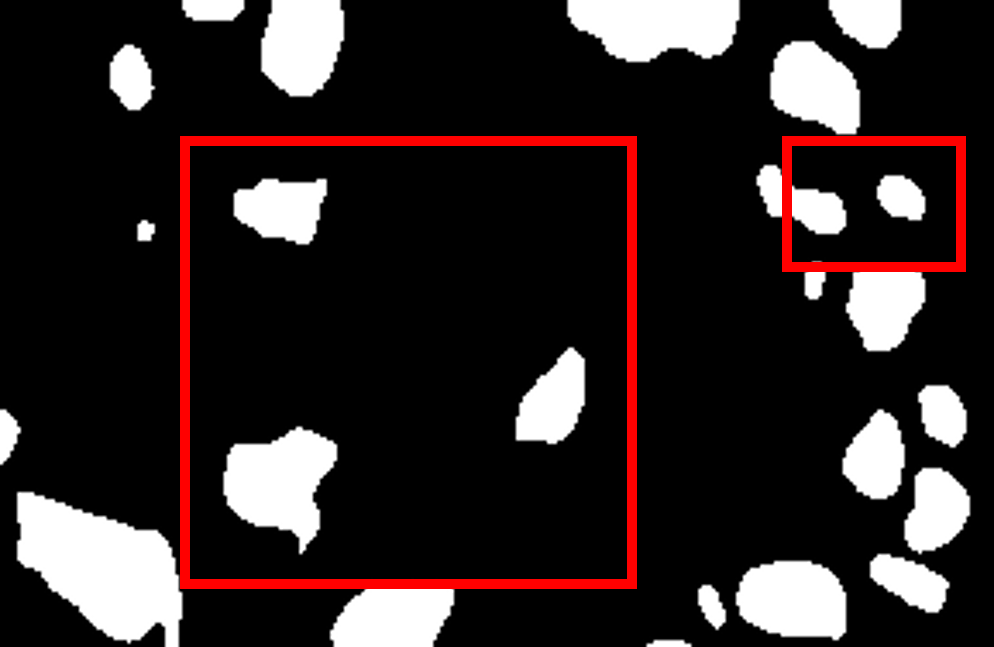}
    \end{subfigure}
    \begin{subfigure}{0.115\textwidth}
        \includegraphics[width=\linewidth]{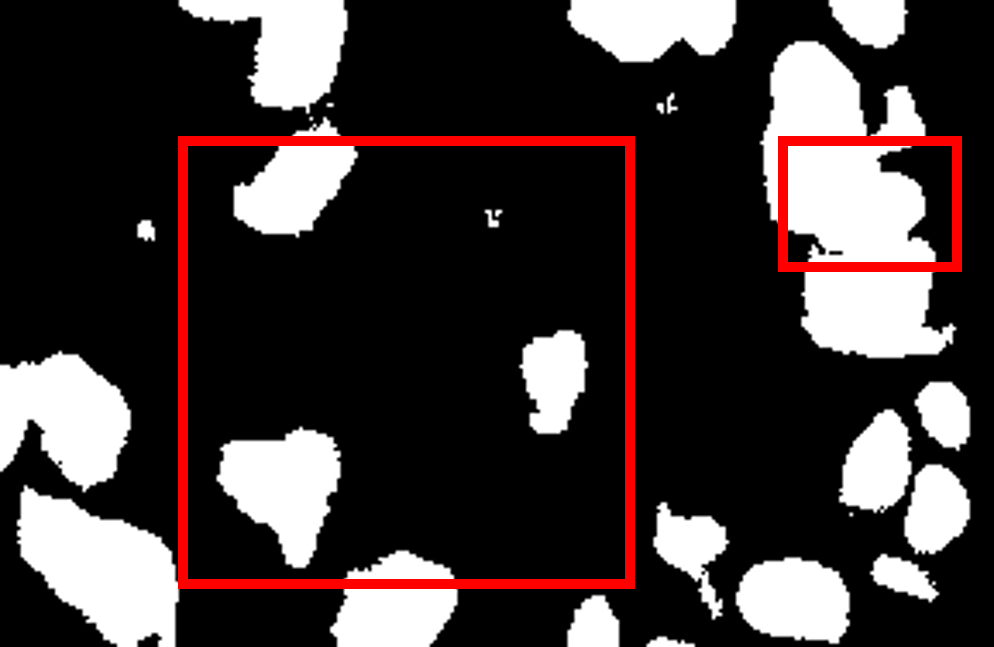}
    \end{subfigure} 
    \begin{subfigure}{0.115\textwidth}
        \includegraphics[width=\linewidth]{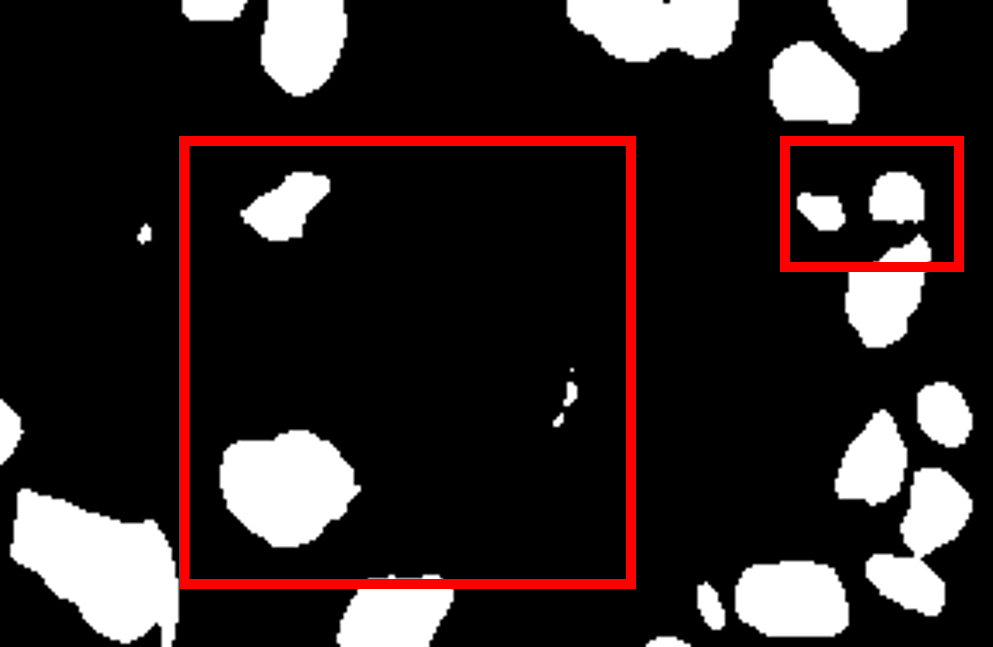}
    \end{subfigure} 
    \begin{subfigure}{0.115\textwidth}
        \includegraphics[width=\linewidth]{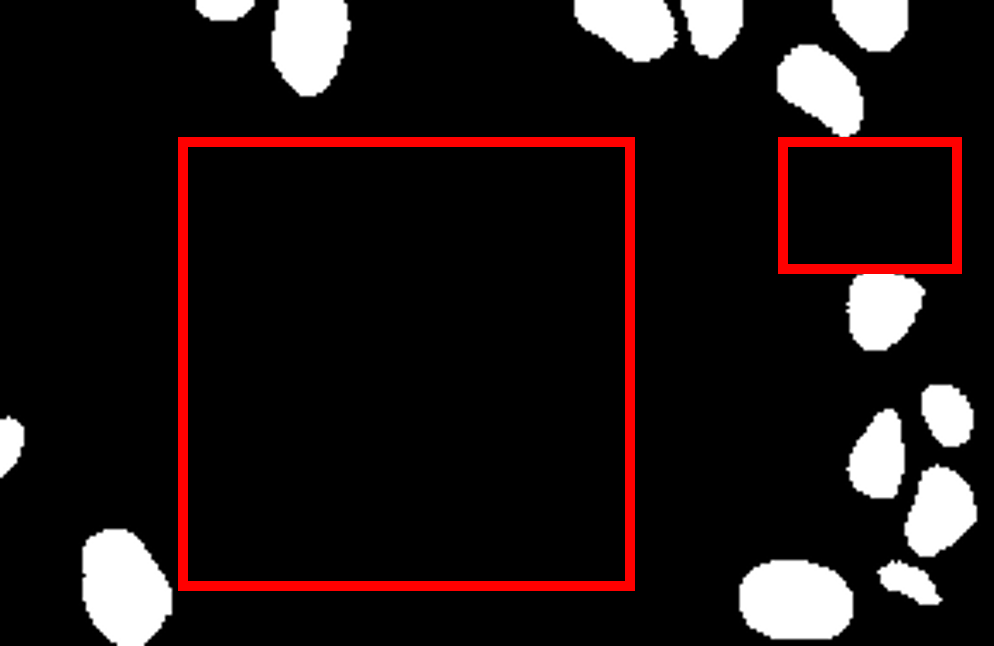}
    \end{subfigure} 

    \begin{subfigure}{0.115\textwidth}
        \includegraphics[width=\linewidth]{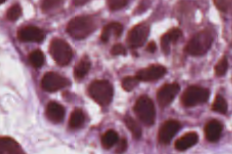}
        \caption{Image}
    \end{subfigure}
    \begin{subfigure}{0.115\textwidth}
        \includegraphics[width=\linewidth]{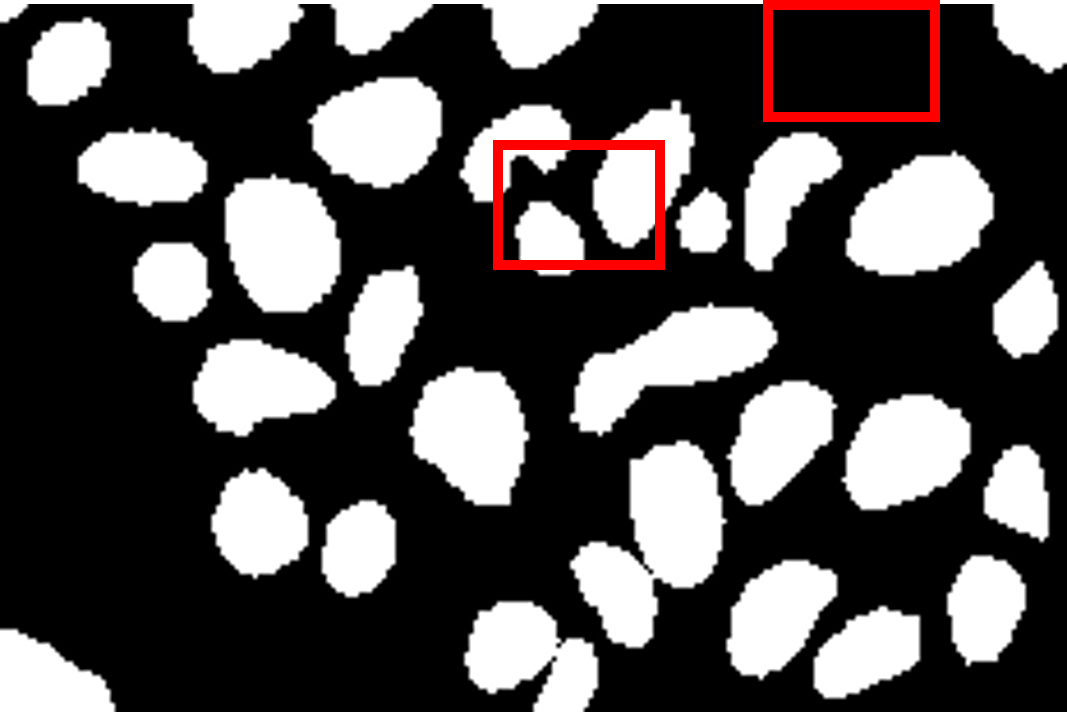}
        \caption{GT}
    \end{subfigure} 
    \begin{subfigure}{0.115\textwidth}
        \includegraphics[width=\linewidth]{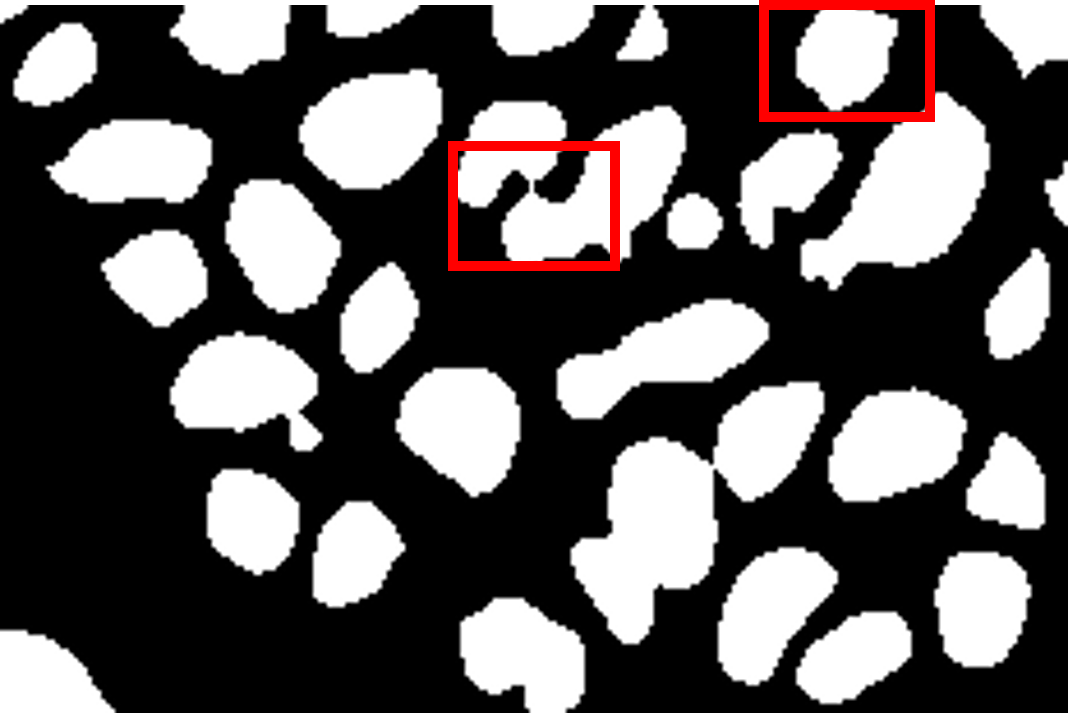}
        \caption{MT}
    \end{subfigure} 
    \begin{subfigure}{0.115\textwidth}
        \includegraphics[width=\linewidth]{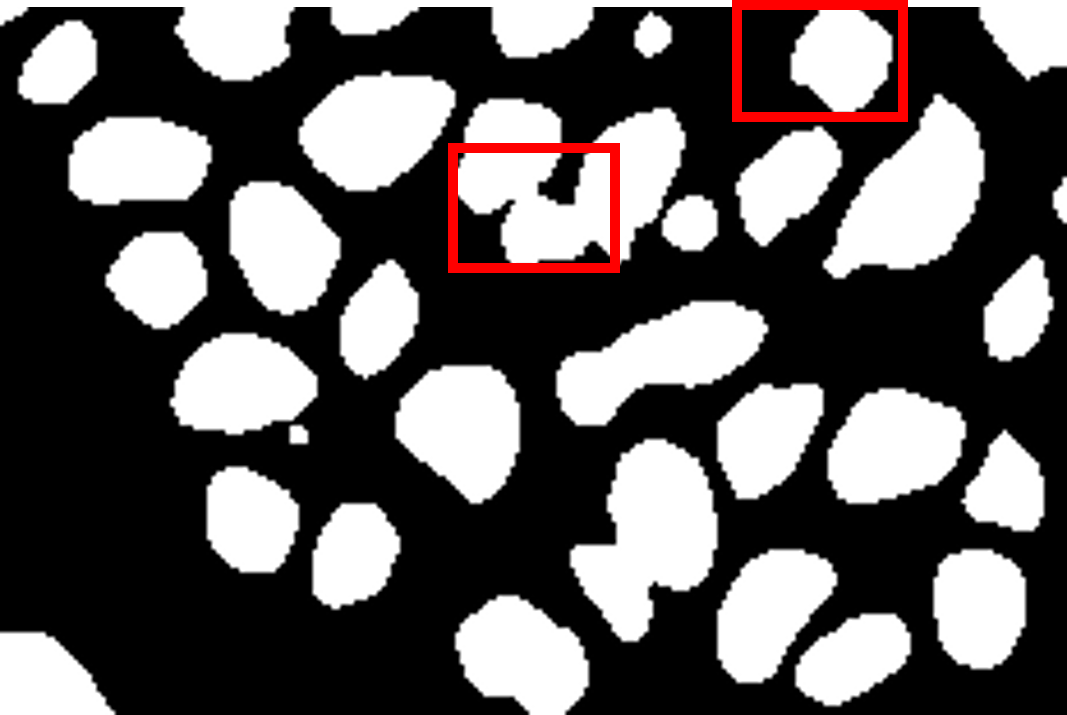}
        \caption{EM}
    \end{subfigure} 
    \begin{subfigure}{0.115\textwidth}
        \includegraphics[width=\linewidth]{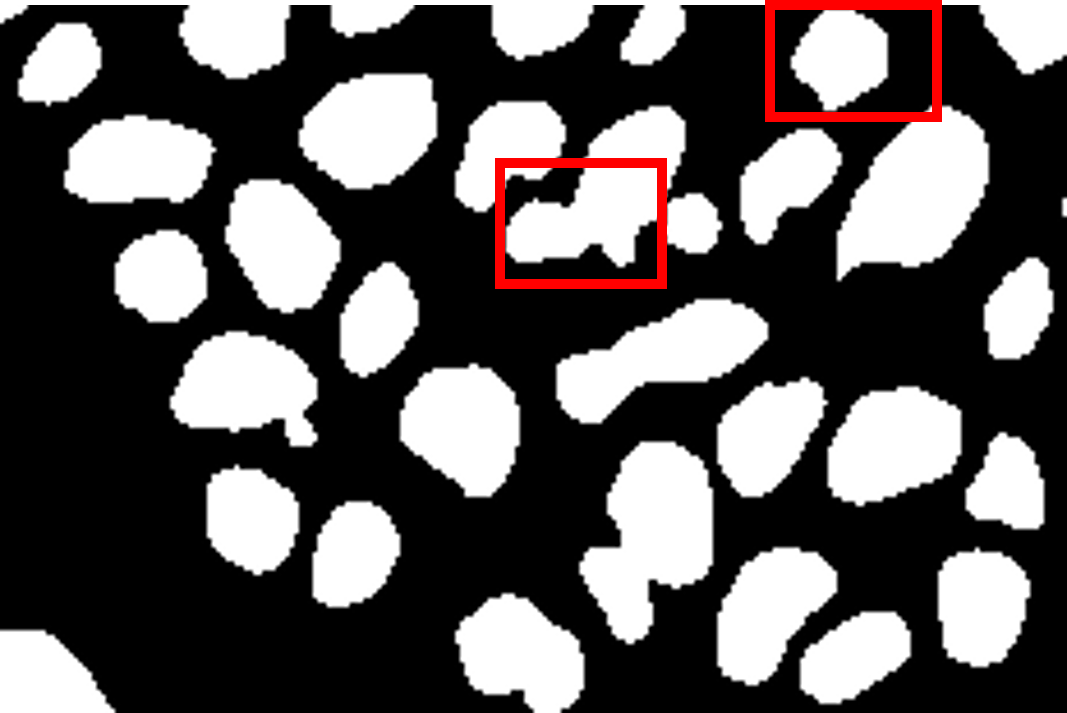}
        \caption{UAMT}
    \end{subfigure}
    \begin{subfigure}{0.115\textwidth}
        \includegraphics[width=\linewidth]{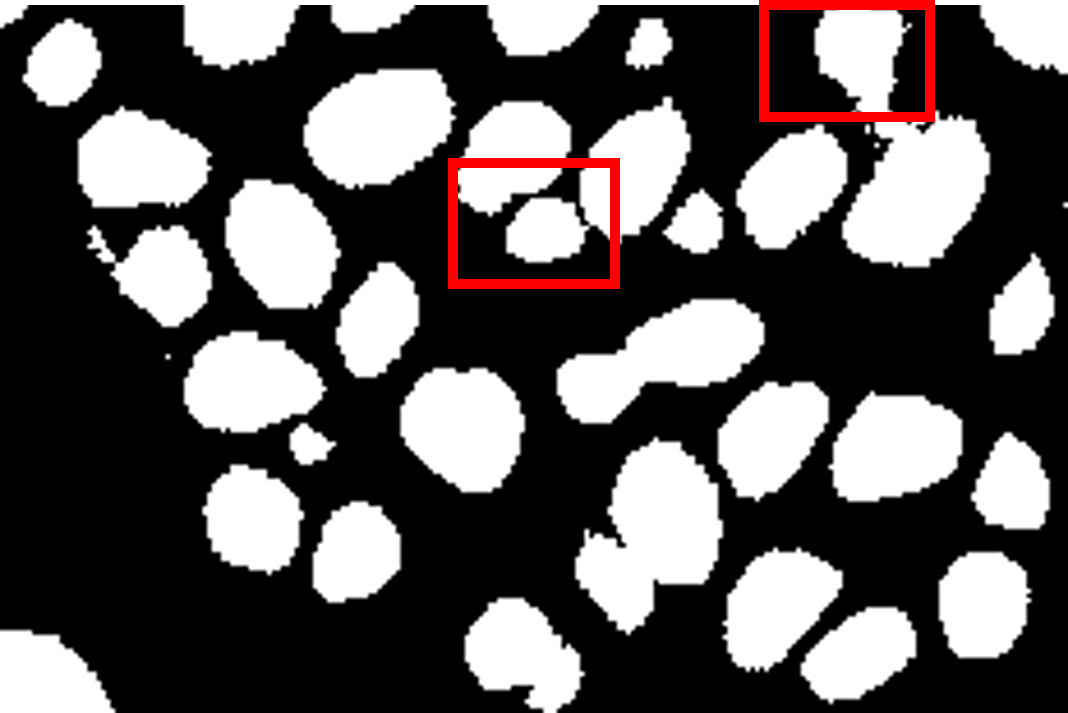}
        \caption{HCE}
    \end{subfigure} 
    \begin{subfigure}{0.115\textwidth}
        \includegraphics[width=\linewidth]{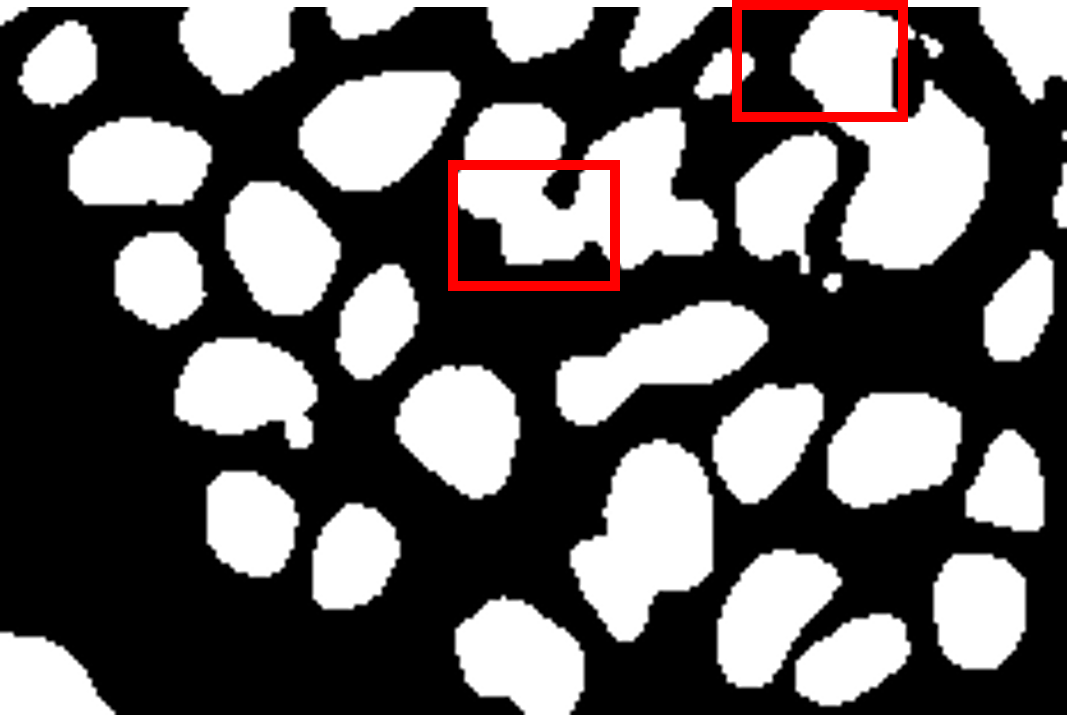}
        \caption{URPC}
    \end{subfigure} 
    \begin{subfigure}{0.115\textwidth}
        \includegraphics[width=\linewidth]{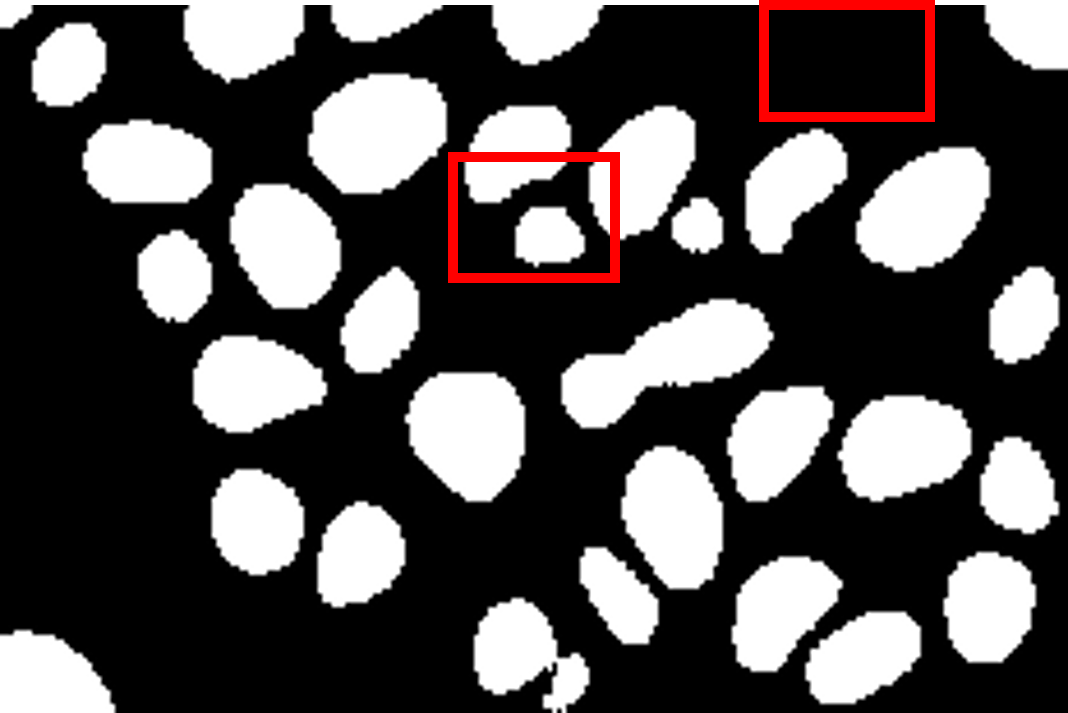}
        \caption{Ours}
    \end{subfigure}
    \caption{Qualitative results on three histopathology image datasets using $20\%$ labeled data for training. Locations prone to topological errors are shown within \textcolor{red}{\textbf{red}} boxes. Row 1: CRAG, Row 2: GlaS, Rows 3 \& 4: MoNuSeg. Zoom in for better views. 
    }
    \label{fig:Qualitative_Results}
\end{figure*}

\subsection{Results: Comparison with SoTA SemiSL methods}
We conduct experiments on different fractions of labeled data, specifically, $10\%$ and $20\%$. Training UNet++~\cite{zhou2018unet++} on $100\%$ of the labeled data is treated as the performance upper bound. To indicate the effectiveness and superiority of our method, we select several SoTA semi-supervised methods for comparison both from pixel and topological perspectives. Quantitative results are shown in \cref{tab:quant}, and qualitative results are shown in \cref{fig:Qualitative_Results}. We discuss more below.

\myparagraph{Quantitative Results.}
For a comprehensive comparison, we select several classical and recent SoTA SemiSL methods like MT~\cite{tarvainen2017mean}, EM~\cite{vu2019advent}, UA-MT~\cite{yu2019uncertainty}, HCE~\cite{jin2022semi_miccai}, URPC~\cite{luo2022semi}, XNet~\cite{zhou2023xnet} and CCT~\cite{ouali2020semi}. Note that HCE is re-implemented by ourselves due to code unavailability. As shown in \cref{tab:quant}, our method not only achieves comparable performance on pixel-wise evaluation metrics but also achieves the best results on all topology-wise metrics. This indicates that our proposed TopoSemiSeg is able to unearth and utilize topological information in unlabeled data well, without sacrificing pixel-level performance.

\myparagraph{Qualitative Results.}
In \cref{fig:Qualitative_Results}, we provide the qualitative results of the methods on $20\%$ labeled data for each dataset. Compared to other SoTA SemiSL methods, our method does better where topological errors are prone to occur, as shown in the \textcolor{red}{\textbf{red}} boxes. The proposed TopoSemiSeg ensures topological integrity: by enforcing signal consistency, we can maintain the thin separation between the densely distributed glands and cells. Additionally, the noise removal component of our loss minimizes the occurrence of false positive cells, as can be seen in Row 3. This is in contrast to the results obtained from the other baseline methods, which contain a discernible presence of noise and unoccupied interspaces in and around the glandular and cellular structures. Our method can effectively address and rectify these issues.
This is because we not only focus on the signal topology which should be preserved, but also remove all the noisy topology during training, thus making the model learn more robust and accurate topological representations from the unlabeled data. More qualitative results are provided in the Supplementary.

\subsection{Ablation Studies}
We conduct experiments to illustrate the effectiveness and robustness of our hyper-parameters selections and experimental settings. All experiments are performed on the CRAG dataset using $20\%$ labeled data. \textbf{To save space, additional ablation studies are provided in~\cref{addi_ablation_study} in the Supplementary.}

\myparagraph{Weight of Noise-aware Topological Consistency loss $\lambda^U_{2}$.}
We study the effect of the weight of the noise-aware topological consistency loss $\lambda^U_{2}$ introduced in \cref{unsupervised_loss}. As shown in \cref{ablation study:topo_loss_weight}, at $\lambda^U_{2} = 0.002$, the model achieves the best Object-level Dice coefficient, Betti Matching Error, and VOI. Additionally, a reasonable range of $\lambda^U_{2}$ always results in improvement. This demonstrates the efficacy and robustness of the proposed method.

\setlength{\tabcolsep}{7pt}
\begin{table}[ht]
\centering
\scriptsize
\caption{Ablation study on loss weight $\lambda^U_{2}$.}
\begin{tabular}{cccccc}
\hline
\multirow{2}{*}{$\lambda^U_{2}$} & Pixel-Wise & \multirow{2}{*}{} & \multicolumn{3}{c}{Topology-Wise} \\ \cline{2-2} \cline{4-6} 
 & Dice\_Obj $\uparrow$  & & Betti Error $\downarrow$ & Betti Matching Error $\downarrow$ & VOI $\downarrow$ \\ \hline
0 & 0.887 &  & 0.306 & 12.125 & 0.783 \\
0.001 & 0.874 &  & 0.217 & 12.175 & 0.736 \\
0.002 & \textbf{0.898} &  & 0.226 & \textbf{8.575} & \textbf{0.709} \\ 
0.005 & 0.889 &  & \textbf{0.213} & 9.875 & 0.739 \\
0.008 & 0.896 &  & 0.235 & 9.700 & 0.722 \\
0.01 & 0.873 &  & 0.277 & 9.725 & 0.754 \\ \hline
\end{tabular}
\label{ablation study:topo_loss_weight}
\end{table}

\myparagraph{Robustness of Persistence Threshold $\phi$.}  
To compute the noise-aware topological consistency loss, we define a persistence threshold $\phi$ to decompose both persistence diagrams into signal and noise parts. We conduct experiments on different values of $\phi$. As we can see from \cref{ablation study:latent factor}, our method is not sensitive to the value of $\phi$ and a wide range of $\phi$ (from $0.5$ to $0.9$) results in improvements on topological metrics. This demonstrates the robustness of our method with respect to perturbations. Moreover, note that $\phi=0$ means without decomposition (direct matching). As in~\cref{ablation study:latent factor}, this significantly hurts the performance. This is because both teacher and student predictions are noisy. Direct matching of diagrams will force the student to learn from many noisy structures, confusing the model and resulting in a significant performance drop. 

\setlength{\tabcolsep}{7pt}
\begin{table}[ht]
\centering
\scriptsize
\caption{Ablation study on persistence threshold $\phi$. }
\begin{tabular}{cccccc}
\hline
\multirow{2}{*}{$\phi$} & Pixel-Wise & \multirow{2}{*}{} & \multicolumn{3}{c}{Topology-Wise} \\ \cline{2-2} \cline{4-6} 
 & Dice\_Obj $\uparrow$  & & Betti Error $\downarrow$ & Betti Matching Error $\downarrow$ & VOI $\downarrow$ \\ \hline
0 & 0.880  &  & 0.317 & 14.025 & 0.762 \\
0.50 & 0.881  &  & 0.241 & 8.950 & 0.753 \\
0.60 & 0.895  &  & 0.219 & 9.600 & 0.725 \\
0.70 & \textbf{0.898} &  & 0.226 & \textbf{8.575} & \textbf{0.709} \\
0.80 & 0.896 &   & \textbf{0.209} & 9.000 & 0.722 \\
0.90 & 0.889 &   & 0.231 & 10.150 & 0.717 \\ \hline
\end{tabular}
\label{ablation study:latent factor}
\end{table}

\myparagraph{Generalizability to Different Backbones.}
We verify the generalizability of our method by performing experiments on three different backbones, UNet~\cite{ronneberger2015u}, PSPNet~\cite{zhao2017pyramid}, and DeepLabV3+~\cite{chen2018encoder}, keeping the same values of hyper-parameters for each. \cref{ablation study:different backbones} shows that our method is robust to backbone selections and can obtain performance improvements with each of them. For those with poor topological performances, like PSPNet~\cite{zhao2017pyramid}, our method significantly reduces the number of topological errors. This proves the effectiveness and generalizability of our method in that it can facilitate capturing topological information from the unlabeled data irrespective of the backbone.

\setlength{\tabcolsep}{7pt}
\begin{table}[ht]
\centering
\scriptsize
\caption{Comparison on different backbones. BE and BME respectively denote Betti Error and Betti Matching Error.}
\begin{tabular}{cccccc}
\hline
\multirow{2}{*}{Method} & Pixel-Wise &  & \multicolumn{3}{c}{Topology-Wise} \\ \cline{2-2} \cline{4-6} 
 & Dice\_Obj$\uparrow$ &  & BE$\downarrow$ & BME$\downarrow$ & VOI$\downarrow$\\ \hline
UNet~\cite{ronneberger2015u} & 0.892 &  & 0.266 & 10.775 & 0.790 \\
UNet~\cite{ronneberger2015u}+Ours & \textbf{0.893} &  & \textbf{0.236} & \textbf{8.700} & \textbf{0.722} \\ \hline
PSPNet~\cite{zhao2017pyramid} & 0.773 &  & 1.809 & 70.625 & 1.337 \\
PSPNet~\cite{zhao2017pyramid}+Ours & \textbf{ 0.775} &  & \textbf{1.021} & \textbf{44.150} & \textbf{1.040} \\ \hline
DeepLabV3+~\cite{chen2018encoder} & 0.883 &  & 0.293 & 14.000 & 0.725 \\
DeepLabV3+~\cite{chen2018encoder}+Ours & \textbf{0.891} &  & \textbf{0.265} & \textbf{11.725} & \textbf{0.713} \\ \hline
UNet++~\cite{zhou2018unet++} & 0.887 &  & 0.306 & 12.125 & 0.783  \\
UNet++~\cite{zhou2018unet++}+Ours & \textbf{0.898} &  & \textbf{0.226} & \textbf{8.575} & \textbf{0.709} \\ \hline
\end{tabular}
\label{ablation study:different backbones}
\end{table}

\myparagraph{Ablation study on loss components.}
We conduct an ablation study on the signal consistency loss and noise removal loss to verify their effectiveness.  
The signal topology $\mathcal{L}^{\text{U}}_{\text{topo-cons}}$ is used to guide the student to learn the meaningful structures from the teacher. Meanwhile, we use the noise removal loss $\mathcal{L}^{\text{U}}_{\text{topo-rem}}$ to directly train the student model to avoid noise in its prediction. 
In~\cref{table:decomposition}, we present the ablation study of these two components and from the performance, we can see that both components are necessary for us to enhance the segmentation quality.

\setlength{\tabcolsep}{7pt}
\begin{table}[ht]
\centering
\scriptsize
\caption{Ablation study on loss components.}
\begin{tabular}{cccccc}
\hline
\textbf{$\mathcal{L}^{\text{U}}_{\text{topo-cons}}$} & \textbf{$\mathcal{L}^{\text{U}}_{\text{topo-rem}}$} & Dice\_Obj $\uparrow$&  BE $\downarrow$& BME $\downarrow$& VOI $\downarrow$\\ \cline{1-6}
\xmark & \xmark & 0.887 &  0.306 & 12.125 & 0.783 \\ \hline
\cmark & \xmark & 0.890 &  0.275 & 10.150 & 0.715 \\ \hline
\xmark & \cmark & 0.888 &  0.230 & 10.525 & 0.723 \\ \hline
\cmark & \cmark & \textbf{0.898} &  \textbf{0.226} & \textbf{8.575} & \textbf{0.709} \\ \hline
\end{tabular}
\label{table:decomposition}
\end{table}

\myparagraph{Limitations.}
In the teacher-student framework, the prediction of the teacher model is not always accurate. On the contrary, the teacher model’s output may be unreliable sometimes. How to alleviate the uncertainty of the teacher model’s topological output will be the one of topics we will be exploring in the future. Further, the current decomposition threshold is set manually and needs to be set adaptively according to the data distribution. Finally, our current implementation is only available for 2D datasets, extending the proposed method to 3D scenarios with acceptable computational cost will be of great value in practice. For the failure cases, we observe that when the objects are extremely densely distributed, our model sometimes cannot split the glands in close proximity.
\section{Conclusion}
This work introduces TopoSemiSeg, the first semi-supervised method that learns topological representation from unlabeled histopathology images. It consists of a novel and differentiable noise-aware topological consistency loss integrated into the teacher-student framework. 
We propose to decompose the calculated persistence diagrams into true signal and noisy components, and respectively formulate signal consistency and noise removal losses from them. 
These losses enforce the model to learn a robust representation of topology from unlabeled data and can be incorporated into any variant of the teacher-student framework. Extensive experiments on several histopathology image datasets show the effectiveness of the proposed method on both pixel- and topology-wise evaluation metrics.

\myparagraph{Acknowledgements.} We thank the anonymous reviewers for their constructive feedback. This research was partially supported by the National Science Foundation (NSF) grant CCF-2144901, the National Institute of General Medical Sciences (NIGMS) grant R01GM148970, and the Stony Brook Trustees Faculty Award.


\clearpage  

%
%
{
\bibliographystyle{splncs04}
\bibliography{egbib}

\begin{thebibliography}{10}
\providecommand{\url}[1]{\texttt{#1}}
\providecommand{\urlprefix}{URL }
\providecommand{\doi}[1]{https://doi.org/#1}

\bibitem{basak2023pseudo}
Basak, H., Yin, Z.: Pseudo-label guided contrastive learning for semi-supervised medical image segmentation. In: CVPR (2023)

\bibitem{berthelot2019mixmatch}
Berthelot, D., Carlini, N., Goodfellow, I., Papernot, N., Oliver, A., Raffel, C.A.: Mixmatch: A holistic approach to semi-supervised learning. In: NeurIPS (2019)

\bibitem{cao2022swin}
Cao, H., Wang, Y., Chen, J., Jiang, D., Zhang, X., Tian, Q., Wang, M.: Swin-unet: Unet-like pure transformer for medical image segmentation. In: ECCV (2022)

\bibitem{chen2021mtans}
Chen, G., Ru, J., Zhou, Y., Rekik, I., Pan, Z., Liu, X., Lin, Y., Lu, B., Shi, J.: Mtans: Multi-scale mean teacher combined adversarial network with shape-aware embedding for semi-supervised brain lesion segmentation. NeuroImage  (2021)

\bibitem{chen2018encoder}
Chen, L.C., Zhu, Y., Papandreou, G., Schroff, F., Adam, H.: Encoder-decoder with atrous separable convolution for semantic image segmentation. In: ECCV (2018)

\bibitem{clough2020topological}
Clough, J.R., Byrne, N., Oksuz, I., Zimmer, V.A., Schnabel, J.A., King, A.P.: A topological loss function for deep-learning based image segmentation using persistent homology. TPAMI  (2020)

\bibitem{cohen2005stability}
Cohen-Steiner, D., Edelsbrunner, H., Harer, J.: Stability of persistence diagrams. In: Proceedings of the twenty-first annual symposium on Computational geometry (2005)

\bibitem{cohen2010lipschitz}
Cohen-Steiner, D., Edelsbrunner, H., Harer, J., Mileyko, Y.: Lipschitz functions have l p-stable persistence. Foundations of Computational Mathematics  (2010)

\bibitem{edelsbrunner2002topological}
Edelsbrunner, Letscher, Zomorodian: Topological persistence and simplification. Discrete \& Computational Geometry  (2002)

\bibitem{edelsbrunner2022computational}
Edelsbrunner, H., Harer, J.L.: Computational topology: an introduction. American Mathematical Society (2022)

\bibitem{fang2020dmnet}
Fang, K., Li, W.J.: Dmnet: difference minimization network for semi-supervised segmentation in medical images. In: MICCAI (2020)

\bibitem{fleming2012colorectal}
Fleming, M., Ravula, S., Tatishchev, S.F., Wang, H.L.: Colorectal carcinoma: Pathologic aspects. Journal of Gastrointestinal Oncology  (2012)

\bibitem{graham2019mild}
Graham, S., Chen, H., Gamper, J., Dou, Q., Heng, P.A., Snead, D., Tsang, Y.W., Rajpoot, N.: Mild-net: Minimal information loss dilated network for gland instance segmentation in colon histology images. MedIA  (2019)

\bibitem{grandvalet2004semi}
Grandvalet, Y., Bengio, Y.: Semi-supervised learning by entropy minimization. In: NeurIPS (2004)

\bibitem{gupta2022learning}
Gupta, S., Hu, X., Kaan, J., Jin, M., Mpoy, M., Chung, K., Singh, G., Saltz, M., Kurc, T., Saltz, J., et~al.: Learning topological interactions for multi-class medical image segmentation. In: ECCV (2022)

\bibitem{gupta2023topology}
Gupta, S., Zhang, Y., Hu, X., Prasanna, P., Chen, C.: Topology-aware uncertainty for image segmentation. In: NeurIPS (2023)

\bibitem{hu2022structure}
Hu, X.: Structure-aware image segmentation with homotopy warping. In: NeurIPS (2022)

\bibitem{hu2019topology}
Hu, X., Li, F., Samaras, D., Chen, C.: Topology-preserving deep image segmentation. In: NeurIPS (2019)

\bibitem{hu2023learning}
Hu, X., Samaras, D., Chen, C.: Learning probabilistic topological representations using discrete morse theory. In: ICLR (2023)

\bibitem{hu2020topology}
Hu, X., Wang, Y., Fuxin, L., Samaras, D., Chen, C.: Topology-aware segmentation using discrete morse theory. In: ICLR (2021)

\bibitem{huang2022semi}
Huang, W., Chen, C., Xiong, Z., Zhang, Y., Chen, X., Sun, X., Wu, F.: Semi-supervised neuron segmentation via reinforced consistency learning. TMI  (2022)

\bibitem{isensee2021nnu}
Isensee, F., Jaeger, P.F., Kohl, S.A., Petersen, J., Maier-Hein, K.H.: nnu-net: a self-configuring method for deep learning-based biomedical image segmentation. Nature Methods  (2021)

\bibitem{jeong2019consistency}
Jeong, J., Lee, S., Kim, J., Kwak, N.: Consistency-based semi-supervised learning for object detection. In: NeurIPS (2019)

\bibitem{jiao2022learning}
Jiao, R., Zhang, Y., Ding, L., Cai, R., Zhang, J.: Learning with limited annotations: a survey on deep semi-supervised learning for medical image segmentation. arXiv preprint arXiv:2207.14191  (2022)

\bibitem{jin2022semi_miccai}
Jin, Q., Cui, H., Sun, C., Zheng, J., Wei, L., Fang, Z., Meng, Z., Su, R.: Semi-supervised histological image segmentation via hierarchical consistency enforcement. In: MICCAI (2022)

\bibitem{jin2022semi}
Jin, Y., Wang, J., Lin, D.: Semi-supervised semantic segmentation via gentle teaching assistant. In: NeurIPS (2022)

\bibitem{kerber2016geometry}
Kerber, M., Morozov, D., Nigmetov, A.: Geometry helps to compare persistence diagrams. In: 2016 Proceedings of the Eighteenth Workshop on Algorithm Engineering and Experiments (ALENEX). SIAM (2016)

\bibitem{kumar2019multi}
Kumar, N., Verma, R., Anand, D., Zhou, Y., Onder, O.F., Tsougenis, E., Chen, H., Heng, P.A., Li, J., Hu, Z., et~al.: A multi-organ nucleus segmentation challenge. TMI  (2019)

\bibitem{lacombe2018large}
Lacombe, T., Cuturi, M., Oudot, S.: Large scale computation of means and clusters for persistence diagrams using optimal transport. In: NeurIPS (2018)

\bibitem{li2023calibrating}
Li, C., Hu, X., Abousamra, S., Chen, C.: Calibrating uncertainty for semi-supervised crowd counting. In: ICCV (2023)

\bibitem{li2020transformation}
Li, X., Yu, L., Chen, H., Fu, C.W., Xing, L., Heng, P.A.: Transformation-consistent self-ensembling model for semisupervised medical image segmentation. TNNLS  (2020)

\bibitem{li2021dual}
Li, Y., Luo, L., Lin, H., Chen, H., Heng, P.A.: Dual-consistency semi-supervised learning with uncertainty quantification for covid-19 lesion segmentation from ct images. In: MICCAI (2021)

\bibitem{luo2022semi}
Luo, X., Wang, G., Liao, W., Chen, J., Song, T., Chen, Y., Zhang, S., Metaxas, D.N., Zhang, S.: Semi-supervised medical image segmentation via uncertainty rectified pyramid consistency. MedIA  (2022)

\bibitem{meilua2003comparing}
Meil{\u{a}}, M.: Comparing clusterings by the variation of information. In: Learning Theory and Kernel Machines: 16th Annual Conference on Learning Theory and 7th Kernel Workshop, COLT/Kernel 2003 (2003)

\bibitem{montironi2005gleason}
Montironi, R., Mazzuccheli, R., Scarpelli, M., Lopez-Beltran, A., Fellegara, G., Algaba, F.: Gleason grading of prostate cancer in needle biopsies or radical prostatectomy specimens: contemporary approach, current clinical significance and sources of pathology discrepancies. BJU International  (2005)

\bibitem{munkres1984elements}
Munkres, J.R.: Elements of algebraic topology (1984)

\bibitem{ouali2020semi}
Ouali, Y., Hudelot, C., Tami, M.: Semi-supervised semantic segmentation with cross-consistency training. In: CVPR (2020)

\bibitem{ronneberger2015u}
Ronneberger, O., Fischer, P., Brox, T.: U-net: Convolutional networks for biomedical image segmentation. In: MICCAI (2015)

\bibitem{seibold2022reference}
Seibold, C.M., Rei{\ss}, S., Kleesiek, J., Stiefelhagen, R.: Reference-guided pseudo-label generation for medical semantic segmentation. In: AAAI (2022)

\bibitem{shi2021inconsistency}
Shi, Y., Zhang, J., Ling, T., Lu, J., Zheng, Y., Yu, Q., Qi, L., Gao, Y.: Inconsistency-aware uncertainty estimation for semi-supervised medical image segmentation. TMI  (2021)

\bibitem{shit2021cldice}
Shit, S., Paetzold, J.C., Sekuboyina, A., Ezhov, I., Unger, A., Zhylka, A., Pluim, J.P., Bauer, U., Menze, B.H.: cldice-a novel topology-preserving loss function for tubular structure segmentation. In: CVPR (2021)

\bibitem{sirinukunwattana2017gland}
Sirinukunwattana, K., Pluim, J.P., Chen, H., Qi, X., Heng, P.A., Guo, Y.B., Wang, L.Y., Matuszewski, B.J., Bruni, E., Sanchez, U., et~al.: Gland segmentation in colon histology images: The glas challenge contest. MedIA  (2017)

\bibitem{sohn2020fixmatch}
Sohn, K., Berthelot, D., Carlini, N., Zhang, Z., Zhang, H., Raffel, C.A., Cubuk, E.D., Kurakin, A., Li, C.L.: Fixmatch: Simplifying semi-supervised learning with consistency and confidence. In: NeurIPS (2020)

\bibitem{stucki2023topologically}
Stucki, N., Paetzold, J.C., Shit, S., Menze, B., Bauer, U.: Topologically faithful image segmentation via induced matching of persistence barcodes. In: ICML (2023)

\bibitem{sudre2017generalised}
Sudre, C.H., Li, W., Vercauteren, T., Ourselin, S., Jorge~Cardoso, M.: Generalised dice overlap as a deep learning loss function for highly unbalanced segmentations. In: Deep Learning in Medical Image Analysis and Multimodal Learning for Clinical Decision Support: Third International Workshop, DLMIA 2017, and 7th International Workshop, ML-CDS 2017, Held in Conjunction with MICCAI 2017 (2017)

\bibitem{tarvainen2017mean}
Tarvainen, A., Valpola, H.: Mean teachers are better role models: Weight-averaged consistency targets improve semi-supervised deep learning results. In: NeurIPS (2017)

\bibitem{thompson2022pseudo}
Thompson, B.H., Di~Caterina, G., Voisey, J.P.: Pseudo-label refinement using superpixels for semi-supervised brain tumour segmentation. In: ISBI (2022)

\bibitem{vu2019advent}
Vu, T.H., Jain, H., Bucher, M., Cord, M., P{\'e}rez, P.: Advent: Adversarial entropy minimization for domain adaptation in semantic segmentation. In: CVPR (2019)

\bibitem{wang2020topogan}
Wang, F., Liu, H., Samaras, D., Chen, C.: Topogan: A topology-aware generative adversarial network. In: ECCV (2020)

\bibitem{wang2022uctransnet}
Wang, H., Cao, P., Wang, J., Zaiane, O.R.: Uctransnet: rethinking the skip connections in u-net from a channel-wise perspective with transformer. In: AAAI (2022)

\bibitem{wang2022ta}
Wang, H., Xian, M., Vakanski, A.: Ta-net: Topology-aware network for gland segmentation. In: WACV (2022)

\bibitem{wang2022ssa}
Wang, X., Yuan, Y., Guo, D., Huang, X., Cui, Y., Xia, M., Wang, Z., Bai, C., Chen, S.: Ssa-net: Spatial self-attention network for covid-19 pneumonia infection segmentation with semi-supervised few-shot learning. MedIA  (2022)

\bibitem{wu2022cross}
Wu, H., Wang, Z., Song, Y., Yang, L., Qin, J.: Cross-patch dense contrastive learning for semi-supervised segmentation of cellular nuclei in histopathologic images. In: CVPR (2022)

\bibitem{wu2022mutual}
Wu, Y., Ge, Z., Zhang, D., Xu, M., Zhang, L., Xia, Y., Cai, J.: Mutual consistency learning for semi-supervised medical image segmentation. MedIA  (2022)

\bibitem{xie2019deep}
Xie, Y., Lu, H., Zhang, J., Shen, C., Xia, Y.: Deep segmentation-emendation model for gland instance segmentation. In: MICCAI (2019)

\bibitem{yang2021topological}
Yang, J., Hu, X., Chen, C., Tsai, C.: A topological-attention convlstm network and its application to em images. In: MICCAI (2021)

\bibitem{yao2022enhancing}
Yao, H., Hu, X., Li, X.: Enhancing pseudo label quality for semi-supervised domain-generalized medical image segmentation. In: AAAI (2022)

\bibitem{you2023rethinking}
You, C., Dai, W., Min, Y., Liu, F., Zhang, X., Feng, C., Clifton, D.A., Zhou, S.K., Staib, L.H., Duncan, J.S.: Rethinking semi-supervised medical image segmentation: A variance-reduction perspective. In: NeurIPS (2023)

\bibitem{you2022simcvd}
You, C., Zhou, Y., Zhao, R., Staib, L., Duncan, J.S.: Simcvd: Simple contrastive voxel-wise representation distillation for semi-supervised medical image segmentation. TMI  (2022)

\bibitem{yu2019uncertainty}
Yu, L., Wang, S., Li, X., Fu, C.W., Heng, P.A.: Uncertainty-aware self-ensembling model for semi-supervised 3d left atrium segmentation. In: MICCAI (2019)

\bibitem{zhang2022boostmis}
Zhang, W., Zhu, L., Hallinan, J., Zhang, S., Makmur, A., Cai, Q., Ooi, B.C.: Boostmis: Boosting medical image semi-supervised learning with adaptive pseudo labeling and informative active annotation. In: CVPR (2022)

\bibitem{zhang2022discriminative}
Zhang, Z., Tian, C., Bai, H.X., Jiao, Z., Tian, X.: Discriminative error prediction network for semi-supervised colon gland segmentation. MedIA  (2022)

\bibitem{zhao2017pyramid}
Zhao, H., Shi, J., Qi, X., Wang, X., Jia, J.: Pyramid scene parsing network. In: CVPR (2017)

\bibitem{zhou2023xnet}
Zhou, Y., Huang, J., Wang, C., Song, L., Yang, G.: Xnet: Wavelet-based low and high frequency fusion networks for fully-and semi-supervised semantic segmentation of biomedical images. In: ICCV (2023)

\bibitem{zhou2018unet++}
Zhou, Z., Rahman~Siddiquee, M.M., Tajbakhsh, N., Liang, J.: Unet++: A nested u-net architecture for medical image segmentation. In: Deep Learning in Medical Image Analysis and Multimodal Learning for Clinical Decision Support: 4th International Workshop, DLMIA 2018, and 8th International Workshop, ML-CDS 2018, Held in Conjunction with MICCAI 2018 (2018)

\end{thebibliography}
}

\clearpage

\title{Semi-supervised Segmentation of Histopathology Images with Noise-Aware Topological Consistency\\--- Supplementary Material ---}  

\titlerunning{Topological Consistency for Semi-supervised Segmentation - Supplementary}

\author{Meilong Xu\inst{1} \and Xiaoling Hu\inst{2} \and Saumya Gupta\inst{1} \and Shahira Abousamra\inst{1} \and Chao Chen\inst{1}}

\authorrunning{M.~Xu et al.}

\institute{Stony Brook University, Stony Brook, NY, USA \and 
Athinoula A. Martinos Center for Biomedical Imaging, \\
Massachusetts General Hospital and Harvard Medical School, MA, USA\\
\email{meixu@cs.stonybrook.edu}}

\maketitle
\setcounter{section}{5}
\setcounter{figure}{5}
\setcounter{table}{5}
\setcounter{footnote}{2}

In the supplementary material, we begin with notations for foreground and background in \cref{sec:fore_back_note}, followed by a detailed introduction to persistent homology in~\cref{supp:bg}. Then, we describe the correspondence between persistent dots and the likelihood map in \cref{sec:mapping}. 
Next, we discuss the differentiability of the noise-aware topological consistency loss in \cref{sec:differentiability}.
In \cref{sec:data_detailed}, we provide detailed descriptions of the datasets, followed by implementation details in \cref{sec:impl}. We also provide the reference of our baselines in \cref{sec:baseline_ref}. In \cref{sec:evalmet_detailed}, we describe the evaluation metrics in detail. More qualitative results are given in \cref{sec:addi_qua_results}. Finally, additional ablation studies and results are provided in~\cref{addi_ablation_study}.

\section{Notes on Foreground and Background}
\label{sec:fore_back_note}
Here, we provide some notations about foreground and background in our paper. Our algorithm uses black as the foreground and white as the background as can be seen in~\cref{fig:illustration}-~\cref{fig:overall_pipeline} of the main paper and \cref{fig:pdlh} of the Supplementary. For better visualization, however, we display the segmentation results and ground truth with white as the foreground in~\cref{fig:Motivation_sample} and~\cref{fig:Qualitative_Results} of the main paper and \cref{fig:Qualitative_ResultsSup} of the Supplementary.

\section{Background: Persistent Homology}
\label{supp:bg}
In algebraic topology~\cite{munkres1984elements}, \textit{homology classes} account for topological structures in all dimensions. 0-, 1-, and 2-dimensional structures describe connected components, loops/holes, and cavities/voids, respectively. For binary images, the number of $d$-dimensional topological structures is called the \emph{$d$-dimensional Betti number}, $\beta_d$.\footnote{Technically, $\beta_d$ counts the dimension of the $d$-dimensional homology group. The number of distinct homology classes/topological structures is exponential to $\beta_d$.} 
Despite the well-understood topological space for a binary image, the theory does not directly extend to real-world scenarios with continuous, noisy data. 
For example, in image analysis, we need a principled tool to reason about the topology from a continuous likelihood map. To bridge this gap, the theory of \textit{persistent homology} was invented in the early 2000s~\cite{edelsbrunner2002topological}. 

Persistent homology has emerged as a powerful tool for analyzing the topology of various kinds of real-world data, including images. In the image segmentation task, we apply persistent homology to the likelihood map of a deep neural network to reason about its topology. Given an image in the 2D domain \mbox{$I \subseteq \mathbb{R}^2$}, we use a network to generate a likelihood map $f$. The segmentation map is obtained by thresholding $f$ at a certain threshold $c$ (usually $0.5$). We  define a \emph{sublevel set}: $S_c:=\{(m,n)\in I \: | \: f(m,n)\leq c\}$. With all different threshold values sorted in an increasing order ($c_1 < c_2 < \cdots < c_n$), we obtain a filtration, i.e., a series of growing sublevel sets: $ \varnothing \subseteq S_{c_1} \subseteq S_{c_2} \subseteq ... \subseteq S_{c_n} = I$. As the threshold $c$ increases, topology of the sublevel set changes. New topological structures appear while old ones disappear.

Persistent homology tracks the evolution of all topological structures, such as connected components and loops. All the topological structures and their birth/death times are captured in a so-called \textit{persistence diagram}, providing a multi-scale topological representation (See \cref{fig:illustration}). 

A persistence diagram (PD) consists of multiple dots in a 2-dimensional plane. These dots are called \textit{persistent dots}. Given a continuous-valued likelihood map function $f$, we have its persistence diagram $Dgm(f)$. Each persistent dot $p\in Dgm(f)$ represents a topological structure. Its two coordinates denote the birth and death filtration values for the corresponding topological structure, i.e., $p=(b, d)$, where $b=birth(p)$ and $d=death(p)$. We can calculate the persistence diagrams for outputs of both the student and the teacher models, in order to compare the two likelihood maps from a topological perspective.

\section{Mapping Persistent Dots to the Likelihood}
\label{sec:mapping}

\begin{figure}[ht]
\centering
    \includegraphics[width=0.8\linewidth]{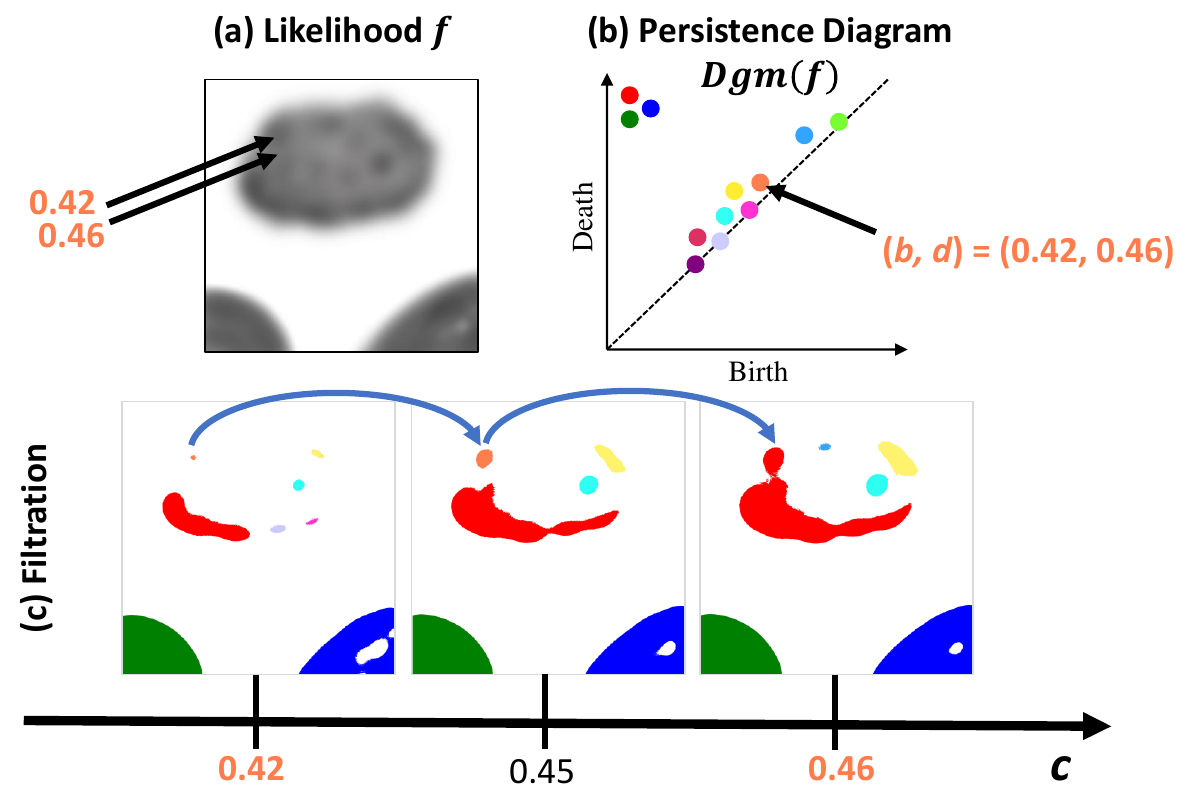}
    \caption{\textbf{(a)} A predicted likelihood map $f$, and \textbf{(b)} the corresponding persistence diagram $Dgm(f)$. Consider the \textcolor{orange}{\textbf{orange}} persistent dot having birth $b$ and death $d$ times as $(b, d) = (0.42, 0.46)$. We show the corresponding filtration in \textbf{(c)} for these specific birth/death times. At birth $b = 0.42$, the connected component corresponding to the \textcolor{orange}{\textbf{orange}} is born. At death $d = 0.46$, this connected component dies as it gets absorbed into the older \textcolor{red}{\textbf{red}} connected component. Note that we only show 0-dim persistent dots pertaining to connected components in $Dgm(f)$.
    }
    \label{fig:pdlh}
\end{figure}

In \cref{fig:pdlh}, we show how persistent dots in the persistence diagram can ultimately be mapped to pixels/voxels in the likelihood map. Consequently, the loss functions defined in~\cref{eq:topo-cons-v1}-~\cref{twoloss} of the main paper are differentiable: the penalty applied to the persistent dots is ultimately a penalty on the pixels/voxels of the likelihood. Hence backpropagation can take place: our proposed losses are differentiable.

In \cref{fig:pdlh}, we give an example of a likelihood $f$ in \cref{fig:pdlh}(a), and focus on the \textcolor{orange}{\textbf{orange}} persistent dot in \cref{fig:pdlh}(b); let us call it $p$. It's coordinate in the persistence diagram $Dgm(f)$ is nothing but its birth $b$ and death $d$ given by $(b, d) = (0.42, 0.46)$. 

There are precisely two pixels in the likelihood that capture the lifetime of this persistent dot $p$. We call them \textit{critical} pixels. We denote the location of these critical pixels in $f$ using black arrows in \cref{fig:pdlh}(a). These two critical pixels have the values $0.42$ and $0.46$ respectively. We now map the likelihood to the persistence diagram below.

In the filtration \cref{fig:pdlh}(c), when the threshold is $0.42$, the critical pixel of the same value gets included into the binary map. It is a connected component on its own and is denoted by \textcolor{orange}{\textbf{orange}} in \cref{fig:pdlh}(c) when $c = 0.42$. This marks the \textit{birth} of the connected component corresponding to the persistent dot $p$. At threshold $c = 0.45$, we see this \textcolor{orange}{\textbf{orange}} connected component grows larger as more pixels get introduced into the binary map. Finally, at $c = 0.46$, the second critical pixel is introduced which joins the \textcolor{orange}{\textbf{orange}} connected component to the older \textcolor{red}{\textbf{red}} connected component. This marks the \textit{death} of the connected component corresponding to $p$ as it gets absorbed into the older \textcolor{red}{\textbf{red}} connected component. Hence, the persistent dot $p \in Dgm(f) $'s birth and death values each correspond to a single pixel location in the likelihood $f$.

Now, this persistent dot gets matched to the diagonal according to the bijection $\gamma^{*}$ introduced in~\cref{subsec:topo_unsup_reg}. Consequently, the loss described in~\cref{eq:topo-rem-v1} pushes $p$ towards the diagonal. This means $p$ is a noisy structure and we would like to suppress/remove it. On pushing it to the diagonal, we force the birth and death times to be the same: the moment this structure is born, it should be automatically included in the older connected component. Hence it ceases to exist as a standalone connected component across any and all filtration values and is thus effectively removed as noise.

\section{Differentiability of the Topology-Aware Losses.}
\label{sec:differentiability}
Both $\mathcal{L}_{\text{topo-cons}}^{U}$ and $\mathcal{L}_{\text{topo-rem}}^{U}$ are differentiable, as \cref{eq:topo-cons-v2} and \cref{eq:topo-rem-v1} are both written as polynomials of the likelihood map $f_s$ at certain critical pixels. Here it is crucial to assume the critical pixels, $x_p^b$ and $x_p^d$, remain constant locally.
This is because the likelihood map is a piecewise linear function determined by the function values at a discrete set of pixel locations.  Assuming without loss of generality that all pixels have distinct values, we can show that within a small neighborhood of the likelihood $f_s$, the order of all pixels in $f_s$ remains the same. Therefore, the algorithmic computation of persistent homology will associate the same set of critical pixels with each persistent dot $x$ in the diagram. In other words, we can assume $x_p^b$ and $x_p^d$ remain constant.

\section{Details of the Datasets}
\label{sec:data_detailed}
\begin{enumerate}
    \item \textbf{Colorectal Adenocarcinoma Gland (CRAG)}~\cite{graham2019mild} is a collection of $213$ H\&E stained colorectal adenocarcinoma image tiles captured at $20\times$ magnification, with full instance-level annotation. Most of the images are of the size $1512\times1516$. It is officially divided into a training set with 173 samples and a test set with 40 samples. In our experiments, we separate the training set into $153$ images for training and $20$ images for validation. For $10\%$ and $20\%$ labeled data splitting, we randomly select $16$ and $31$ images with labels respectively, for training.
    \item \textbf{Gland Segmentation in Colon Histology Images Challenge (GlaS)} is introduced in~\cite{sirinukunwattana2017gland} and comprises of $165$ images derived from 16 H\&E stained histological sections of stage T$3$ or T$4$ colorectal adenocarcinoma. The dataset is officially separated into a training set with $85$ samples and a test set with $80$ samples. In our experiments, we divide the training set into $68$ images for training and $17$ images for validation. For $10\%$ and $20\%$ labeled data splitting, we randomly select $7$ and $14$ images with labels for training.
    \item \textbf{Multi-Organ Nuclei Segmentation (MoNuSeg)}~\cite{kumar2019multi} contains $44$ H\&E stained images of size $1000\times1000$ from seven organs. It consists of two sets, $30$ images containing $21,623$ nuclei for training and $14$ images for testing. In our experiments, we choose $20\%$ training data ($6$ images) as the validation set, and for $10\%$ and $20\%$ labeled data splitting, we randomly select $3$ and $5$ images with labels respectively for training.
\end{enumerate}

\section{Implementation Details}
\label{sec:impl}
We train our model in two stages. The first stage is pre-training, using only $\mathcal{L}^{S}$ and $\mathcal{L}^{U}_{pixel}$ to train the network for several iterations. For CRAG and GlaS, we pre-train the model for $12000$ iterations; for MoNuSeg, we pre-train the model for $2000$ iterations. The second stage is fine-tuning using our topological consistency loss. We fine-tune the model for $500$ epochs using~\cref{final_loss} as the overall training objective. While training, we use UNet++~\cite{zhou2018unet++} as our backbone for both student and teacher networks, and we adopt the Adam optimizer solver to train the model. The proposed algorithm is implemented on the PyTorch platform. The training hyper-parameters are set as follows: for CRAG and GlaS, the batch size is $16$, and the learning rate is $5e-4$. For MoNuSeg, the batch size is $8$, and the learning rate is $1e-4$. We first apply random cropping on both labeled and unlabeled data. The cropping size is $256\times 256$ for CRAG and GlaS and $416\times416$ for MoNuSeg. After random cropping, we apply random rotation and flipping for weak augmentations, and for strong augmentations, we apply color change and morphological shift. The EMA decay rate $\alpha$ and $\lambda^U_{2}$ are set to $0.999$ and $0.002$ respectively. Introduced in~\cite{li2020transformation}, the weight factor of pixel-wise consistency loss is calculated by the Gaussian ramp-up function $\lambda^{U}_{1}=k*e^{-5*(1-\frac{\tau}{T})^2}$, where $k=0.1$ and $T$ is the total number of iterations. $\lambda_1^{L}$ and $\lambda_2^{L}$ in $\mathcal{L}^{S}$ are all set to $0.5$. The persistence threshold $\phi$ for decomposing the persistence diagrams is $0.7$. All the experiments are conducted on an NVIDIA RTX $A6000$ GPU with $48$ GB RAM.

\section{Baseline Reference}
\label{sec:baseline_ref}
In our experiments, some baselines are based on the implementations of publicly available repositories. Here, we provide our baselines' source for reference and appreciate their efforts on the public code. 

MT~\cite{tarvainen2017mean}, EM~\cite{vu2019advent}, UA-MT~\cite{yu2019uncertainty}, and URPC~\cite{luo2022semi} are based on the implementations from: \href{https://github.com/HiLab-git/SSL4MIS}{https://github.com/HiLab-git/SSL4MIS}.

XNet~\cite{zhou2023xnet} is based on the implementations from: \href{https://github.com/guspan-tanadi/XNetfromYanfeng-Zhou}{https://github.com/guspan-tanadi/XNetfromYanfeng-Zhou}.

CCT~\cite{ouali2020semi} is based on the implementations from: 

\href{https://github.com/yassouali/CCT}{https://github.com/yassouali/CCT}.

HCE~\cite{jin2022semi_miccai} is implemented by ourselves due to the lack of code.

\section{Evaluation Metrics}
\label{sec:evalmet_detailed}
We select three widely used pixel-wise evaluation metrics, \textbf{Object-level Dice coefficient (Dice\_{Obj})}~\cite{xie2019deep}, \textbf{Intersection over Union (IoU)} and \textbf{Pixel-wise accuracy}. 
Object-level Dice coefficient mainly measures the similarity between two segmented objects, and this is especially useful in pathology imaging, where accurately segmenting individual anatomical structures is crucial. IoU provides a measure of how well the predicted segmentation or detected object aligns with the ground truth. Pixel-wise accuracy evaluates how many pixels in the segmentation maps are correctly classified. 
The larger these three metrics are, the better the segmentation performance is.

Topology-relevant metrics mainly measure the structural accuracy. We also select three topological evaluation metrics, \textbf{Betti Error}~\cite{hu2019topology}, \textbf{Betti Matching Error}~\cite{stucki2023topologically}, and \textbf{Variation of Information (VOI)}~\cite{meilua2003comparing}.
For the Betti error, we split the prediction and ground truth into patches in a sliding-window fashion and calculate the average absolute discrepancy between their 0-dimensional Betti number. The size of the window is $256\times256$.
Betti matching error considers the spatial location of the features within their respective images and can be regarded as a variant of Betti error.
VOI mainly measures the distance between two clusterings.
The smaller these metrics are, the better the segmentation performance is.

\section{Additional Qualitative Results}
\label{sec:addi_qua_results}
\begin{figure*}[ht]
    \centering
    \begin{subfigure}{0.115\textwidth}
        \includegraphics[width=\linewidth]{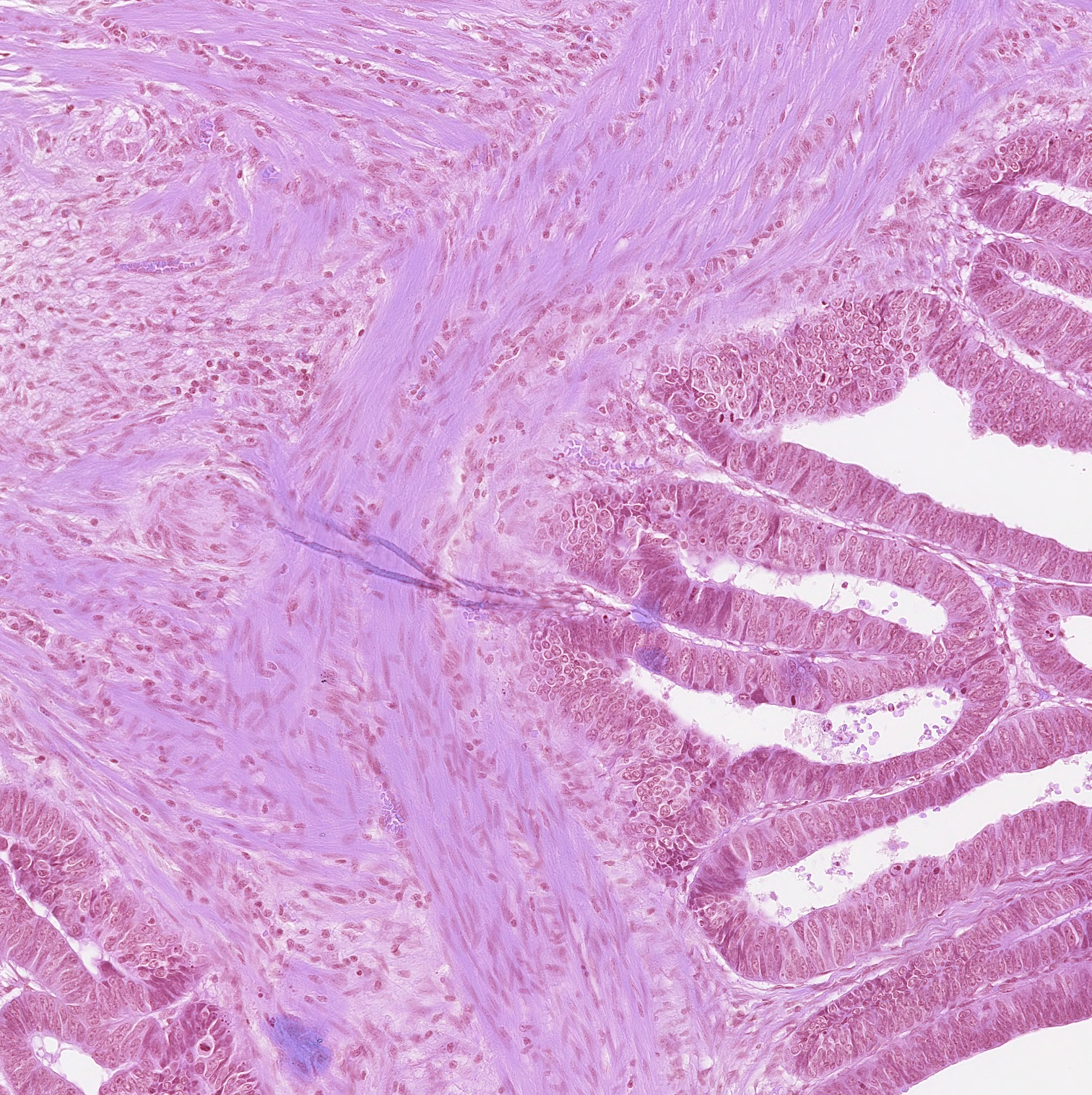}
    \end{subfigure} 
    \begin{subfigure}{0.115\textwidth}
        \includegraphics[width=\linewidth]{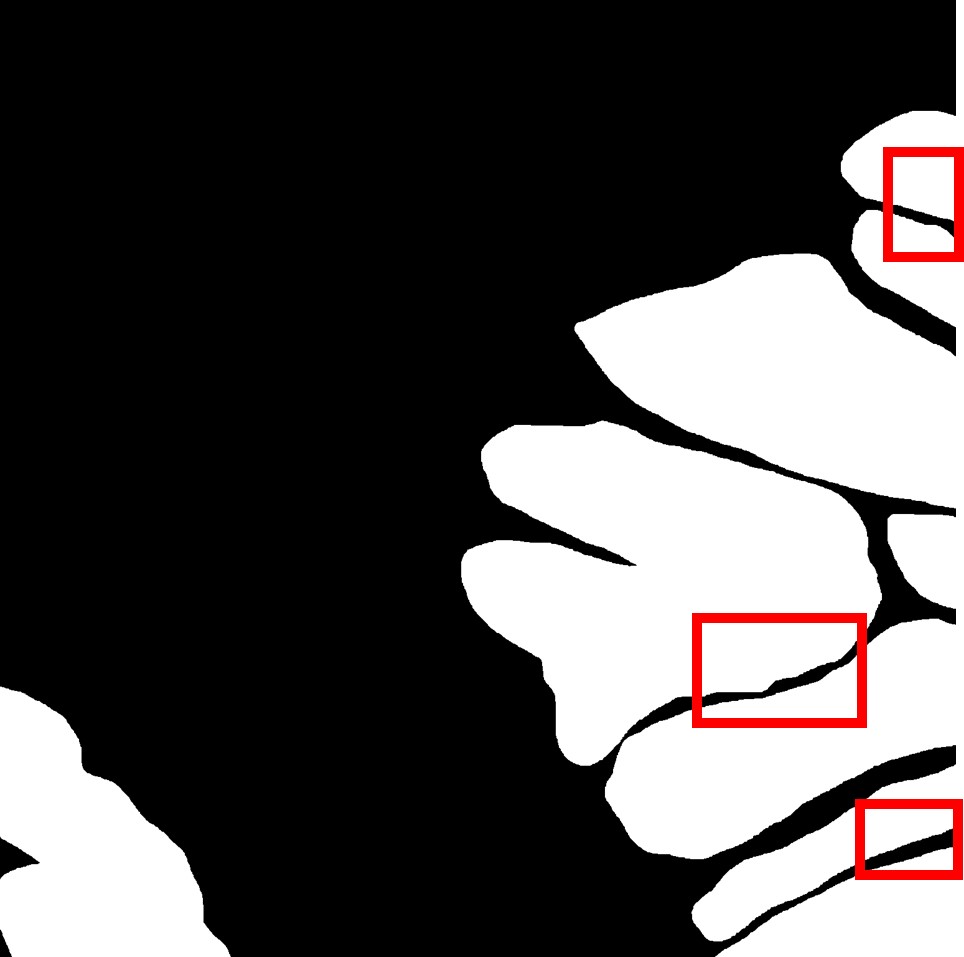}
    \end{subfigure}
    \begin{subfigure}{0.115\textwidth}
        \includegraphics[width=\linewidth]{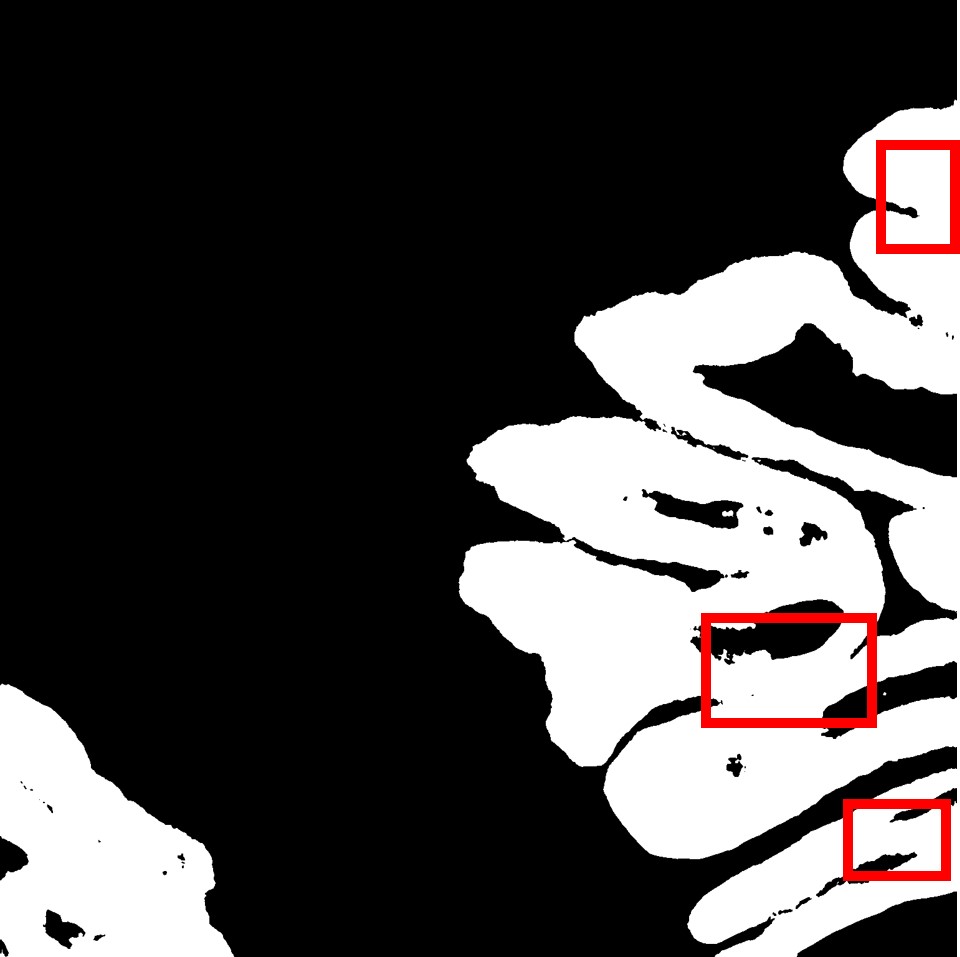}
    \end{subfigure}
    \begin{subfigure}{0.115\textwidth}
        \includegraphics[width=\linewidth]{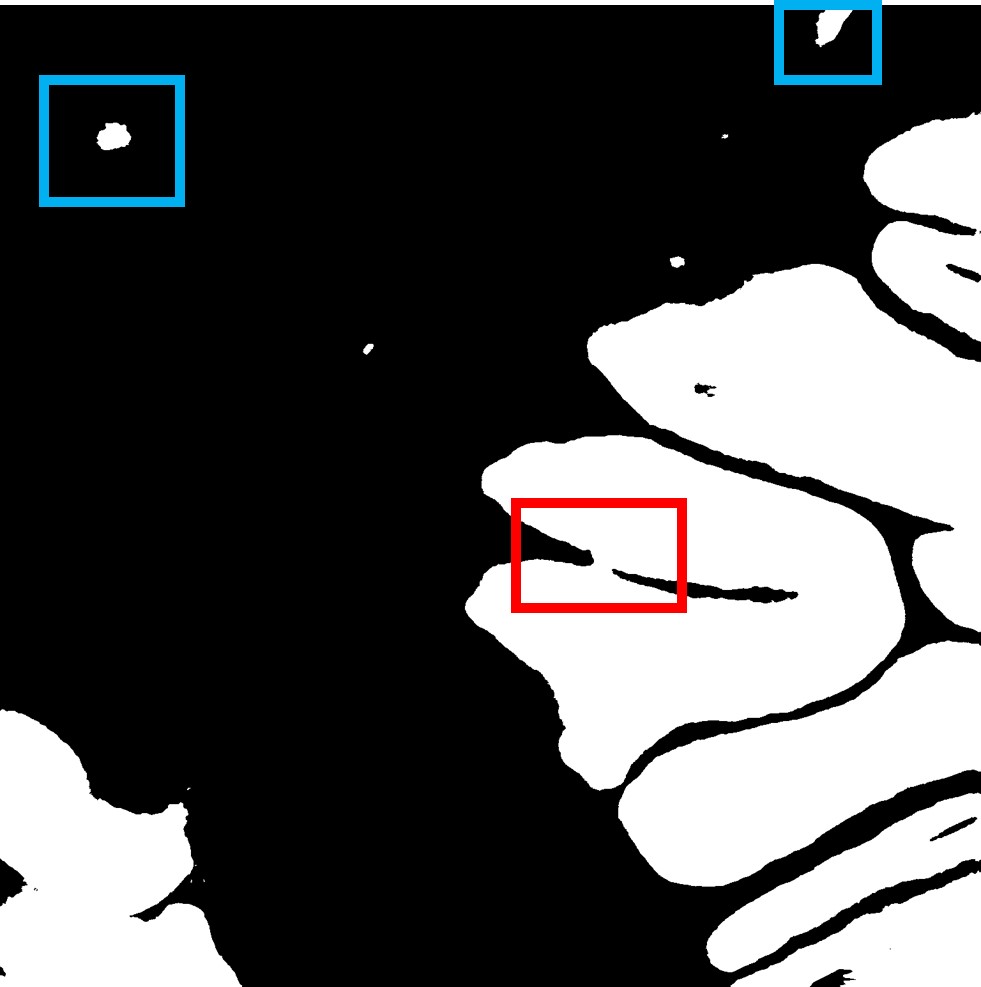}
    \end{subfigure}
    \begin{subfigure}{0.115\textwidth}
        \includegraphics[width=\linewidth]{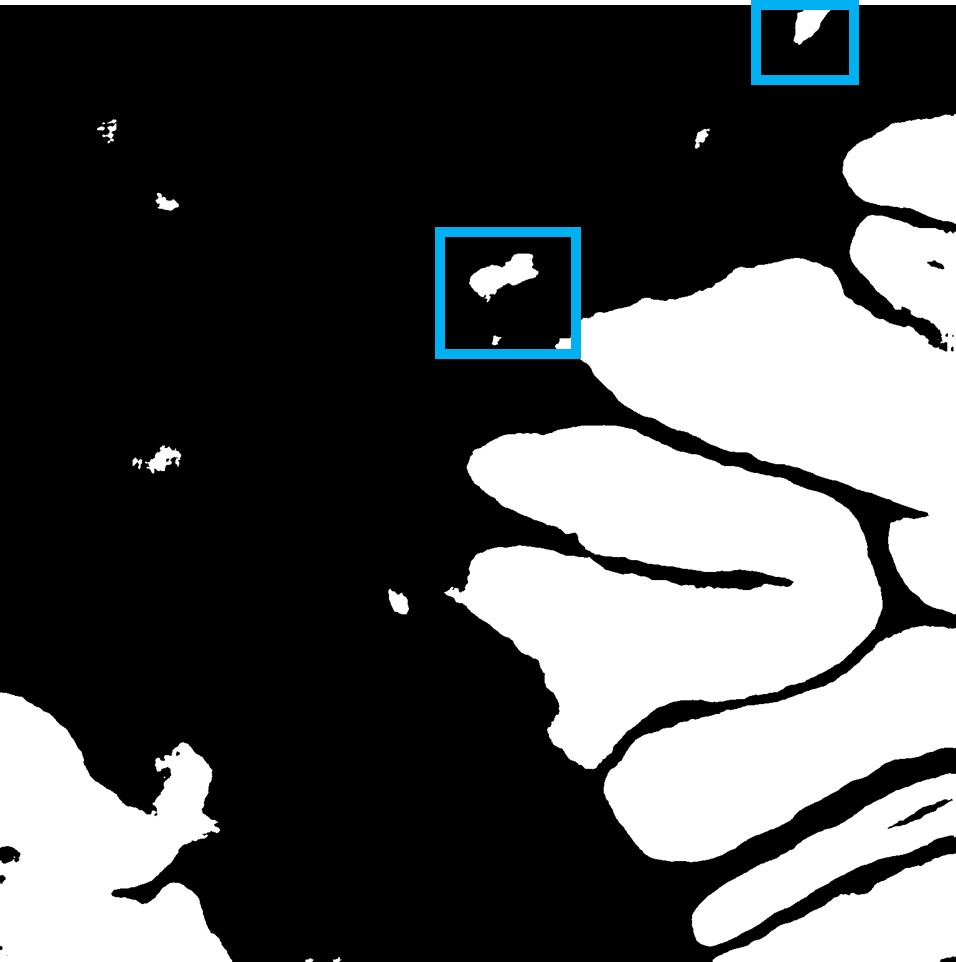}
    \end{subfigure}
    \begin{subfigure}{0.115\textwidth}
        \includegraphics[width=\linewidth]{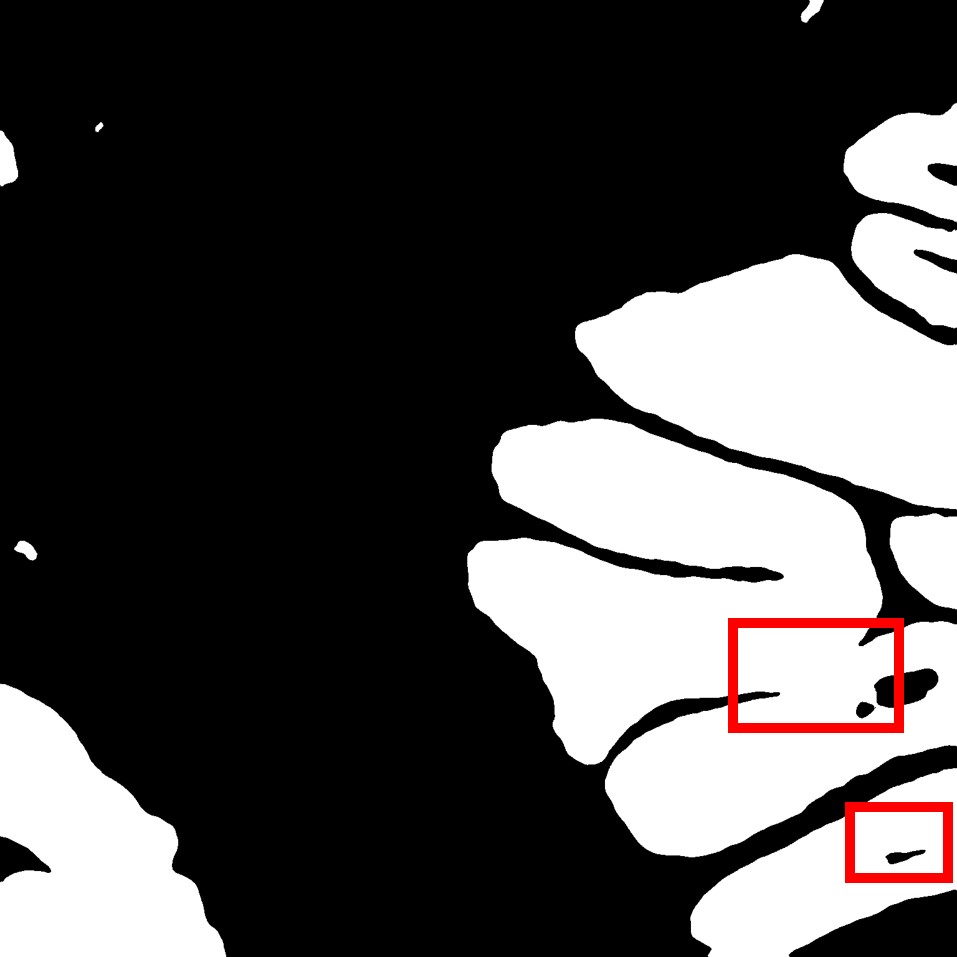}    
    \end{subfigure} 
    \begin{subfigure}{0.115\textwidth}
        \includegraphics[width=\linewidth]{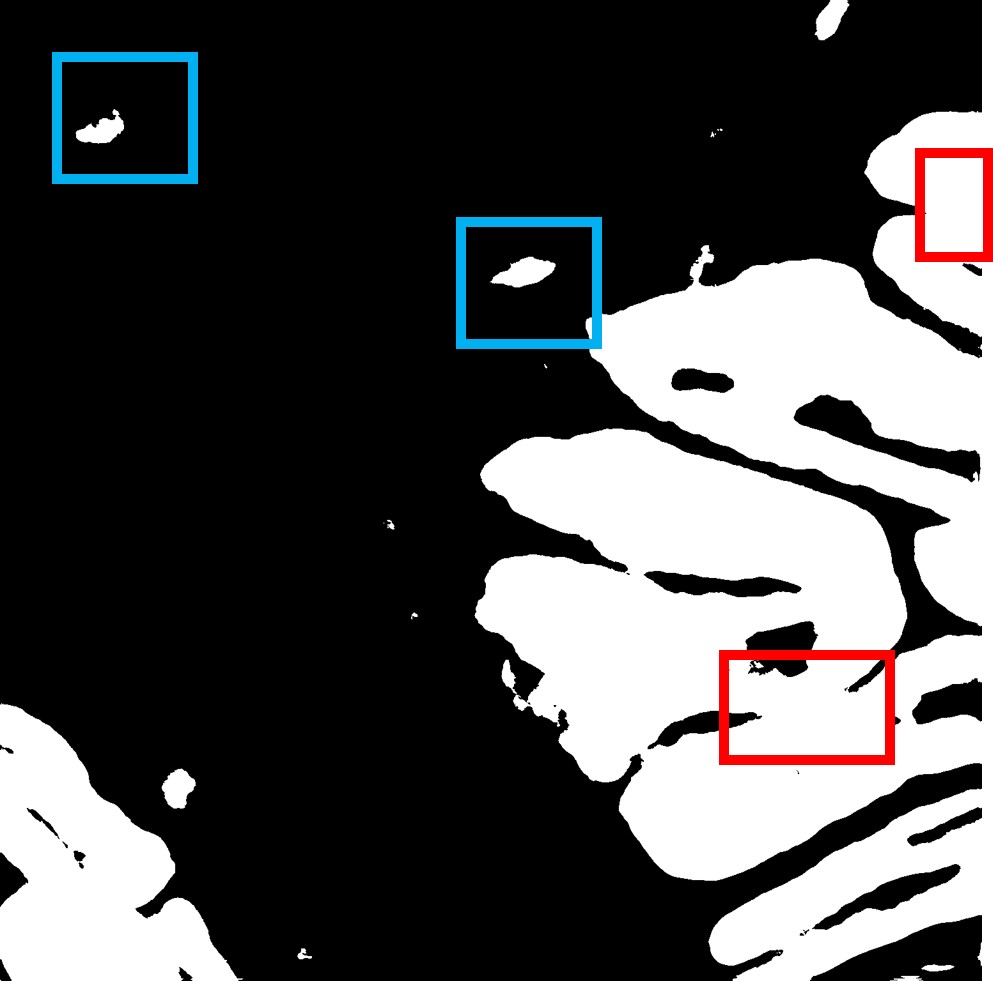}
    \end{subfigure}
    \begin{subfigure}{0.115\textwidth}
        \includegraphics[width=\linewidth]{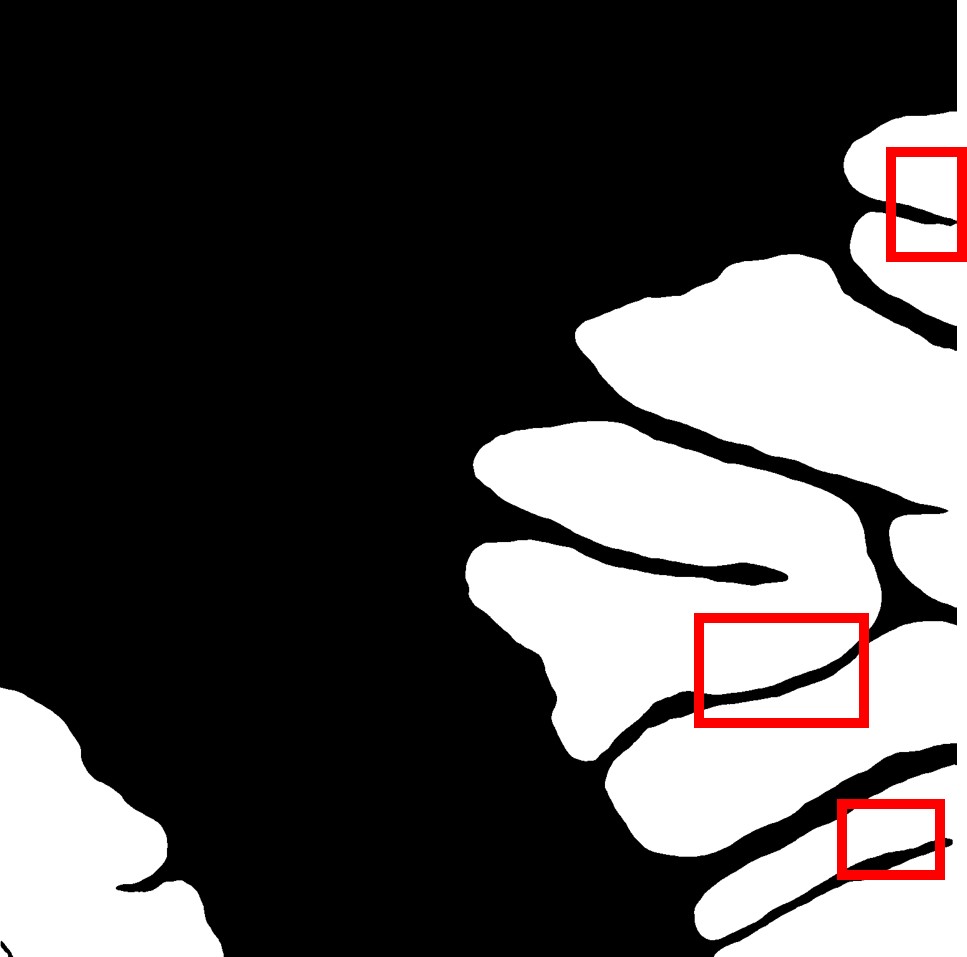}
    \end{subfigure}

    \begin{subfigure}{0.115\textwidth}
        \includegraphics[width=\linewidth]{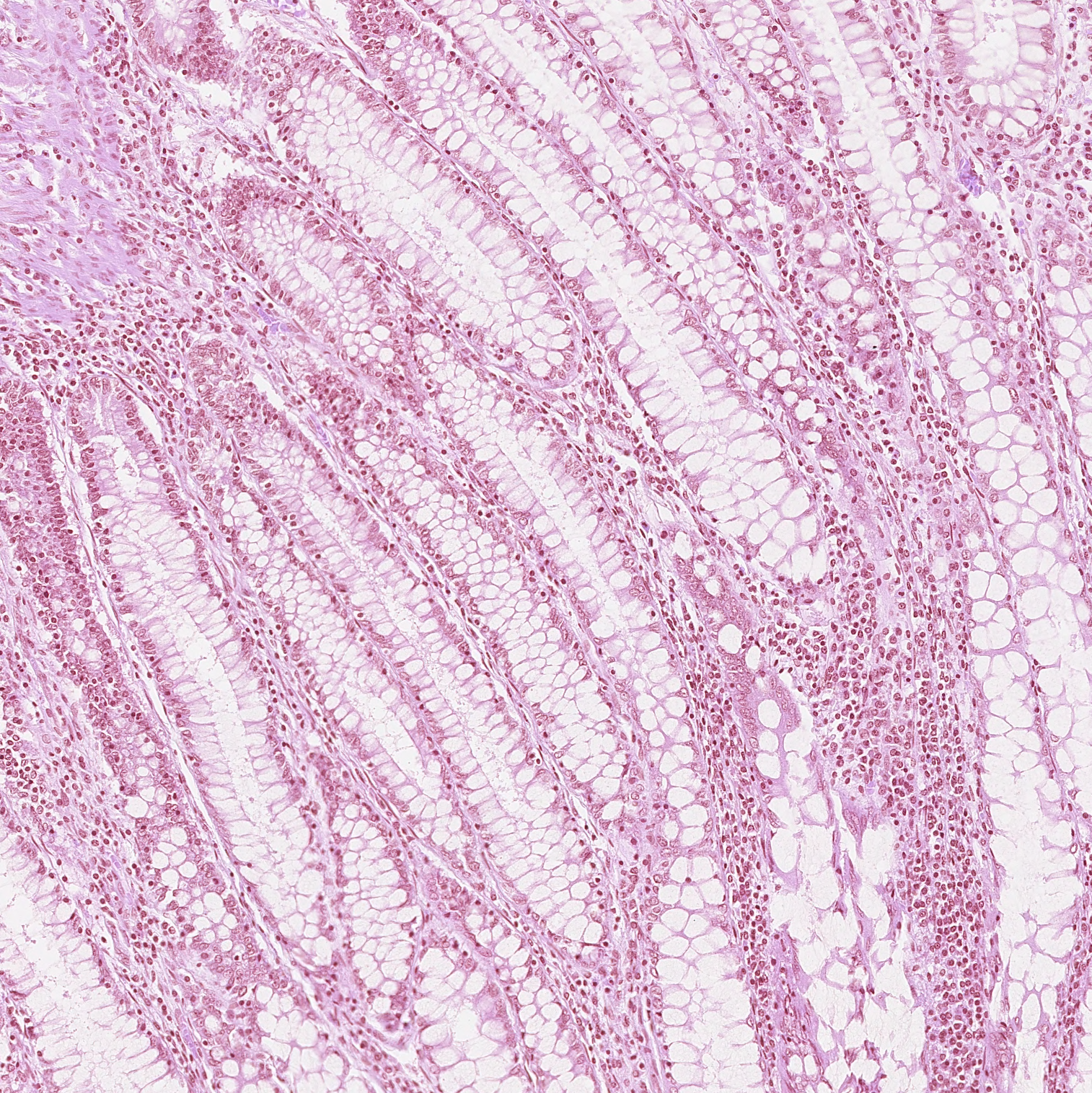}
    \end{subfigure} 
    \begin{subfigure}{0.115\textwidth}
        \includegraphics[width=\linewidth]{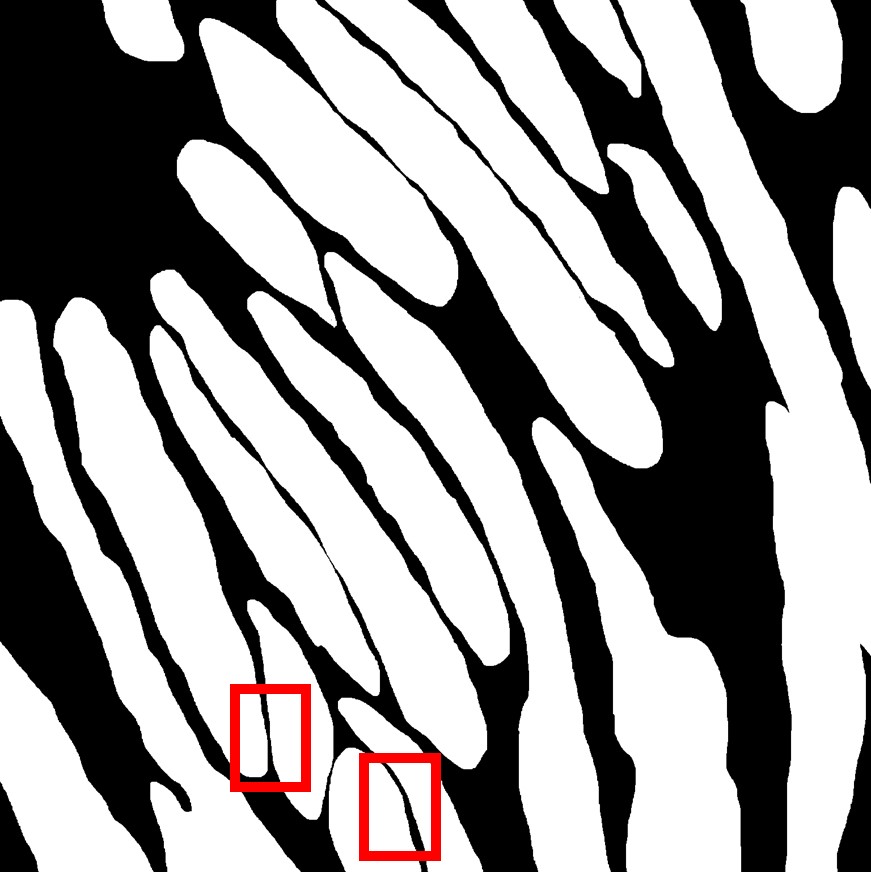}
    \end{subfigure}
    \begin{subfigure}{0.115\textwidth}
        \includegraphics[width=\linewidth]{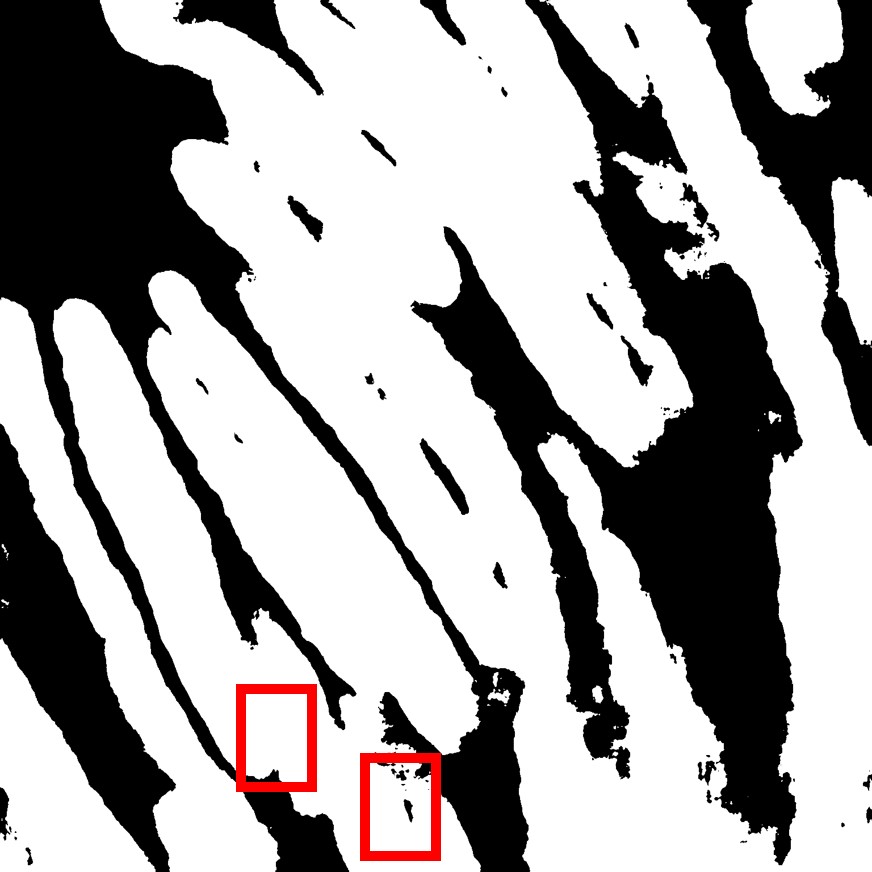}
    \end{subfigure}
    \begin{subfigure}{0.115\textwidth}
        \includegraphics[width=\linewidth]{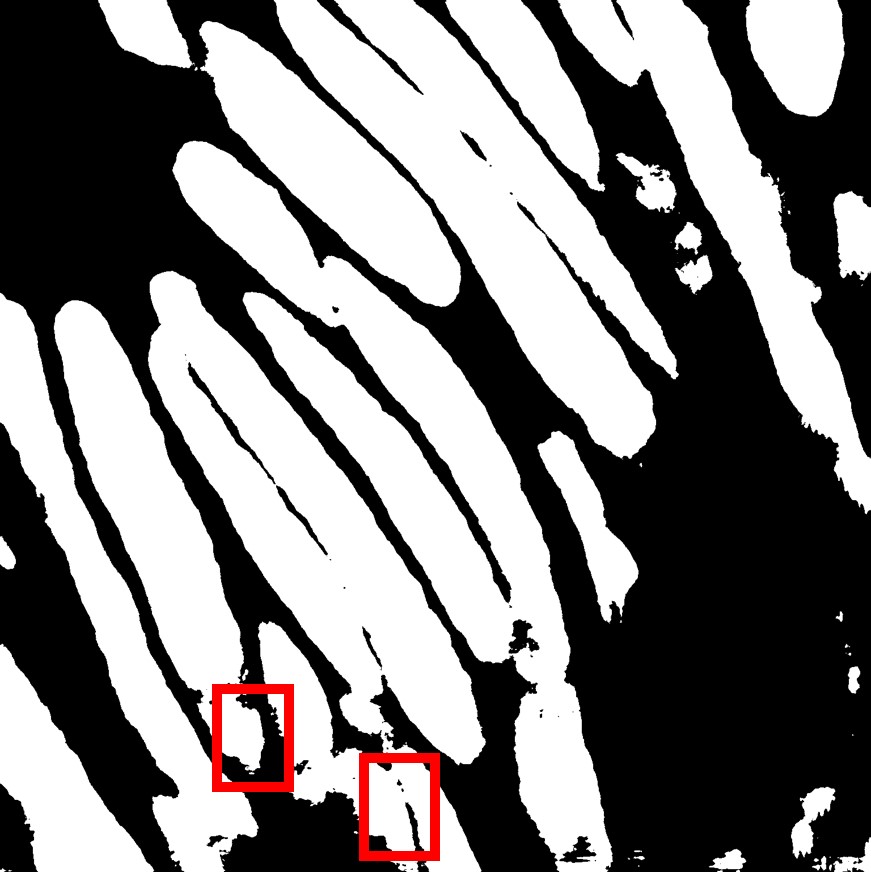}
    \end{subfigure}
    \begin{subfigure}{0.115\textwidth}
        \includegraphics[width=\linewidth]{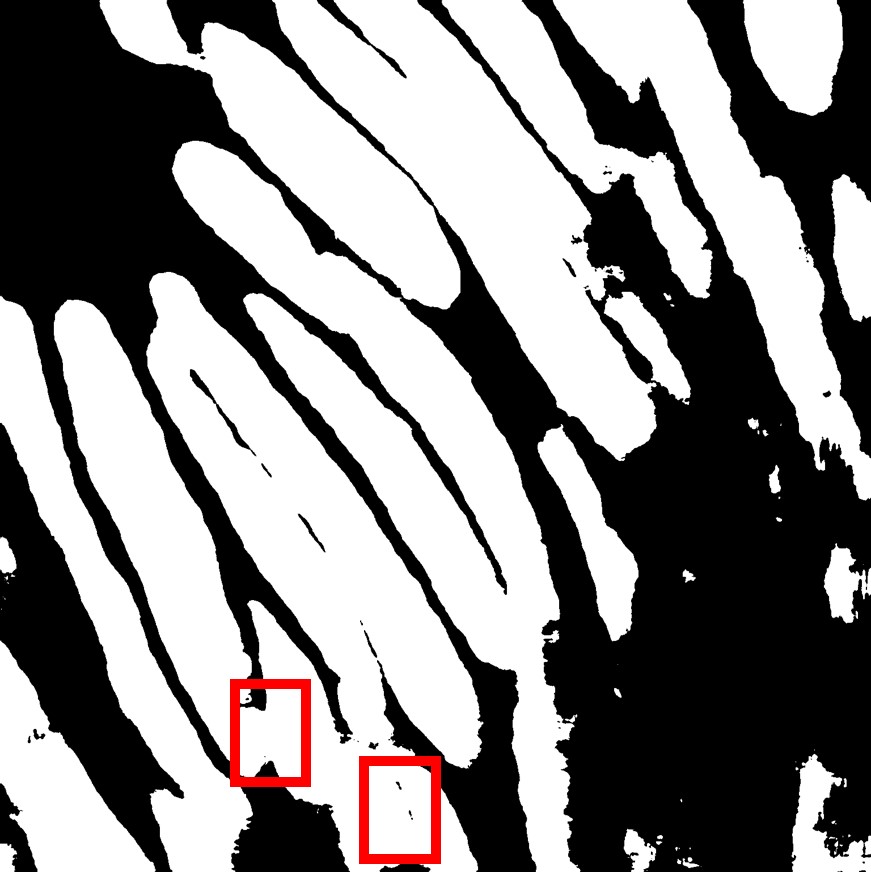}
    \end{subfigure}
    \begin{subfigure}{0.115\textwidth}
        \includegraphics[width=\linewidth]{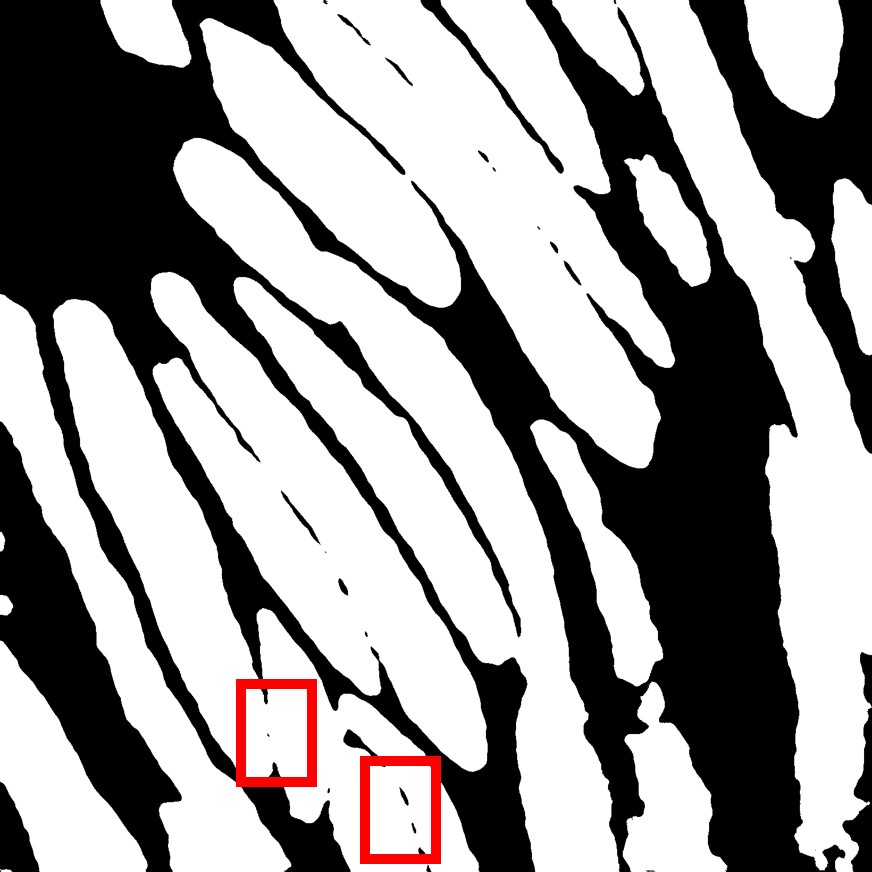}    
    \end{subfigure} 
    \begin{subfigure}{0.115\textwidth}
        \includegraphics[width=\linewidth]{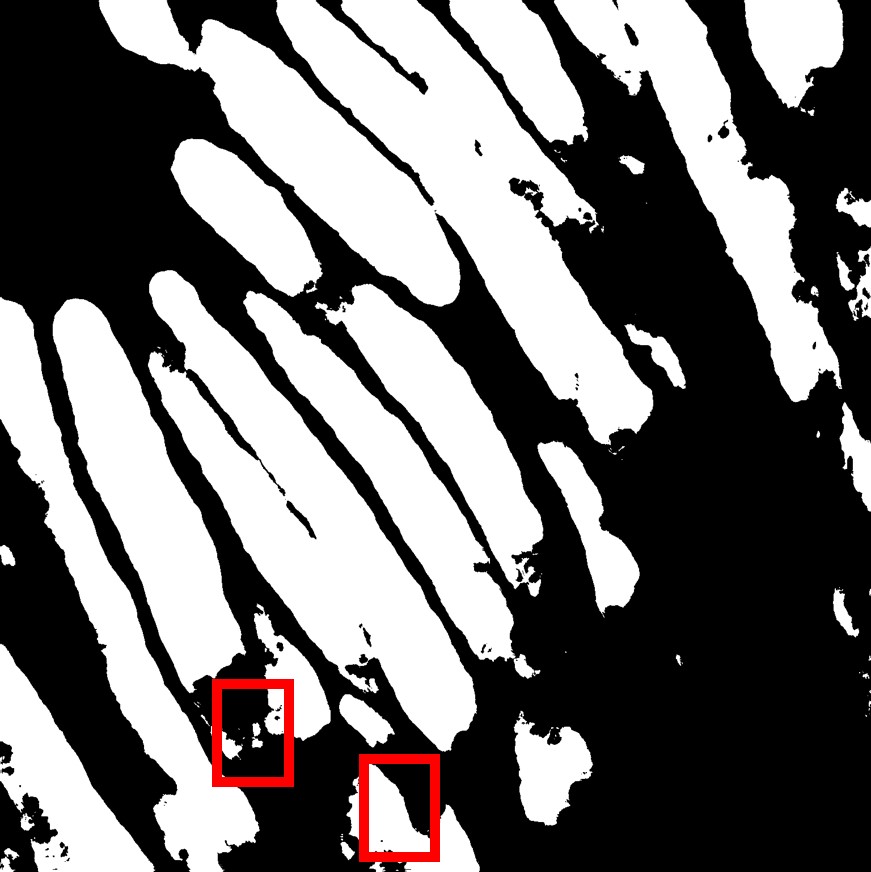}
    \end{subfigure}
    \begin{subfigure}{0.115\textwidth}
        \includegraphics[width=\linewidth]{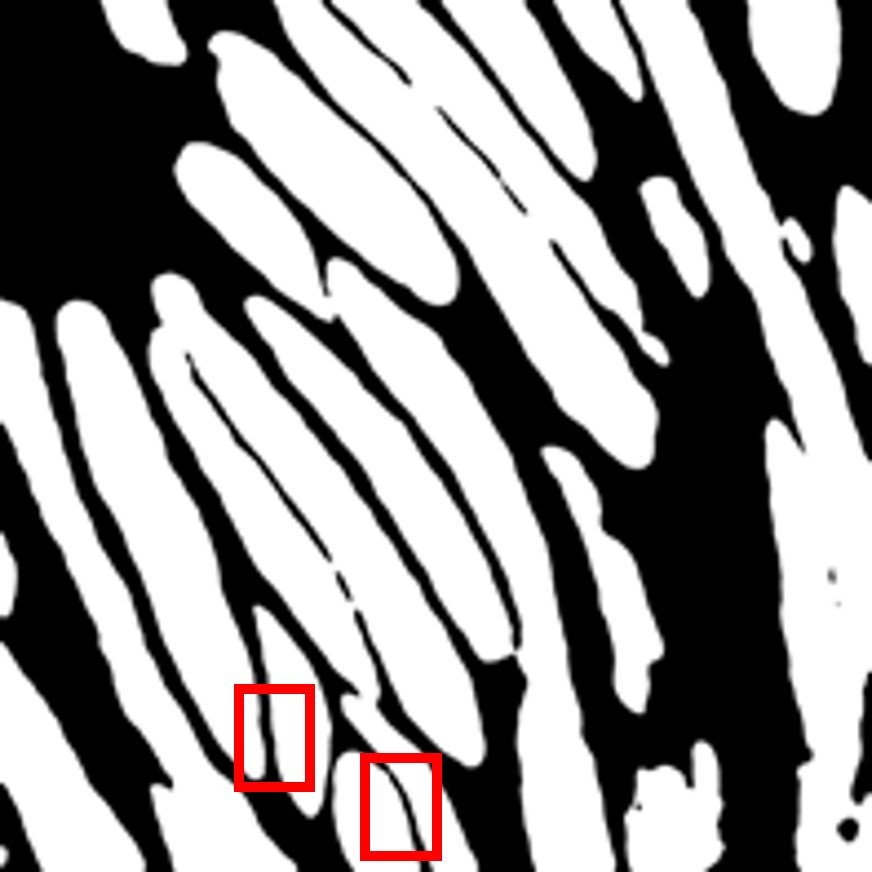}
    \end{subfigure}

    \begin{subfigure}{0.115\textwidth}
        \includegraphics[width=\linewidth]{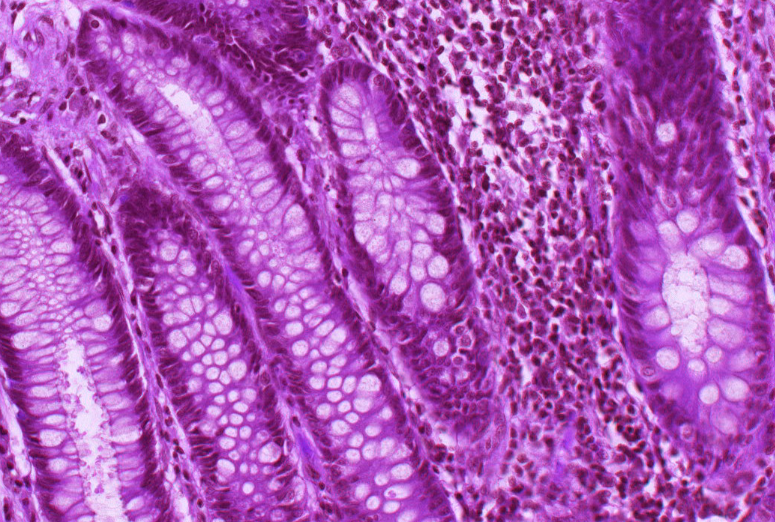}
    \end{subfigure} 
    \begin{subfigure}{0.115\textwidth}
        \includegraphics[width=\linewidth]{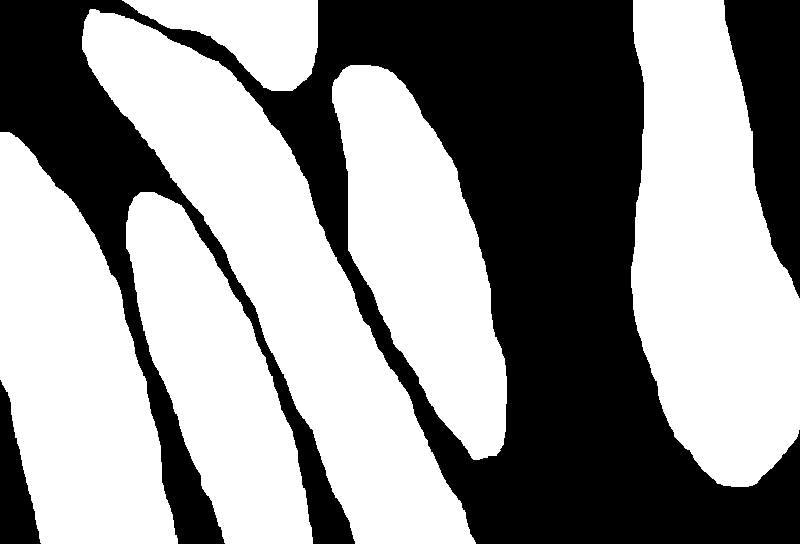}
    \end{subfigure}
    \begin{subfigure}{0.115\textwidth}
        \includegraphics[width=\linewidth]{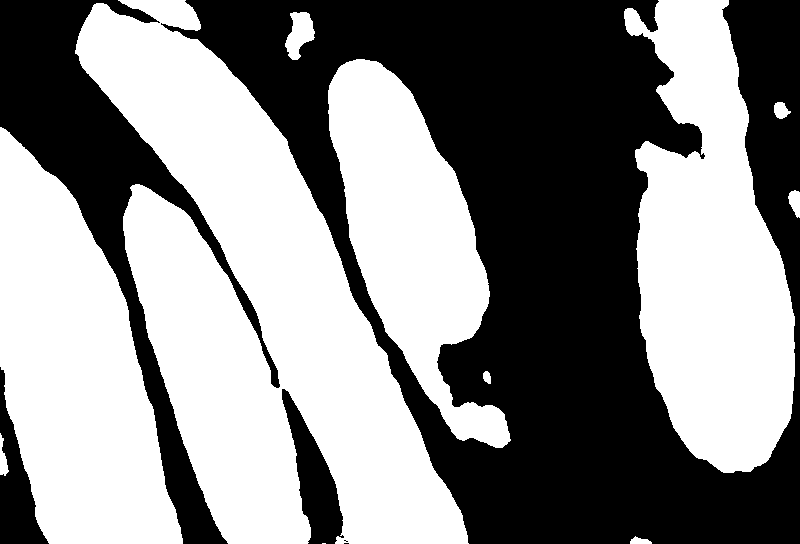}
    \end{subfigure}
    \begin{subfigure}{0.115\textwidth}
        \includegraphics[width=\linewidth]{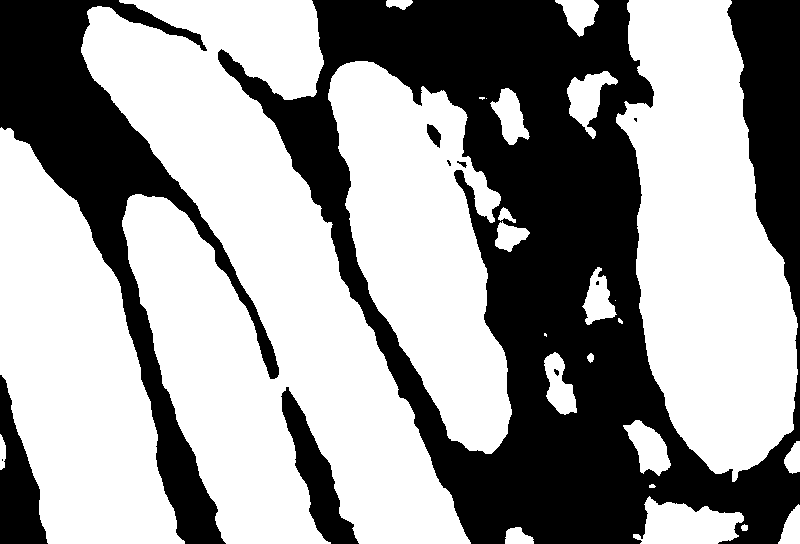}
    \end{subfigure}
    \begin{subfigure}{0.115\textwidth}
        \includegraphics[width=\linewidth]{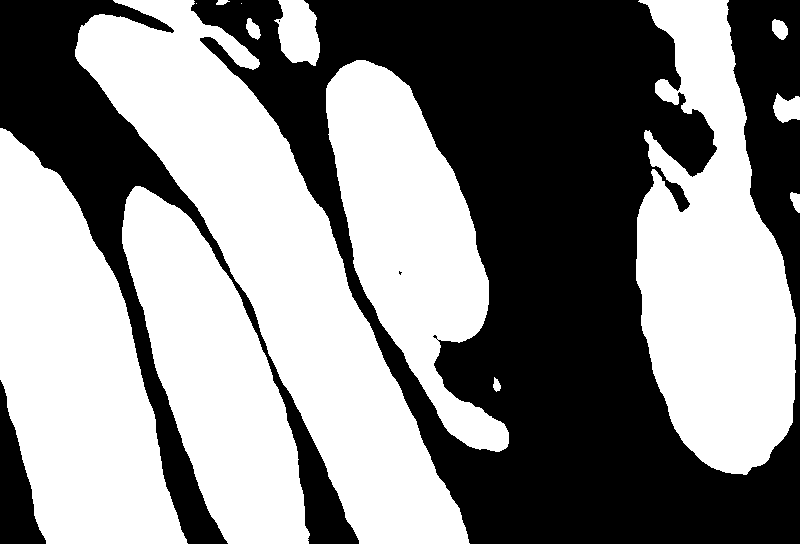}
    \end{subfigure}
    \begin{subfigure}{0.115\textwidth}
        \includegraphics[width=\linewidth]{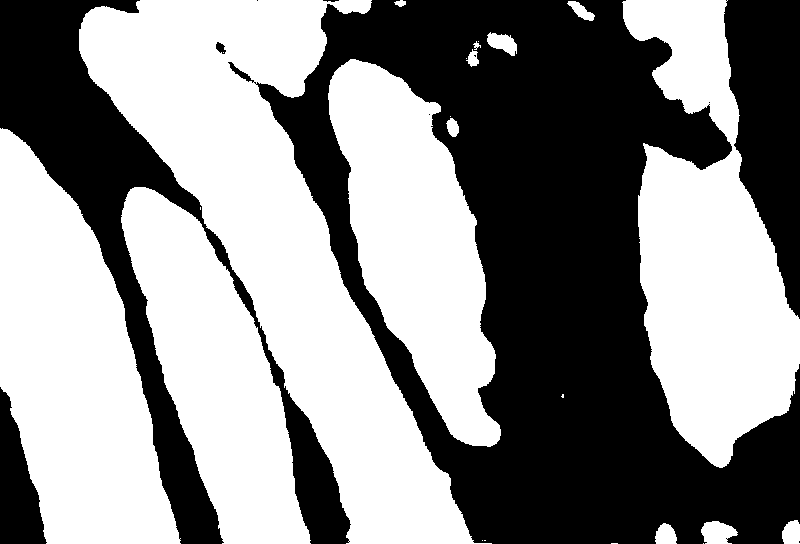}    
    \end{subfigure} 
    \begin{subfigure}{0.115\textwidth}
        \includegraphics[width=\linewidth]{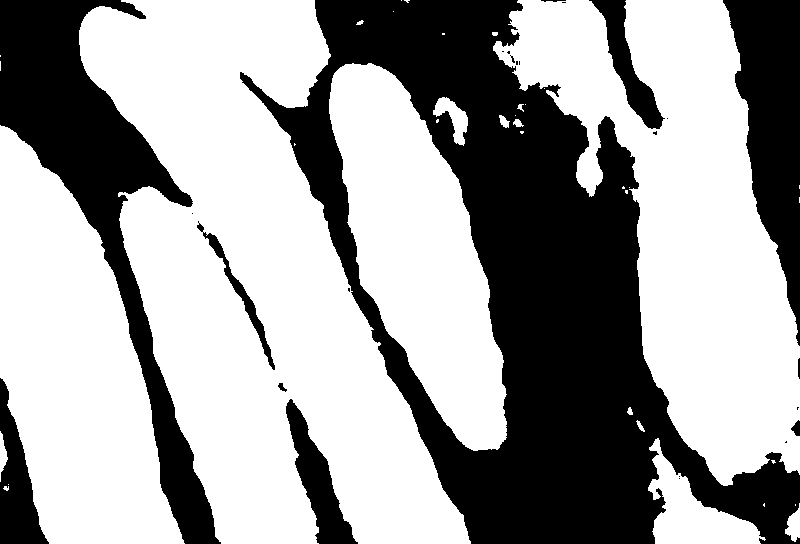}
    \end{subfigure}
    \begin{subfigure}{0.115\textwidth}
        \includegraphics[width=\linewidth]{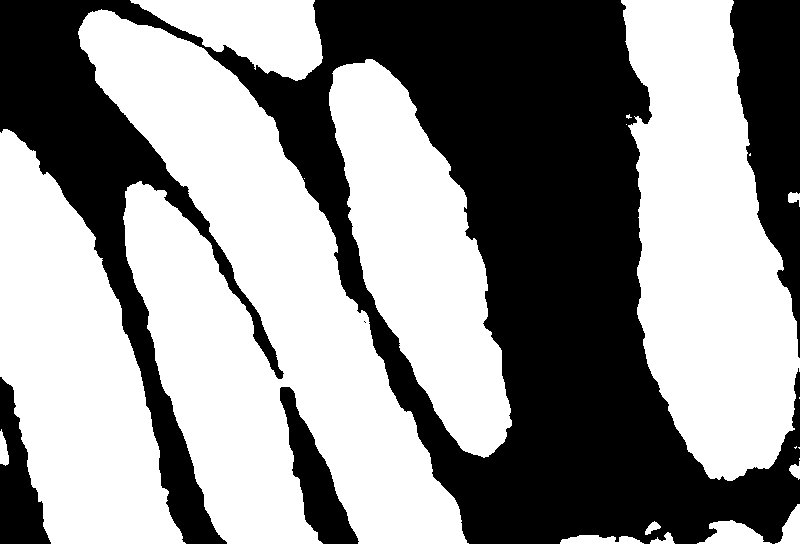}
    \end{subfigure}

    \begin{subfigure}{0.115\textwidth}
        \includegraphics[width=\linewidth]{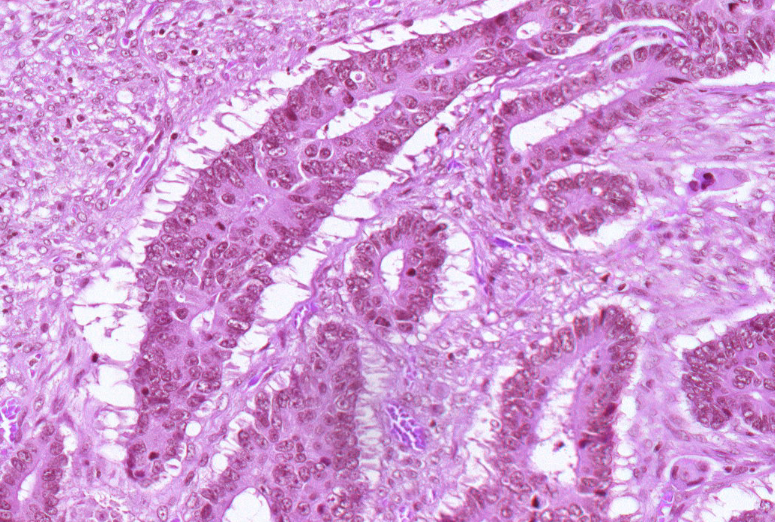}
    \end{subfigure} 
    \begin{subfigure}{0.115\textwidth}
        \includegraphics[width=\linewidth]{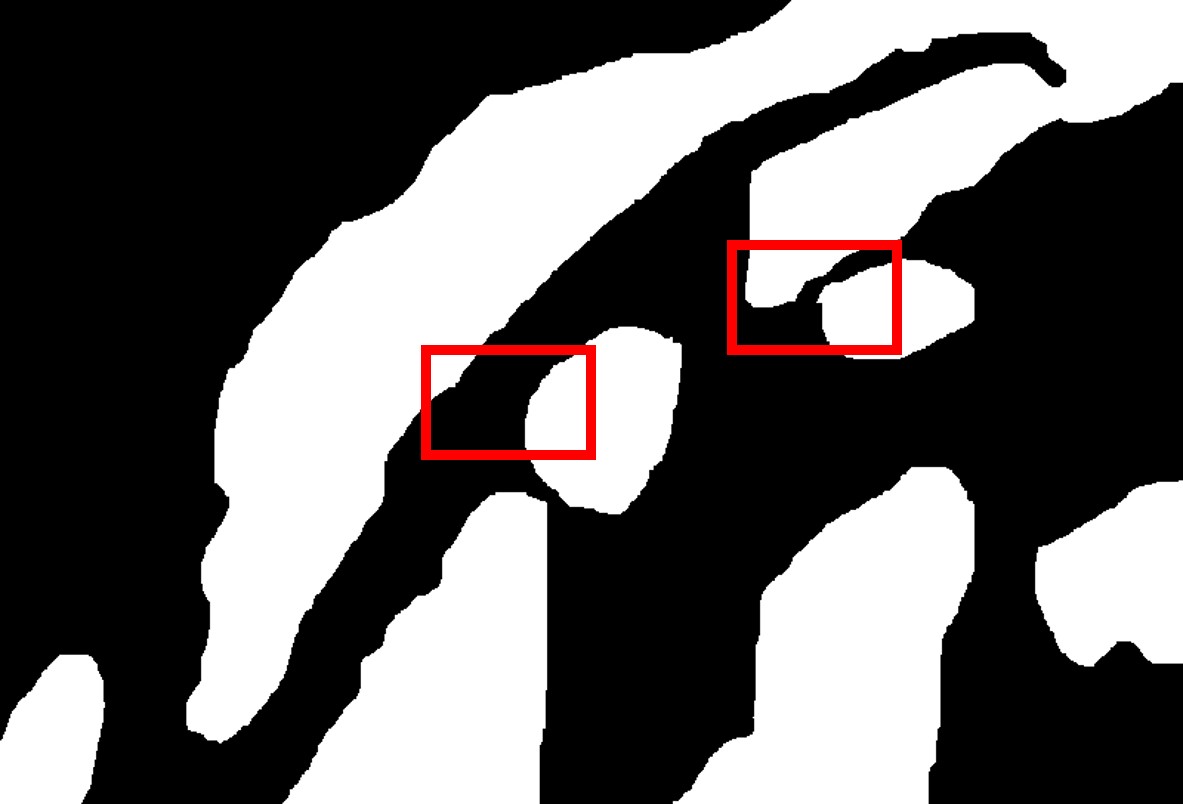}
    \end{subfigure}
    \begin{subfigure}{0.115\textwidth}
        \includegraphics[width=\linewidth]{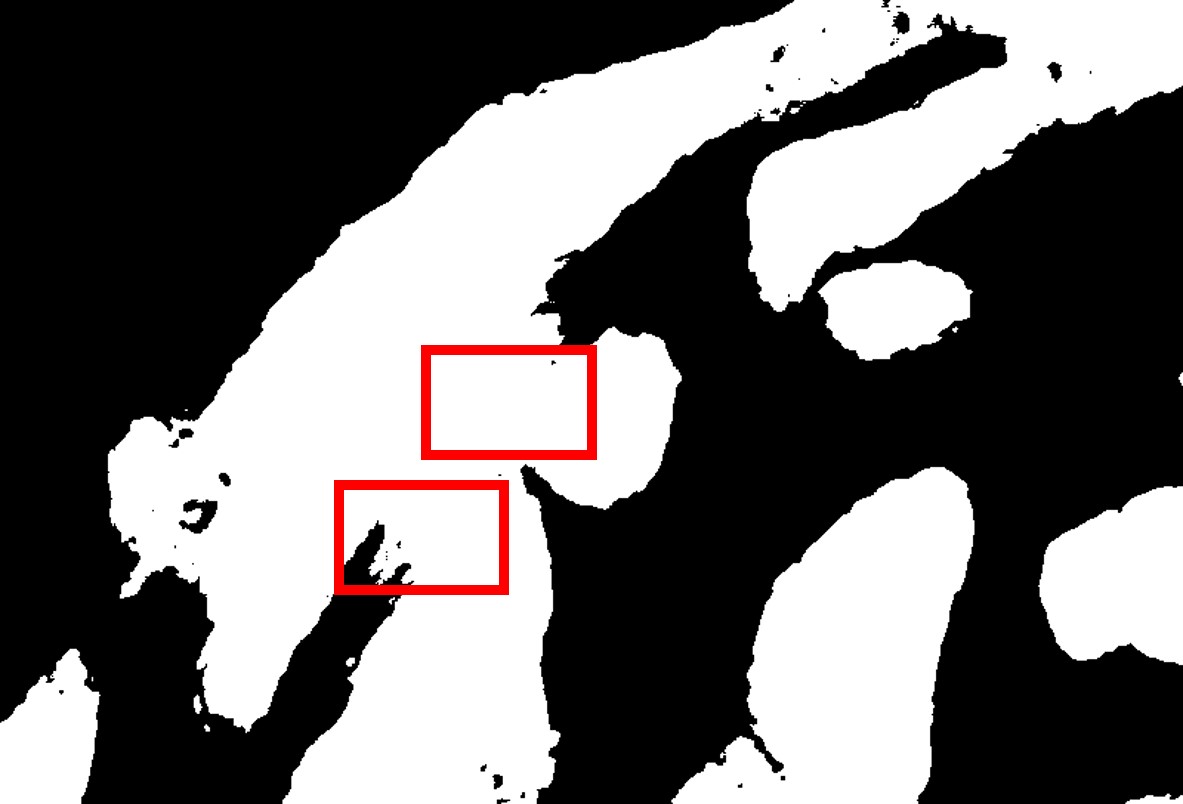}
    \end{subfigure}
    \begin{subfigure}{0.115\textwidth}
        \includegraphics[width=\linewidth]{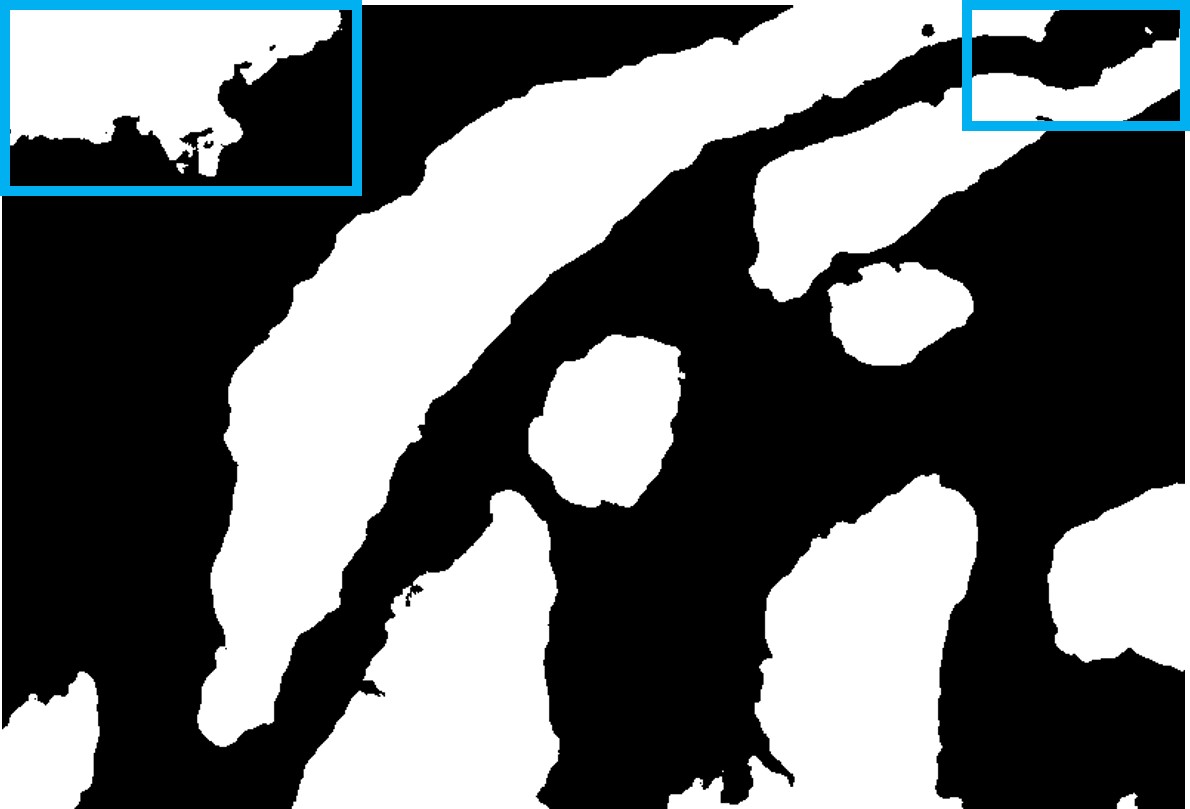}
    \end{subfigure}
    \begin{subfigure}{0.115\textwidth}
        \includegraphics[width=\linewidth]{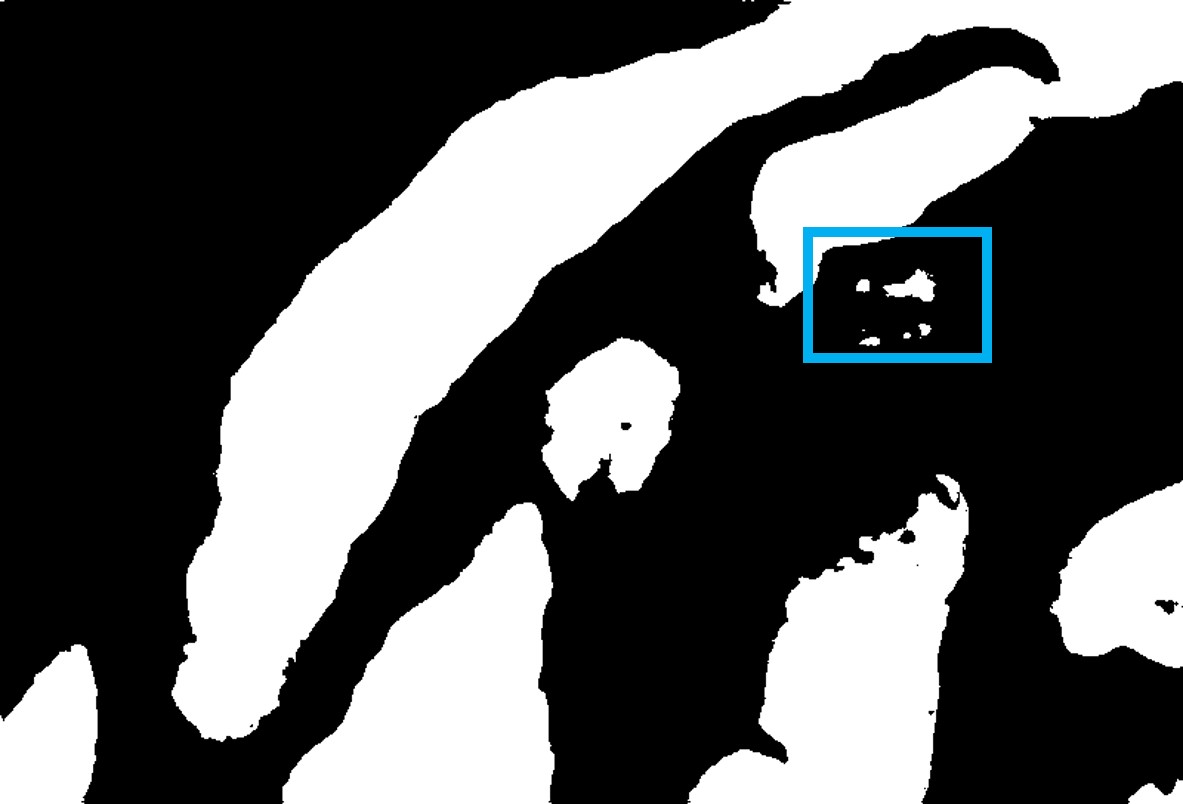}
    \end{subfigure}
    \begin{subfigure}{0.115\textwidth}
        \includegraphics[width=\linewidth]{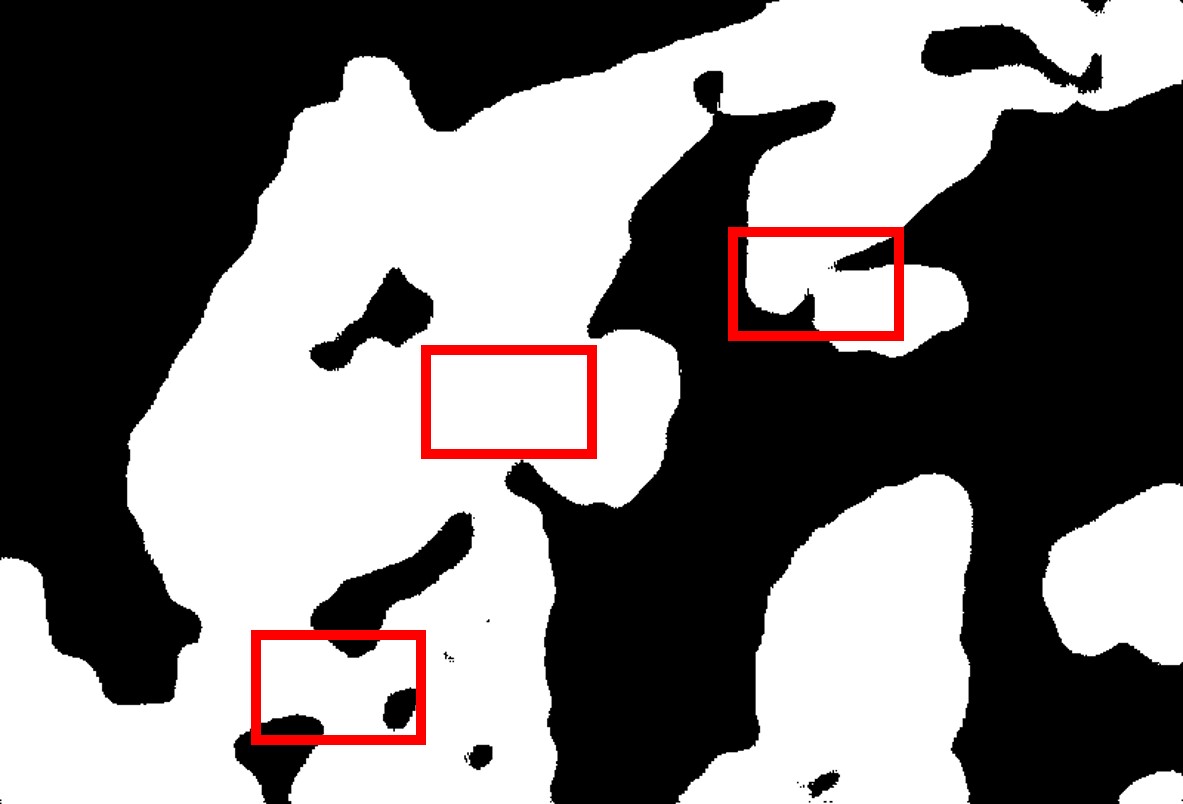}    
    \end{subfigure} 
    \begin{subfigure}{0.115\textwidth}
        \includegraphics[width=\linewidth]{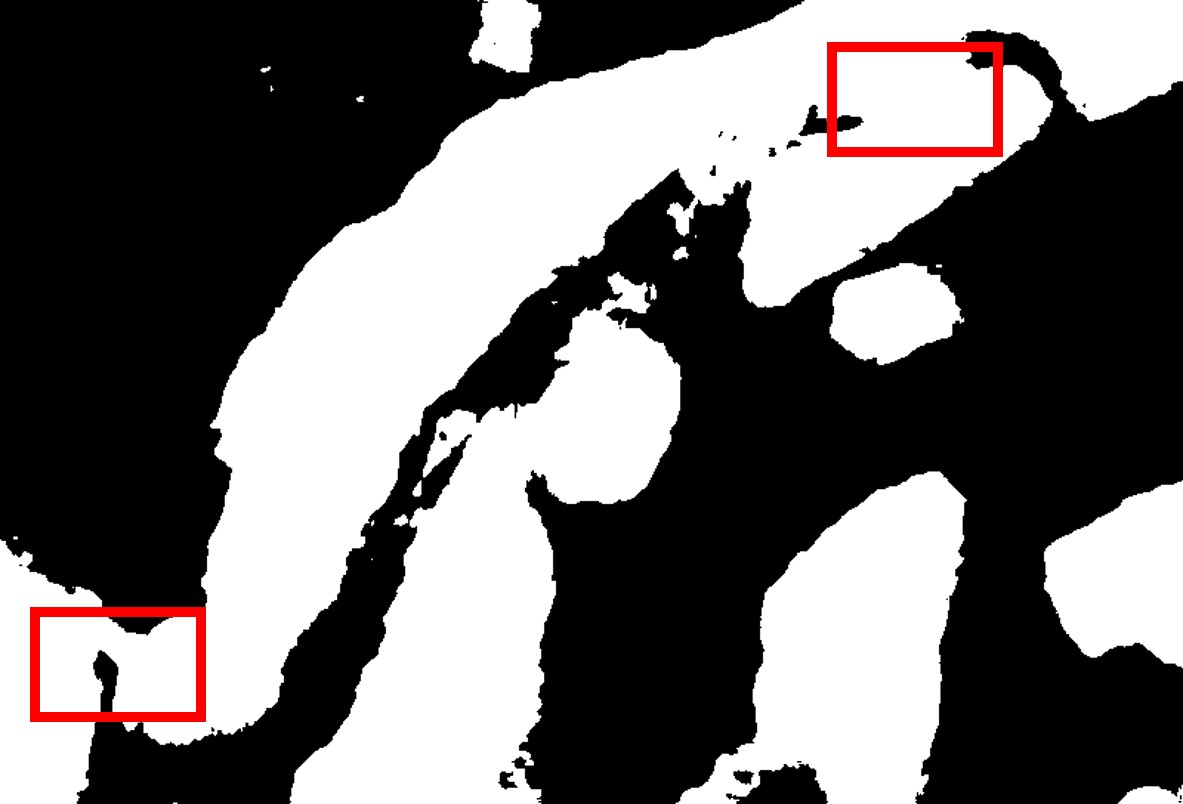}
    \end{subfigure}
    \begin{subfigure}{0.115\textwidth}
        \includegraphics[width=\linewidth]{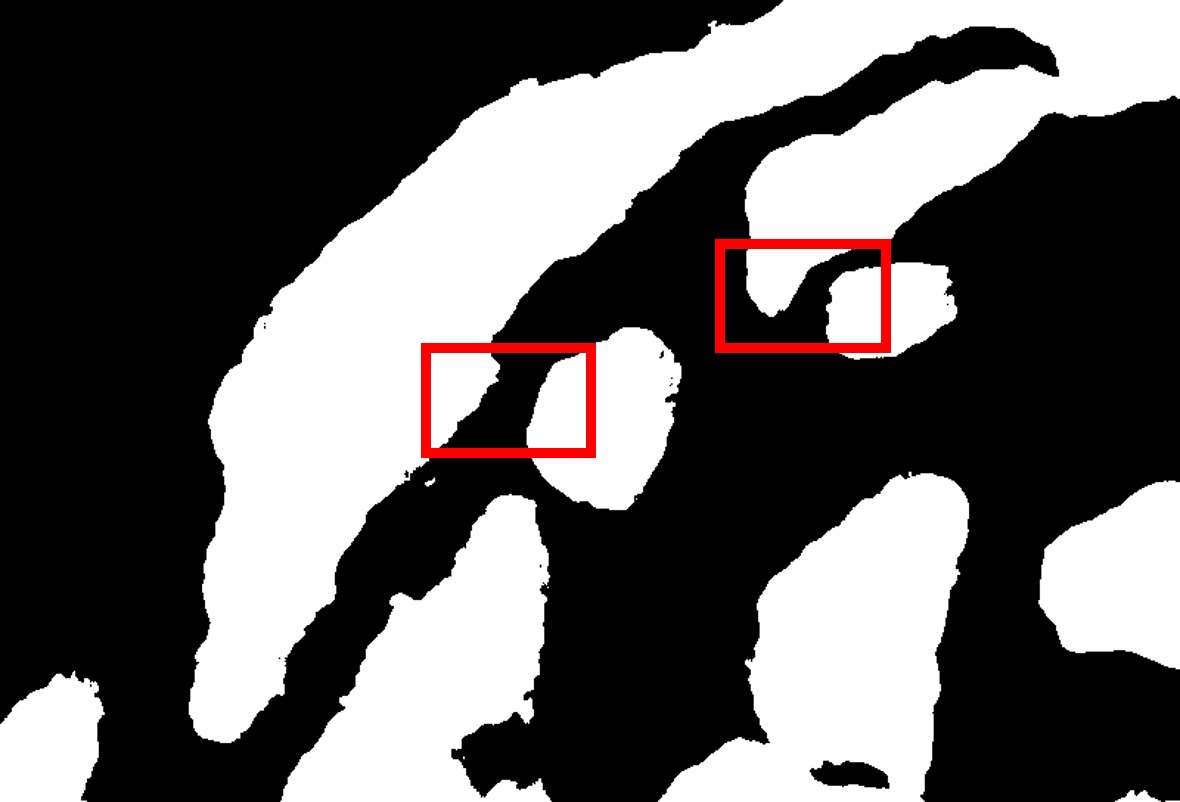}
    \end{subfigure}

    \begin{subfigure}{0.115\textwidth}
        \includegraphics[width=\linewidth]{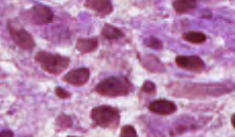}
    \end{subfigure}
    \begin{subfigure}{0.115\textwidth}
        \includegraphics[width=\linewidth]{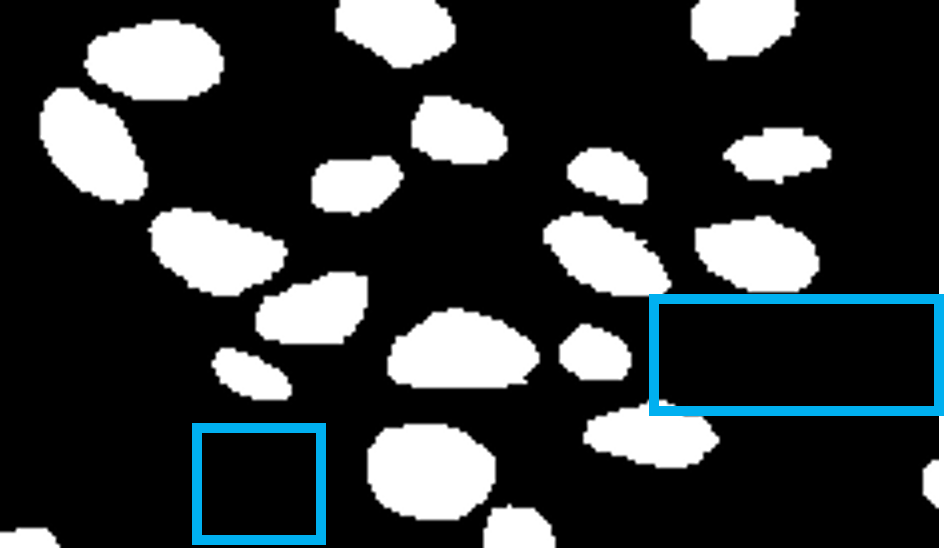}
    \end{subfigure} 
    \begin{subfigure}{0.115\textwidth}
        \includegraphics[width=\linewidth]{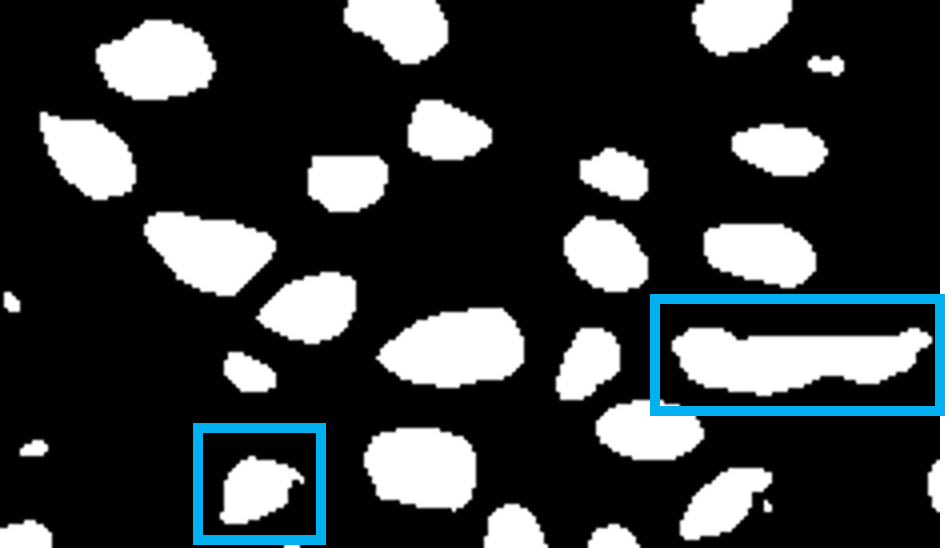}
    \end{subfigure} 
    \begin{subfigure}{0.115\textwidth}
        \includegraphics[width=\linewidth]{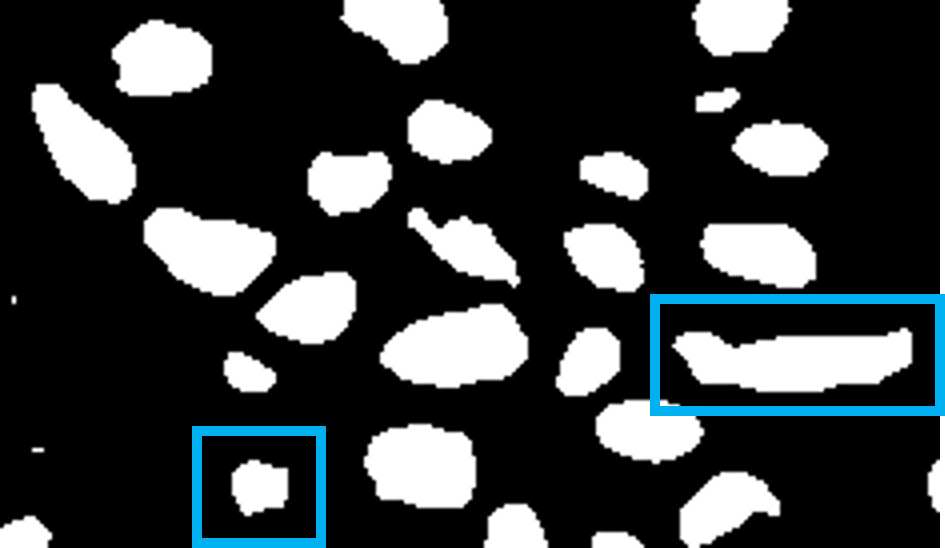}
    \end{subfigure} 
    \begin{subfigure}{0.115\textwidth}
        \includegraphics[width=\linewidth]{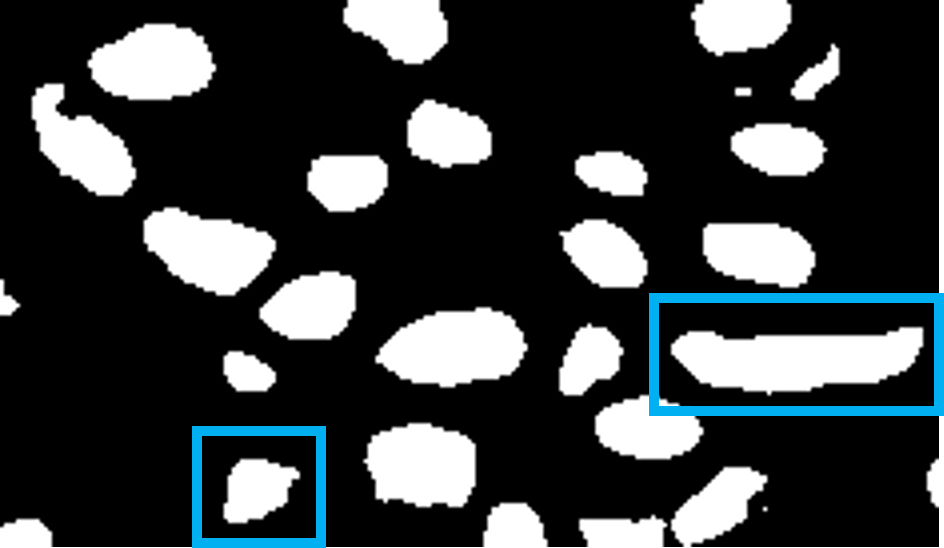}
    \end{subfigure}
    \begin{subfigure}{0.115\textwidth}
        \includegraphics[width=\linewidth]{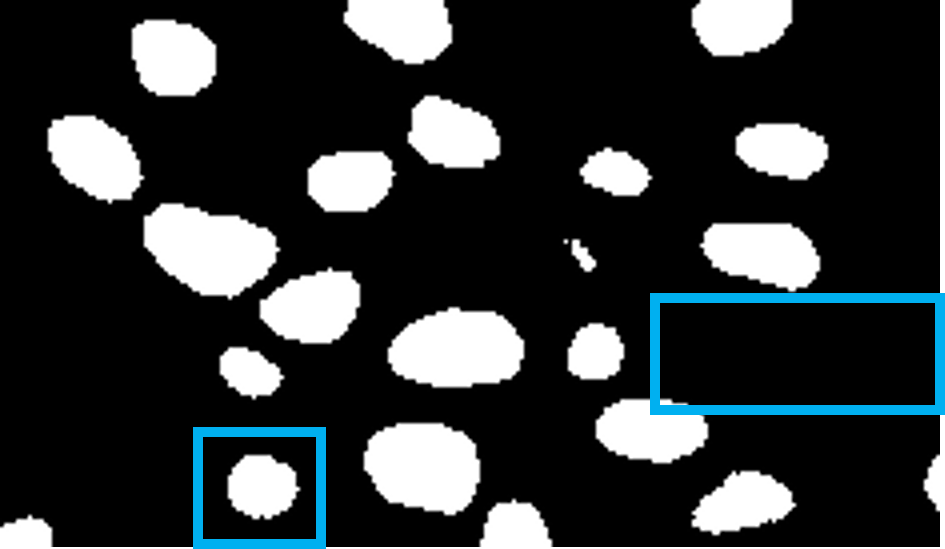}
    \end{subfigure} 
    \begin{subfigure}{0.115\textwidth}
        \includegraphics[width=\linewidth]{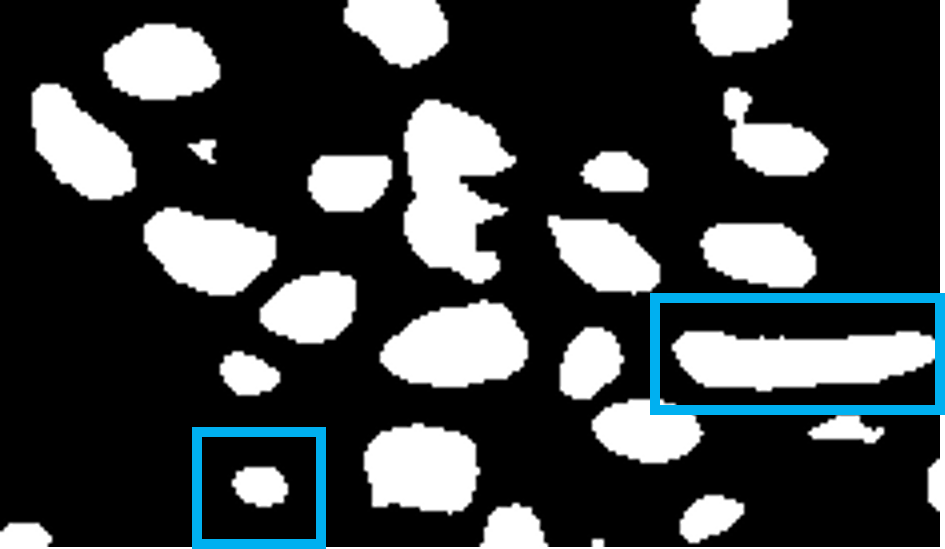}
    \end{subfigure} 
    \begin{subfigure}{0.115\textwidth}
        \includegraphics[width=\linewidth]{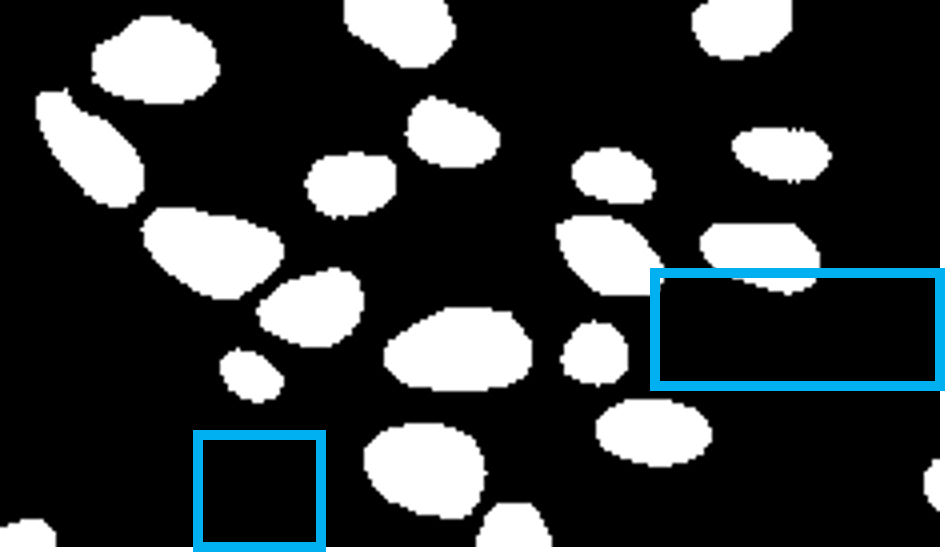}
    \end{subfigure} 

    \begin{subfigure}{0.115\textwidth}
        \includegraphics[width=\linewidth]{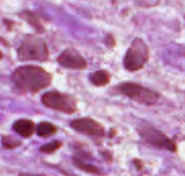}
        \caption{Image}
    \end{subfigure}
    \begin{subfigure}{0.115\textwidth}
        \includegraphics[width=\linewidth]{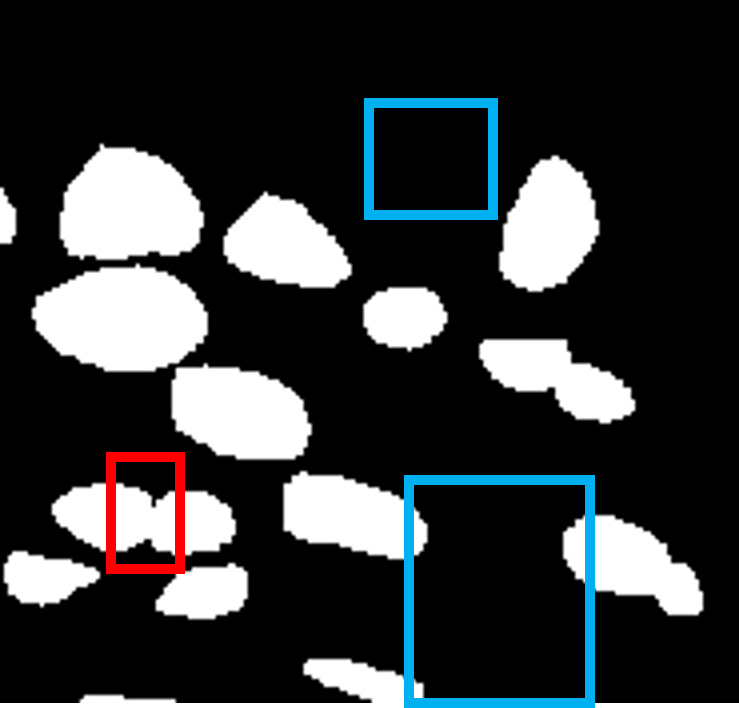}
        \caption{GT}
    \end{subfigure} 
    \begin{subfigure}{0.115\textwidth}
        \includegraphics[width=\linewidth]{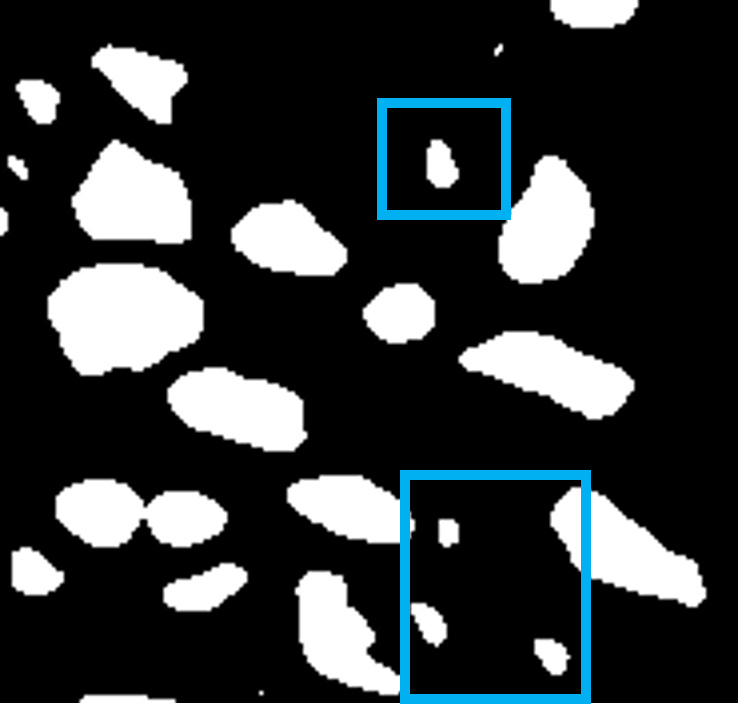}
        \caption{MT}
    \end{subfigure} 
    \begin{subfigure}{0.115\textwidth}
        \includegraphics[width=\linewidth]{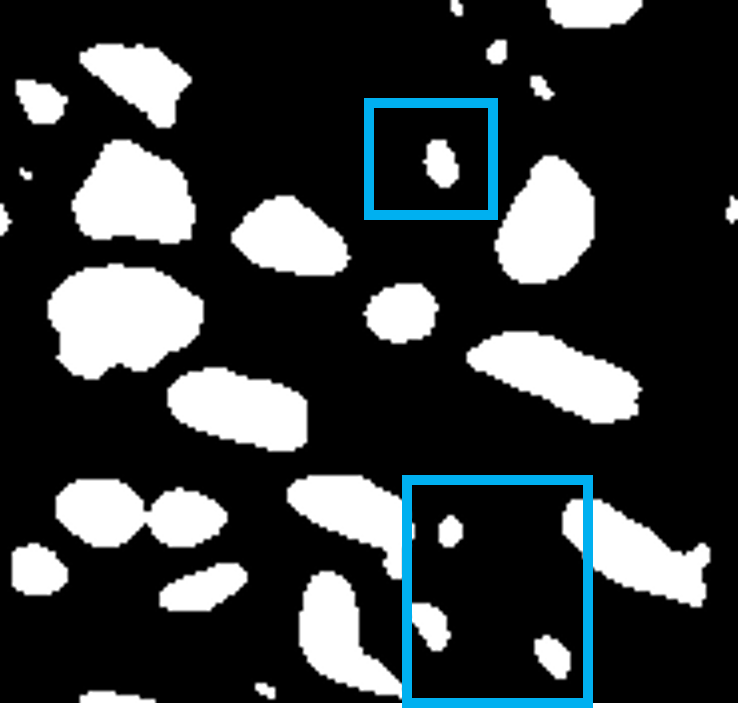}
        \caption{EM}
    \end{subfigure} 
    \begin{subfigure}{0.115\textwidth}
        \includegraphics[width=\linewidth]{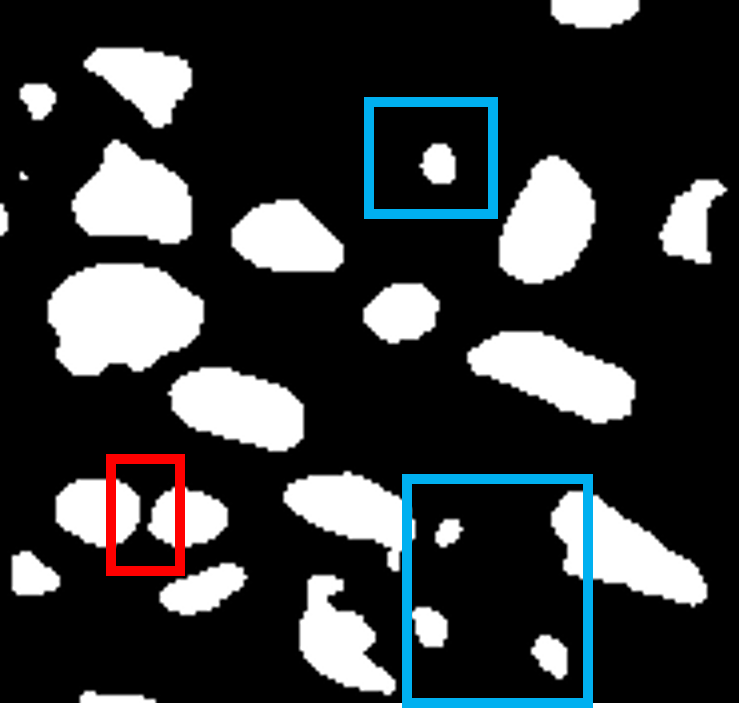}
        \caption{UAMT}
    \end{subfigure}
    \begin{subfigure}{0.115\textwidth}
        \includegraphics[width=\linewidth]{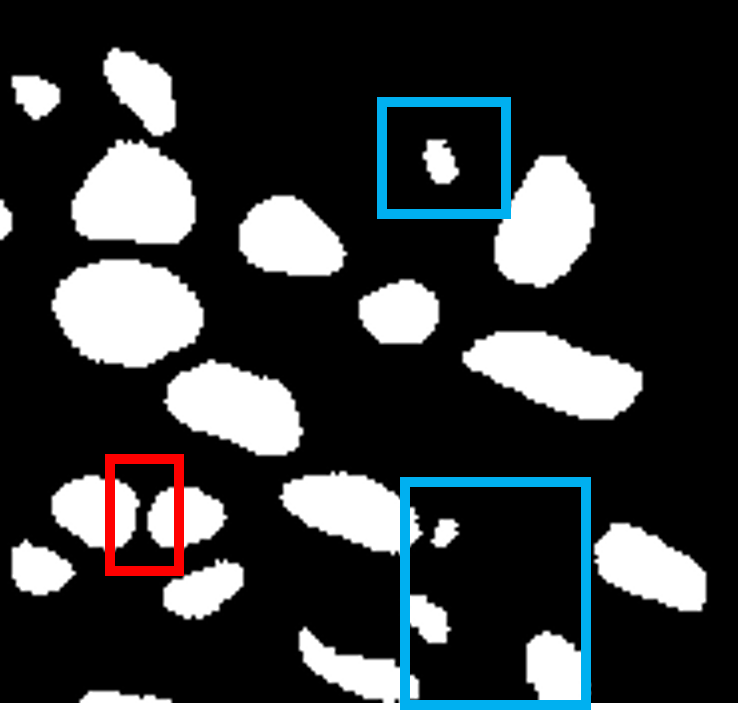}
        \caption{HCE}
    \end{subfigure} 
    \begin{subfigure}{0.115\textwidth}
        \includegraphics[width=\linewidth]{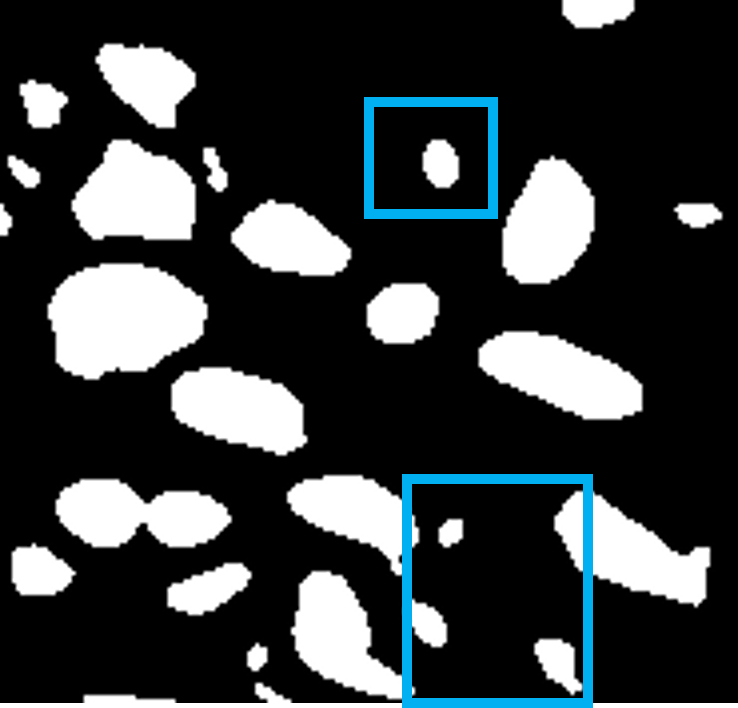}
        \caption{URPC}
    \end{subfigure} 
    \begin{subfigure}{0.115\textwidth}
        \includegraphics[width=\linewidth]{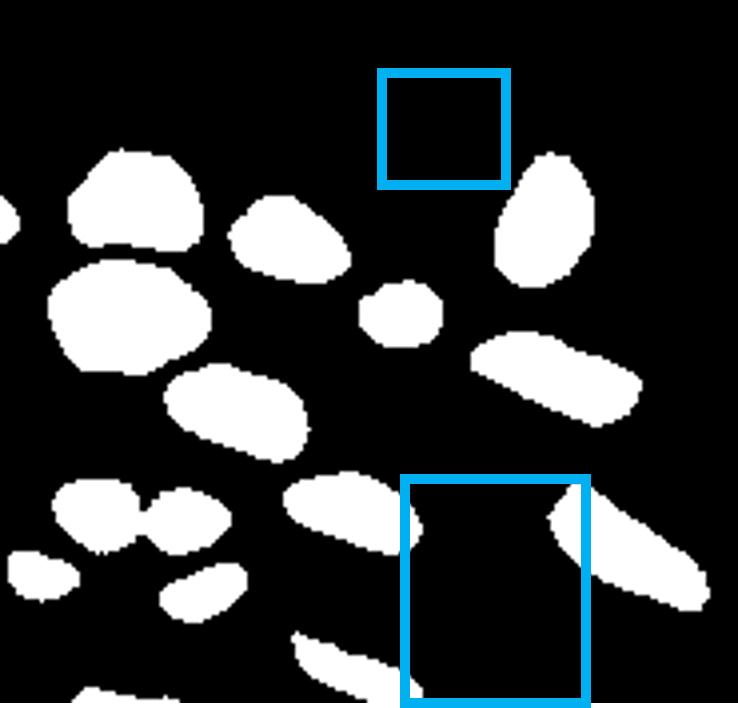}
        \caption{Ours}
    \end{subfigure}
    \caption{Additional qualitative results. The \textcolor{red}{\textbf{red}} boxes indicate the regions that are prone to topological errors such as incorrect merging or separating adjacent glands; the \textcolor{CornflowerBlue}{\textbf{blue}} boxes indicate false positive gland predictions or missing glands. 
    Rows 1-2: CRAG. Rows 3-4: GlaS. Rows 5-6: MoNuSeg.
    Zoom in for better views.
    }
    \label{fig:Qualitative_ResultsSup}
\end{figure*}

Here, we provide more qualitative results in~\cref{fig:Qualitative_ResultsSup} further to verify the effectiveness and superiority of our proposed method. 

\section{Additional Ablation Study}
\label{addi_ablation_study}
In this section, we conduct additional ablation studies to further demonstrate the effectiveness of our TopoSemiSeg.

\myparagraph{Ablation study on both $\mathcal{L}^{\text{S}}_{\text{topo}}$ and $\mathcal{L}^{\text{U}}_{\text{topo}}$.} To further validate the effectiveness of our method, we conduct additional experiments on $100\%$ and $20\%$ labeled data with supervised and unsupervised topological constraints. 
$\mathcal{L}^{\text{S}}_{\text{topo}}$ and $\mathcal{L}^{\text{U}}_{\text{topo}}$ resp. 
denote the topology-based loss for labeled and unlabeled data ($\lambda^S_{topo}$ and $\lambda^U_{topo}$ are resp. weights).
The results are shown in~\cref{table:ablation_both_topo}. In the top half of~\cref{table:ablation_both_topo}, most of the topology-wise performance has improved, with a slight loss in pixel-level performance. Also, with sufficient labeled data, only adding the $\mathcal{L}^{\text{S}}_{\text{topo}}$ performs better.
In the bottom half of~\cref{table:ablation_both_topo}, applying 
$\mathcal{L}^{\text{S}}_{\text{topo}}$ and $\mathcal{L}^{\text{U}}_{\text{topo}}$ simultaneously gives mixed results on the metrics, without any significant change in overall performance. It can also be seen that the method we proposed makes good use of unlabeled data.

\setlength{\tabcolsep}{7pt}
\begin{table}[]
\scriptsize
\centering
\caption{Ablation study on \textbf{$\mathcal{L}^{\text{S}}_{\text{topo}}$} and \textbf{$\mathcal{L}^{\text{U}}_{\text{topo}}$}.}
\begin{tabular}{cccccccc}
\hline
Labeled Ratio (\%) &  $\lambda^S_{topo}$ & $\lambda^U_{topo}$ & Dice\_Obj $\uparrow$& BE $\downarrow$& BME $\downarrow$& VOI $\downarrow$\\ \hline
100\% & 0 & 0 & \textbf{0.928} & 0.149 & 5.650 & 0.547 \\ \hline
100\% & 0.002 & 0 & 0.913 & \textbf{0.141} & \textbf{5.150} & \textbf{0.532} \\ \hline
100\% & 0 & 0.002 & 0.912 & 0.146 & 5.178 & 0.539 \\ \hline
100\% & 0.002 & 0.002 & 0.922 & 0.153 & 5.239 & 0.543 \\ \hline \hline
20\% & 0.001 & 0.001 & 0.893 & 0.218 & 12.850 & 0.727 \\ \hline
20\% & 0.002 & 0.002 & 0.895 & \textbf{0.189} & 8.725 & 0.723 \\ \hline
20\% & 0.005 & 0.005 & 0.876 & 0.246 & 10.825 & 0.787 \\ \hline
20\% & 0 & 0.002 & \textbf{0.898} & 0.226 & \textbf{8.575} & \textbf{0.709} \\ \hline
\end{tabular}
\label{table:ablation_both_topo}
\end{table}

\myparagraph{Ablation Study on EMA decay $\alpha$.} 
The EMA decay $\alpha$ plays an important role in the teacher-student framework where it provides a smoothing effect over the parameters of the model. A higher decay (closer to 1) gives more weight to the historical parameters, leading to a more stable representation of the student model's knowledge over time. However, too high EMA decay may result in the teacher model lagging too far behind the student model due to the rapidly changing or non-stationary environments, failing to capture the latest patterns of the data. So to verify the effectiveness of our selected $\alpha$, we conduct an ablation study on EMA decay. The results are shown in~\cref{table:ablation_ema_decay}. From the results we can see when $\alpha=0.999$, our model performs the best.

\setlength{\tabcolsep}{15pt}
\begin{table}[]
\centering
\scriptsize
\caption{Ablation study on EMA decay $\alpha$.}
\begin{tabular}{cccccc}
\hline
$\alpha$ & Dice\_Obj $\uparrow$&  BE $\downarrow$& BME $\downarrow$& VOI $\downarrow$\\ \cline{1-5} 
0.99 & 0.887 &  0.249 & 11.525 & 0.734 \\ \hline
0.999 & \textbf{0.898} &  \textbf{0.226} & \textbf{8.575} & \textbf{0.709} \\ \hline
0.9999 & 0.873 &  0.252 & 11.850 & 0.752 \\ \hline
\end{tabular}
\label{table:ablation_ema_decay}
\end{table}

\myparagraph{Ablation study on data augmentation.} 
Our method relies on the assumption that, for the model to be robust, its predictions should not change significantly for small perturbations of the input data in terms of topology. Hence, data augmentation and its hyper-parameter selections are crucial for our method. In~\cref{table:data_aug_hyperparam}, we report the results of the ablation study on data augmentations, and in the last row we also report using strong augmentations on labeled data. The \textit{italicized numbers} are our selected hyper-parameters. We conduct experiments on hyper-parameters of strong augmentations, specifically, brightness and contrast. We provide results on several combinations of hyper-parameter values.
The results indicate that our method is robust to the choice of data augmentations' hyper-parameters. 

\begin{table}[]
\scriptsize
\centering
\caption{Ablation study on data augmentations.}
\begin{tabular}{ccccccc}
\hline
Brightness & Contrast & Dice\_Obj $\uparrow$&  BE $\downarrow$& BME $\downarrow$& VOI $\downarrow$\\ \cline{1-6} 
\textit{0.3} & \textit{0.1} & 0.898 &  \textbf{0.226} & 8.575 & \textbf{0.709} \\ \hline
0.3 & 0.5 & 0.897 &  0.233 & \textbf{8.000} & 0.720 \\ \hline
0.1 & 0.1 & 0.887 &  0.255 & 11.550 & 0.736 \\ \hline
0.5 & 0.1 & \textbf{0.900} &  0.227 & 8.237 & 0.715 \\ \hline \hline
\multicolumn{2}{c}{strong aug. on labeled data} & 0.883 &  0.238 & 8.025 & 0.717 \\ \hline
\end{tabular}
\label{table:data_aug_hyperparam}
\end{table}

\myparagraph{Ablation study on labeled sampling bias \& Retain noise and remove signal.} 
Here, we conduct the experiments to alleviate the potential sampling bias and report the results in~\cref{table:addi_ablation_studies}.. On GlaS dataset, $20\%$ labeled samples do perform better than $10\%$. In addition, we provide the results that we retain the noise part and remove the signal part in the last $2$ rows of~\cref{table:addi_ablation_studies}. As expected, the performance drops significantly. Removing the signal dots causes the prediction to intentionally overlook the true structures while retaining the noisy dots causes it to include erroneous structures. 
This result, together with our ablation study (\cref{table:decomposition}), shows how our signal/noise decomposition helps the model learn even without GT annotation. 

\setlength{\tabcolsep}{20pt}
\begin{table}[]
\centering
\caption{The first $2$ rows: the results that rerun the experiments $5$ times with different labeled training samples on GlaS dataset. The last $2$ rows: the ablation study on retaining the noise and removing the signal.}
\begin{tabular}{ccc}
\hline
Method & Dice\_obj~$\uparrow$ & BME~$\downarrow$ \\ \hline
Ours (10\%) & 0.876$\pm$0.035 & 9.885$\pm$0.825 \\ \hline
Ours (20\%) & \textbf{0.893$\pm$0.007} & \textbf{9.384$\pm$0.479} \\ \hline \hline
Noise\cmark Signal \xmark &  0.866 & 21.325 \\ \hline
Ours & \textbf{0.898} & \textbf{8.575} \\ \hline
\end{tabular}
\label{table:addi_ablation_studies}
\end{table}

\myparagraph{Consistent Comparisons.}
To ensure a consistent comparison, we add the results of XNet~\cite{zhou2023xnet} for MoNuSeg dataset, CCT~\cite{ouali2020semi} for CRAG and GlaS dataset and FixMatch~\cite{sohn2020fixmatch} for CRAG dataset in~\cref{table:consistent_results}. Our method consistently outperforms these $3$ methods. Noted that FixMatch simply selects trustworthy pseudo-labels by thresholding the classifier's confidence. Many samples are discarded. Instead, we use all images, using persistence thresholding to select true topology signal to learn (with theoretical and empirical guarantees).

\setlength{\tabcolsep}{7pt}
\begin{table}[]
\caption{The results of XNet, CCT and FixMatch. }
\centering
\scriptsize
\begin{tabular}{ccccccc}
\hline
Dataset & Labeled Ratio (\%) & Method & Dice\_Obj $\uparrow$& BE $\downarrow$& BME $\downarrow$& VOI $\downarrow$ \\ \hline
\multicolumn{1}{c|}{\multirow{4}{*}{CRAG}} & \multicolumn{1}{c|}{\multirow{2}{*}{10\%}} & CCT & 0.853 & 1.954 & 40.210 & 0.864 \\
\multicolumn{1}{c|}{} & \multicolumn{1}{c|}{} & Ours & \textbf{0.884} & \textbf{0.227} & \textbf{10.475} & \textbf{0.758} \\ \cline{2-7} 
\multicolumn{1}{c|}{} & \multicolumn{1}{c|}{\multirow{2}{*}{20\%}} & CCT & 0.872 & 1.262 & 25.420 & 0.773 \\
\multicolumn{1}{c|}{} & \multicolumn{1}{c|}{} & FixMatch & 0.868 & 1.706 & 30.680 & 0.855 \\
\multicolumn{1}{c|}{} & \multicolumn{1}{c|}{} & Ours & \textbf{0.898} & \textbf{0.226} & \textbf{8.575} & \textbf{0.709} \\ \hline
\multicolumn{1}{c|}{\multirow{4}{*}{GlaS}} & \multicolumn{1}{c|}{\multirow{2}{*}{10\%}} & CCT & 0.864 & 0.862 & 16.645 & 0.932 \\
\multicolumn{1}{c|}{} & \multicolumn{1}{c|}{} & Ours & \textbf{0.878} & \textbf{0.551} & \textbf{8.300} & \textbf{0.811} \\ \cline{2-7} 
\multicolumn{1}{c|}{} & \multicolumn{1}{c|}{\multirow{2}{*}{20\%}} & CCT & 0.876 & 0.761 & 13.125 & 0.834 \\
\multicolumn{1}{c|}{} & \multicolumn{1}{c|}{} & Ours & \textbf{0.895} & \textbf{0.510} & \textbf{9.825} & \textbf{0.808} \\ \hline
\multicolumn{1}{c|}{\multirow{4}{*}{MoNuSeg}} & \multicolumn{1}{c|}{\multirow{2}{*}{10\%}} & XNet & 0.762 & 7.152 & 220.405 & 0.842 \\
\multicolumn{1}{c|}{} & \multicolumn{1}{c|}{} & Ours & \textbf{0.783} & \textbf{6.661} & \textbf{196.357} & \textbf{0.789} \\ \cline{2-7} 
\multicolumn{1}{c|}{} & \multicolumn{1}{c|}{\multirow{2}{*}{20\%}} & XNet & 0.776 & 6.750 & 198.525 & 0.831 \\
\multicolumn{1}{c|}{} & \multicolumn{1}{c|}{} & Ours & \textbf{0.793} &\textbf{ 4.250} & \textbf{188.642} & \textbf{0.787} \\ \hline
\end{tabular}
\label{table:consistent_results}
\end{table}

\myparagraph{Comparison to fully-sup. baselines.}
To better demonstrate that our method can effectively unearth the topological information from the unlabeled data, we make a comparison with two fully-supervised methods:~\cite{hu2019topology} and~\cite{clough2020topological}. We use these two losses only on $20\%$ labeled training data and report the results in~\cref{tab:comparison_with_fully}. Our TopoSemiSeg consistently outperforms these baselines because we utilize the topological information from the unlabeled data.

\setlength{\tabcolsep}{20pt}
\begin{table}[]
\caption{Comparison to fully-sup. baselines.}
\centering
\begin{tabular}{ccc}
\hline
Method & Dice\_obj $\uparrow$ & BME $\downarrow$\\ \hline
TopoLoss~\cite{hu2019topology} & 0.865 & 19.925 \\ \hline
TopoLoss~\cite{clough2020topological} & 0.857 & 24.625 \\ \hline
Ours & \textbf{0.898} & \textbf{8.575} \\ \hline
\end{tabular}
\label{tab:comparison_with_fully}
\end{table}

\begin{figure}[htbp]
\centering
    \includegraphics[width=1\linewidth]{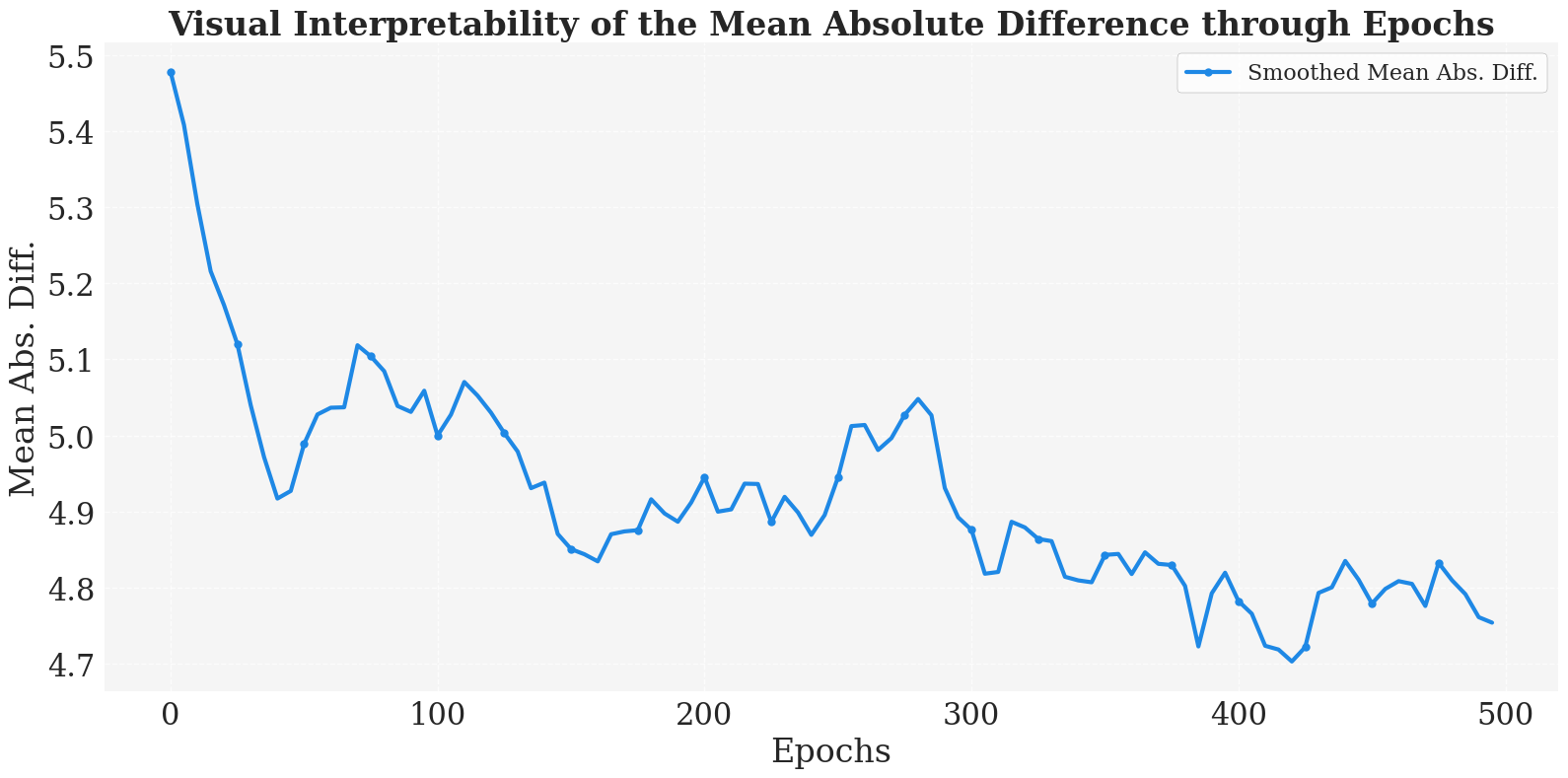}
    \caption{Visual interpretation of the decomposition strategy.}
    \label{fig:visual_exp}
\end{figure}

\myparagraph{Accuracy/guarantee of the decomposition strategy.}
Using a persistence threshold to filter out topological noise is theoretically supported. 
The stability theorem of persistent homology~\cite{cohen2005stability, cohen2010lipschitz} guarantees that topology due to small perturbation has small persistence. This is also demonstrated in~\cref{fig:illustration}. We observe that a proper persistence threshold ensures the model learns true structures and eliminates noise. 
To validate this, we compare the selected signal topology with the ground truth (GT) topology. On CRAG unlabeled training set, we compare the number of selected signal topology of the teacher with the Betti number of the GT. \cref{fig:visual_exp} shows the mean absolute difference between the two decreases during training. Thus, as training continues, the teacher's signal topology approaches GT's. This empirically shows that the decomposition picks up true topology signals, which the student learns from.

\end{document}